\PassOptionsToPackage{inline}{enumitem}%
\PassOptionsToPackage{capitalise}{cleveref}%
\documentclass[citecolor=red,linkcolor=blue]{lmcs}
\pdfoutput=1

\usepackage[utf8]{inputenc}
\usepackage{lastpage}
\lmcsdoi{19}{4}{28}
\lmcsheading{}{\pageref{LastPage}}{}{}%
{Dec.~18,~2021}{Dec.~18,~2023}{}

\pdfoutput=1%

\keywords{session types, $\pi$-calculus, functional programming, deadlock freedom, GV, CP}

\usepackage[english]{babel}

\usepackage[all]{foreign}
\usepackage{rotating}
\usepackage{pdflscape}
\setitemize{noitemsep,topsep=0pt,parsep=0pt,partopsep=0pt}
\usepackage{multicol}
\usepackage{amsmath,amssymb}
\usepackage[strict]{changepage}
\usepackage{tikz}
\usetikzlibrary{positioning}
\usepackage{mdframed}
\usepackage{float}
\let\:\undefined
\usepackage{centernot}
\usepackage{stmaryrd}
\usepackage{cmll}
\usepackage{mathtools}
\usepackage{environ}
\usepackage{tikz-cd}
\usetikzlibrary{arrows,automata,shapes,shapes.geometric}

\allowdisplaybreaks%

\usepackage{namespc}

\newenvironment{case*}[1][]{%
  \vspace{0.5\baselineskip}%
  \par%
  \noindent%
  \textbf{Case}~#1.\itshape\rmfamily}{\par}
\newenvironment{subcase*}[1][]{%
  \vspace{0.5\baselineskip}%
  \par%
  \textbf{Subcase}~#1.\itshape\rmfamily}{\par}

\usepackage{mathpartir}

\usepackage{xcolor}
\newcommand{\tm}[1]{\ensuremath{{\color[HTML]{a40038}#1}}}
\newcommand{\ty}[1]{\ensuremath{{\color[HTML]{00007a}#1}}}
\newcommand{\cs}[1]{\ensuremath{{\color[HTML]{009180}#1}}}
\newcommand{\tmty}[2]{\ensuremath{\tm{#1}:\ty{#2}}}

\newcommand{\sep}{\;\mid\;}
\newcommand{\emptyenv}{\varnothing}
\newcommand{\subst}[4][]{\ifstrempty{#1}{\ensuremath{#2\{#3/#4\}}}{\ensuremath{#2(\{#3/#4\}\cup#1)}}}
\newcommand{\plug}[2]{\ensuremath{#1[#2]}}
\newcommand{\minpr}[0]{\pr}
\providecommand{\pbot}[0]{\ensuremath{\cs{\bot\!}}}
\providecommand{\ptop}[0]{\ensuremath{\cs{\top\!}}}

\newcommand{\defeq}[0]{\triangleq}
\newcommand{\elabarrow}[0]{\triangleq}
\newcommand{\hole}[0]{\ensuremath{\square}}
\DeclareMathOperator{\pr}{pr}
\DeclareMathOperator{\fv}{fv}

\DeclareMathAlphabet{\rel}{OMS}{cmsy}{m}{n}
\DeclareMathAlphabet{\conf}{OMS}{cmsy}{m}{n}

\newcommand{\substarrow}[2]{\xlongrightarrow{\scriptscriptstyle\{\tm{#1}/\tm{#2}\}}}
\DeclarePairedDelimiter{\cpgvT}{\llparenthesis}{\rrparenthesis}
\DeclarePairedDelimiter{\cpgvM}{\llparenthesis}{\rrparenthesis_{\!\scriptscriptstyle{M}}}
\DeclarePairedDelimiter{\cpgvC}{\llparenthesis}{\rrparenthesis_{\!\scriptscriptstyle{\conf{C}}}}
\newcommand{\cpgvMarrow}[0]{\xlongrightarrow{\;\raisebox{2pt}{\ensuremath{{\cpgvM{\cdot}}}}\;}}
\newcommand{\cpgvCarrow}[0]{\xlongrightarrow{\;\raisebox{2pt}{\ensuremath{{\cpgvC{\cdot}}}}\;}}

\namespace*{pcp}{
  \providecommand{\tyone}[1][]{\ensuremath{\mathbf{1}^{#1}}}
  \providecommand{\tynil}[1][]{\ensuremath{\mathbf{0}^{#1}}}
  \providecommand{\tytop}[1][]{\ensuremath{\top^{#1}}}
  \providecommand{\tybot}[1][]{\ensuremath{\bot^{#1}}}
  \providecommand{\typlus}[3][]{\ensuremath{{#2}\mathbin{\oplus^{#1}}{#3}}}
  \providecommand{\tywith}[3][]{\ensuremath{{#2}\mathbin{\with^{#1}}{#3}}}
  \providecommand{\tytens}[3][]{\ensuremath{{#2}\mathbin{\otimes^{#1}}{#3}}}
  \providecommand{\typarr}[3][]{\ensuremath{{#2}\mathbin{\parr^{#1}}{#3}}}
  \providecommand{\co}[1]{\ensuremath{#1^\bot}}
  \providecommand{\res}[4][]{\ensuremath{(\nu{#2}^{#1}{#3}){#4}}}
  \providecommand{\ppar}[2]{\ensuremath{#1\parallel#2}}
  \providecommand{\halt}[0]{\ensuremath{\mathbf{0}}}
  \providecommand{\link}[3][]{\lablink[#1]{#2}{#3}}
  \providecommand{\send}[3]{\ensuremath{\labsend{#1}{#2}.#3}}
  \providecommand{\usend}[3]{\ensuremath{#1\langle{#2}\rangle.#3}}
  \providecommand{\recv}[3]{\ensuremath{\labrecv{#1}{#2}.#3}}
  \providecommand{\close}[2]{\ensuremath{\labclose{#1}.#2}}
  \providecommand{\wait}[2]{\ensuremath{\labwait{#1}.#2}}
  \providecommand{\inl}[2]{\ensuremath{\labselinl{#1}.{#2}}}
  \providecommand{\inr}[2]{\ensuremath{\labselinr{#1}.{#2}}}
  \providecommand{\offer}[3]{\ensuremath{{#1}\triangleright\{\text{inl}:#2;\text{inr}:#3\}}}
  \providecommand{\absurd}[1]{\ensuremath{{#1}\triangleright\{\}}}
  \providecommand{\seq}[2]{\ensuremath{\tm{#1}\vdash{#2}}}

  \providecommand{\red}[0]{\ensuremath{\Longrightarrow}}

  \providecommand{\lablink}[3][]{\ensuremath{{#2}{\leftrightarrow}^{#1}{#3}}}
  \providecommand{\labsend}[2]{\ensuremath{#1[#2]}}
  \providecommand{\labrecv}[2]{\ensuremath{#1(#2)}}
  \providecommand{\labclose}[1]{\labsend{#1}{}}
  \providecommand{\labwait}[1]{\labrecv{#1}{}}
  \providecommand{\labselinl}[1]{\ensuremath{{#1}\triangleleft\text{inl}}}
  \providecommand{\labselinr}[1]{\ensuremath{{#1}\triangleleft\text{inr}}}

  \providecommand{\sched}{\ensuremath{Sched}}
  \providecommand{\agent}[1]{\ensuremath{Agent_{#1}}}
  \providecommand{\proc}[1]{\ensuremath{Proc_{#1}}}
}{}
\newcommand{\pcp}[1]{\namespace*{pcp}{}{#1}}

\namespace*{pgv}{
  \providecommand{\tyunit}[0]{\ensuremath{\mathbf{1}}}
  \providecommand{\tyvoid}[0]{\ensuremath{\mathbf{0}}}
  \providecommand{\typrod}[2]{\ensuremath{{#1}\mathbin{\times}{#2}}}
  \providecommand{\tysum}[2]{\ensuremath{{#1}\mathbin{+}{#2}}}
  \providecommand{\tylolli}[3][]{\ensuremath{{#2}\mathbin{\multimap^{#1}}{#3}}}
  \providecommand{\co}[1]{\ensuremath{\overline{#1}}}
  \providecommand{\tysend}[3][]{\ensuremath{!^{#1}{#2}.{#3}}}
  \providecommand{\tyrecv}[3][]{\ensuremath{?^{#1}{#2}.{#3}}}
  \providecommand{\tyends}[1][]{\ensuremath{\mathbf{end}_{!}^{#1}}}
  \providecommand{\tyendr}[1][]{\ensuremath{\mathbf{end}_{?}^{#1}}}
  \providecommand{\tyselect}[3][]{\ensuremath{{#2}\mathbin{\oplus^{#1}}{#3}}}
  \providecommand{\tyoffer}[3][]{\ensuremath{{#2}\mathbin{\with^{#1}}{#3}}}
  \providecommand{\tyselectemp}[1][]{\ensuremath{{\oplus}^{#1}\{\}}}
  \providecommand{\tyofferemp}[1][]{\ensuremath{{\with}^{#1}\{\}}}
  \providecommand{\andthen}[2]{\ensuremath{#1;#2}}
  \providecommand{\letbind}[3]{\ensuremath{\mathbf{let}\;#1\mathbin{=}#2\;\mathbf{in}\;#3}}
  \providecommand{\pair}[2]{\ensuremath{(#1,#2)}}
  \providecommand{\letpair}[4]{\ensuremath{\letbind{\pair{#1}{#2}}{#3}{#4}}}
  \providecommand{\labinl}[0]{\ensuremath{\mathbf{inl}}}
  \providecommand{\labinr}[0]{\ensuremath{\mathbf{inr}}}
  \providecommand{\inl}[1]{\ensuremath{\labinl\;#1}}
  \providecommand{\inr}[1]{\ensuremath{\labinr\;#1}}
  \providecommand{\casesum}[5]{\ensuremath{\mathbf{case}\;#1\;\left\{\inl{#2}\mapsto{#3};\;\inr{#4}\mapsto{#5}\right\}}}
  \providecommand{\unit}[0]{\ensuremath{()}}
  \providecommand{\letunit}[2]{\andthen{#1}{#2}}
  \providecommand{\absurd}[1]{\ensuremath{\mathbf{absurd}\;#1}}
  \providecommand{\link}[0]{\ensuremath{\mathbf{link}}}
  \providecommand{\new}[0]{\ensuremath{\mathbf{new}}}
  \providecommand{\halt}[0]{\ensuremath{\mathbf{halt}}}
  \providecommand{\spawn}[0]{\ensuremath{\mathbf{spawn}}}
  \providecommand{\send}[0]{\ensuremath{\mathbf{send}}}
  \providecommand{\recv}[0]{\ensuremath{\mathbf{recv}}}
  \providecommand{\fork}[0]{\ensuremath{\mathbf{fork}}}
  \providecommand{\wait}[0]{\ensuremath{\mathbf{wait}}}
  \providecommand{\close}[0]{\ensuremath{\mathbf{close}}}
  \providecommand{\select}[1]{\ensuremath{\mathbf{select}\;#1}}
  \providecommand{\offer}[5]{\ensuremath{\mathbf{offer}\;#1\;\{\inl{#2}\mapsto{#3};\inr{#4}\mapsto{#5}\}}}
  \providecommand{\offeremp}[1]{\ensuremath{\mathbf{offer}\;#1\;\{\}}}
  \providecommand{\ppar}[2]{\ensuremath{#1\parallel#2}}
  \providecommand{\res}[3]{\ensuremath{(\nu#1#2)#3}}
  \providecommand{\main}[0]{\ensuremath{\bullet}}
  \providecommand{\child}[0]{\ensuremath{\circ}}
  \providecommand{\tseq}[4][]{\ensuremath{#2\vdash^{\cs{#1}}\tmty{#3}{#4}}}
  \providecommand{\cseq}[3][]{\ensuremath{#2\vdash^{\tm{#1}}\tm{#3}}}
  \providecommand{\tred}[0]{\ensuremath{\longrightarrow_{M}}}
  \providecommand{\cred}[0]{\ensuremath{\longrightarrow_{\conf{C}}}}
  \providecommand{\notcred}[0]{\ensuremath{\mathbin{\centernot{\longrightarrow}_{\!\!\conf{C}}}}}
  \newcommand{\sched}{\ensuremath{\mathbf{sched}}}
  \newcommand{\agent}[1]{\ensuremath{\mathbf{agent}_{#1}}}
  \newcommand{\proc}[1]{\ensuremath{\mathbf{proc}_{#1}}}
  \newcommand{\echo}[0]{\ensuremath{\mathbf{echo}}}
}{}
\newcommand{\pgv}[1]{\namespace*{pgv}{}{#1}}

\usepackage{varioref}
\usepackage[colorlinks]{hyperref}
\usepackage{cleveref}
\usepackage{doi}
\usepackage[strings]{underscore}

\newcommand{\renewtheorem}[1]{%
  \expandafter\let\csname #1\endcsname\relax
  \expandafter\let\csname c@#1\endcsname\relax
  \expandafter\let\csname end#1\endcsname\relax
  \newtheorem{#1}%
}

\theoremstyle{plain}

\renewtheorem{thm}{Theorem}[section]
\renewtheorem{cor}[thm]{Corollary}
\renewtheorem{lem}[thm]{Lemma}
\renewtheorem{slem}[thm]{Sublemma}
\renewtheorem{prop}[thm]{Proposition}
\renewtheorem{asm}[thm]{Assumption}

\theoremstyle{definition}

\renewtheorem{rem}[thm]{Remark}
\renewtheorem{rems}[thm]{Remarks}
\renewtheorem{exa}[thm]{Example}
\renewtheorem{exas}[thm]{Examples}
\renewtheorem{defi}[thm]{Definition}
\renewtheorem{conv}[thm]{Convention}
\renewtheorem{conj}[thm]{Conjecture}
\renewtheorem{prob}[thm]{Problem}
\renewtheorem{oprob}[thm]{Open Problem}
\renewtheorem{oprobs}[thm]{Open Problems}
\renewtheorem{algo}[thm]{Algorithm}
\renewtheorem{obs}[thm]{Observation}
\renewtheorem{desc}[thm]{Description}
\renewtheorem{fact}[thm]{Fact}
\renewtheorem{qu}[thm]{Question}
\renewtheorem{oqu}[thm]{Open Question}
\renewtheorem{pty}[thm]{Property}
\renewtheorem{clm}[thm]{Claim}
\renewtheorem{nota}[thm]{Notation}
\renewtheorem{com}[thm]{Comment}
\renewtheorem{coms}[thm]{Comments}

\crefname{section}{\S.\@\!}{Sections}%
\crefname{subsection}{\S\!\!\,}{\S\!\!\,}%
\crefname{subsubsection}{\S\!\!\,}{\S\!\!\,}%

\begin{document}
\title{Prioritise the Best Variation}
\thanks{Supported by the EU HORIZON 2020 MSCA RISE project 778233 ``Behavioural Application Program Interfaces'' (BehAPI)}
%
\author[W.~Kokke]{Wen Kokke\lmcsorcid{0000-0002-1662-0381}}
\address{University of Strathclyde, Glasgow, UK}
\email{wen.kokke@strath.ac.uk}

\author[O.~Dardha]{Ornela Dardha\lmcsorcid{0000-0001-9927-7875}}
\address{School of Computing Science, University of Glasgow, UK}	
\email{ornela.dardha@glasgow.ac.uk}  

\begin{abstract}
  Binary session types guarantee communication safety and session fidelity, but \emph{alone} they cannot rule out deadlocks arising from the interleaving of {different} sessions.

  In Classical Processes (CP)~\cite{wadler14}---a process calculus based on classical linear logic---deadlock freedom is guaranteed by combining channel creation and parallel composition under the same logical cut rule. Similarly, in Good Variation (GV)~\cite{wadler15,lindleymorris15}---a linear concurrent $\lambda$-calculus---deadlock freedom is guaranteed by combining channel creation and thread spawning under the same operation, called fork.

  In both CP and GV, deadlock freedom is achieved at the expense of expressivity, as the only processes allowed are tree-structured. This is true more broadly than for CP and GV, and it holds for all works in the research line based on Curry-Howard correspondences between linear logic and session types, starting with Caires and Pfenning \cite{cairespfenning10}.
  To overcome the limitations of tree-structured processes, Dardha and Gay~\cite{dardhagay18} define Priority CP (PCP), which allows cyclic-structured processes and restores deadlock freedom by using \emph{priorities}, in line with Kobayashi and Padovani~\cite{kobayashi06,padovani14}.

  Following PCP, we present Priority GV (PGV), a~variant of GV which decouples channel creation from thread spawning. Consequently, we type cyclic-structured processes and restore deadlock freedom by using priorities. We show that our type system is sound by proving subject reduction and progress. We define an encoding from PCP to PGV and prove that the encoding preserves typing and is sound and complete with respect to the operational semantics.
\end{abstract}
\maketitle

{\section{Introduction}
Session types~\cite{honda93,takeuchihonda94,hondavasconcelos98} are a type formalism that ensures \emph{communication channels} are used according to their protocols, much like, \eg, data types ensure that functions are used according to their signature. Session types have been studied in many settings. Most notably, they have been defined for the $\pi$-calculus~\cite{honda93,takeuchihonda94,hondavasconcelos98}, a foundational calculus for communication and concurrency, and the concurrent $\lambda$-calculi~\cite{gayvasconcelos10}, including the main focus of our paper: Good Variation~\cite[GV]{wadler15,lindleymorris15}.

GV is a concurrent $\lambda$-calculus with \emph{binary} session types, where each channel is shared between exactly two processes. Binary session types guarantee two crucial properties: \emph{communication safety}---\eg, if the protocol says to transmit an integer, you transmit an integer---and \emph{session fidelity}---\eg, if the protocol says send, you send. A third crucial property is \emph{deadlock freedom}, which ensures that processes do not have cyclic dependencies---\eg, when two processes wait for each other to send a value. Binary session types {alone} are insufficient to rule out deadlocks arising from interleaved sessions, but several additional techniques have been developed to guarantee deadlock freedom in session-typed $\pi$-calculus and concurrent $\lambda$-calculus.

In the $\pi$-calculus literature, there have been several developments of Curry-Howard correspondences between session-typed $\pi$-calculus and linear logic~\cite{girard87}: Caires and Pfenning's $\pi$DILL~\cite{cairespfenning10} corresponds to dual intuitionistic linear logic~\cite{barber96}, and Wadler's Classical Processes~\cite[CP]{wadler14} corresponds to classical linear logic~\cite[CLL]{girard87}. Both calculi guarantee deadlock freedom, which they achieve by the combination of binary session types with a restriction on the way processes can be connected by channels: two processes can share at most one channel, and, more generally, information transmitted between any two processes must pass through one unique series of channels and intermediate processes. We refer to such processes as \emph{tree-structured}, because the \emph{communication graph}---where the vertices are ready processes and two vertices are connected by an edge if and only if the corresponding processes share a channel---is a tree. For $\pi$DILL, CP, and GV, tree-structure follows from a syntactic restriction: the combination of name restriction and parallel composition into a single syntactic construct, corresponding to the logical cut.

There are many downsides to combining name restriction and parallel composition, such as lack of modularity, difficulty typing structural congruence and formulating label-transition semantics. GV, specifically, struggles with a complicated metatheory due to the mismatch between its term language---where restriction and parallel composition are combined---and its configuration language---where they are not. There have been various approaches to decoupling restriction and parallel composition.
Hypersequent CP~\cite[HCP]{MP18,kokkemontesi19popl,kokkemontesi19tlla}, Hypersequent GV~\cite{fowleretal21}, and Linear Compositional Choreographies~\cite{CarboneMS18} decouple them, but maintain the tree-structure of processes and a correspondence to linear logic, \eg, while the typing rules for HCP are no longer exactly the proof rules for CLL, every typing derivations in HCP is isomorphic a proof in CLL, and vice versa.
Priority CP~\cite[PCP]{dardhagay18extended} weakens the correspondence to CLL. PCP is a non-conservative extension of CLL: every proof in CLL can be translated to a typing derivation in PCP~\cite{DardhaP22}, but PCP can prove strictly more theorems, including (partial) bijections between several CLL connectives. In exchange, PCP has a much more expressive language which allows cyclic-structured processes. PCP decouples CP's cut rule into two separate constructs: one for parallel composition via a mix rule, and one for name restriction via a cycle rule. To restore deadlock freedom, PCP uses \emph{priorities}~\cite{kobayashi06,padovani14}. Priorities encode the \emph{order of actions} and rule out bad cyclic interleavings. Dardha and Gay~\cite{dardhagay18extended} prove cycle-elimination for PCP, adapting the cut-elimination proof for classical linear logic, and deadlock freedom follows as a corollary.

CP and GV are related via a pair of translations which satisfy simulation~\cite{lindleymorris16}, and which can be tweaked to satisfy operational correspondence. The two calculi share the same strong guarantees. GV achieves deadlock freedom via a similar syntactic restriction: it combines channel creation and thread spawning into a single operation, called ``fork'', which is related to the cut construct in CP. Unfortunately, as with CP, this syntactic restriction has its downsides.

Our aim is to develop a more expressive version of GV while maintaining deadlock freedom. While process calculi have their advantages, \eg, their succinctness compared to concurrent $\lambda$-calculi, we chose to work with GV for several reasons. In general, concurrent $\lambda$-calculi support higher-order functions, and have a capability for abstraction not usually present in process calculi. Within a concurrent $\lambda$-calculus, one can derive extensions of the communication capabilities of the language via well-understood extensions of the functional fragment, \eg, we can derive internal/external choice from sum types. Concurrent $\lambda$-calculi, among other languages, maintain a clear separation between the program which the user writes and the configurations which represent the state of the system as it evaluates the program. However, our main motivation is that results obtained for $\lambda$-calculi transfer more easily to real-world functional programming languages. Case in point: we easily adapted the type system of PGV to Linear Haskell~\cite{bernardyboespflug18}, which gives us a library for deadlock-free session-typed programming~\cite{kokkedardha21hs}.
The benefit of working specifically with GV, as opposed to other concurrent $\lambda$-calculi, is its relation to CP~\cite{wadler14}, and its formal properties, including deadlock freedom.

We thus pose our research question for GV:

\vspace{1em}
\textbf{RQ:}
\emph{Can we design a more expressive GV which guarantees deadlock freedom for cyclic-structured processes?}
\vspace{1em}

We follow the line of work from CP to Priority CP, and present Priority GV (PGV), a~variant of GV which decouples channel creation from thread spawning, thus allowing cyclic-structured processes, but which nonetheless guarantees deadlock freedom via priorities. This closes the circle of the connection between CP and GV~\cite{wadler14}, and their priority-based versions, PCP~\cite{dardhagay18extended} and PGV.
We cannot straightforwardly adapt the priority typing from PCP to PGV, as PGV adds higher-order functions. Instead, the priority typing for PGV follow the work by Padovani and Novara~\cite{padovaninovara15}.

We make the following main contributions:
\begin{enumerate}
  \item \textbf{Priority GV}. We present Priority GV (\cref{sec:pgv}, PGV), a session-typed functional language with priorities, and prove \emph{subject reduction} (\cref{thm:pgv-subject-reduction-confs}) and \emph{progress} (\cref{thm:pgv-closed-progress-confs}).

        We addresses several problems in the original GV language, most notably:
        \begin{enumerate}
          \item PGV does not require the pseudo-type $S^\sharp$;
          \item Structural congruence is type preserving.
        \end{enumerate}
        PGV answers our research question positively as it allows cyclic-structured binary session-typed processes that are deadlock free.
  \item \textbf{Translation from PCP to PGV}.
        We present a \emph{sound and complete encoding} of PCP~\cite{dardhagay18extended} in PGV (\cref{sec:pcp}). We prove the encoding preserves typing (\cref{thm:pcp-to-pgv-confs-preservation}) and satisfies operational correspondence (\cref{thm:pcp-to-pgv-operational-correspondence-soundness} and \cref{thm:pcp-to-pgv-operational-correspondence-completeness}).

        To obtain a tight correspondence, we update PCP, moving away from commuting conversions and reduction as cut elimination towards reduction based on structural congruence, as it is standard in process calculi.
\end{enumerate}

This paper is an improved and extended version of a paper published at FORTE 2021 international conference \cite{kokkedardha21pgv}. We present detailed examples and complete proofs of our technical results.
}
{\section{Priority GV}\label{sec:pgv}
\usingnamespace{pgv}

We present Priority GV (PGV), a session-typed functional language based on GV~\cite{wadler15,lindleymorris15} which uses priorities \`{a} la Kobayashi and Padovani~\cite{kobayashi06,padovaninovara15} to enforce deadlock freedom.
Priority GV is more flexible than GV because it allows processes to share more than one communication channel.

We illustrate this with two programs in PGV, \cref{ex:cycle} and \cref{ex:deadlock}. Each program contains two processes---the main process, and the child process created by $\tm{\spawn}$---which communicate using \emph{two} channels. The child process receives a unit over the channel $\tm{x}/\tm{x'}$, and then sends a unit over the channel $\tm{y}/\tm{y'}$. The main process does one of two things:
\begin{enumerate}[label= (\alph*) ]
  \item in \cref{ex:cycle}, it sends a unit over the channel $\tm{x}/\tm{x'}$, and then waits to receive a unit over the channel $\tm{y}/\tm{y'}$;
  \item in \cref{ex:deadlock}, it does these in the opposite order, which results in a deadlock.
\end{enumerate}
PGV is more expressive than GV: \cref{ex:cycle} is typeable and guaranteed to be deadlock-free in PGV, but is not typeable in GV~\cite{wadler14} and not guaranteed deadlock-free in GV's predecessor~\cite{gayvasconcelos10}. We believe PGV is a non-conservative \emph{extension} of GV, as CP can be embedded in a Kobayashi-style system~\cite{dardhaperez15extended}.

\begin{figure}
  \begin{minipage}{0.5\linewidth}
    \begin{exa}\label{ex:cycle}
      \[\tm{%
          \begin{array}{l}
            \letpair{x}{x'}{\new\;\unit}{}
            \\
            \letpair{y}{y'}{\new\;\unit}{}
            \\
            \spawn\left(\lambda\unit.%
            \begin{array}{l}
                {\letpair{\unit}{x'}{\recv\;{x'}}{}}
                \\
                {\letbind{y}{\send\;{\pair{\unit}{y}}}{}}
                \\
                {\andthen{\wait\;{x'}}{\close\;{y}}}
              \end{array}%
            \right);
            \\
            \underline{\letbind{x}{\send\;{\pair{\unit}{x}}}{}}
            \\
            \underline{\letpair{\unit}{y'}{\recv\;{y'}}{}}
            \\
            {\andthen{\close\;{x}}{\wait\;{y'}}}
          \end{array}
        }\]
    \end{exa}
  \end{minipage}%
  \begin{minipage}{0.5\linewidth}
    \begin{exa}\label{ex:deadlock}
      \[\tm{%
          \begin{array}{l}
            \letpair{x}{x'}{\new\;\unit}{}
            \\
            \letpair{y}{y'}{\new\;\unit}{}
            \\
            \spawn\left(\lambda\unit.%
            \begin{array}{l}
                {\letpair{\unit}{x'}{\recv\;{x'}}{}}
                \\
                {\letbind{y}{\send\;{\pair{\unit}{y}}}{}}
                \\
                {\andthen{\wait\;{x'}}{\close\;{y}}}
              \end{array}%
            \right);
            \\
            \underline{\letpair{\unit}{y'}{\recv\;{y'}}{}}
            \\
            \underline{\letbind{x}{\send\;{\pair{\unit}{x}}}{}}
            \\
            {\andthen{\close\;{x}}{\wait\;{y'}}}
          \end{array}
        }\]
    \end{exa}
  \end{minipage}
\end{figure}

\subsection{Syntax of Types and Terms}
\subsubsection*{Session types}
Session types ($\ty{S}$) are defined by the following grammar:
\[
  \begin{array}{lcl}
    \ty{S}
     & \Coloneqq & \ty{\tysend[\cs{o}]{T}{S}}
    \sep        \ty{\tyrecv[\cs{o}]{T}{S}}
    \sep        \ty{\tyends[\cs{o}]}
    \sep        \ty{\tyendr[\cs{o}]}
  \end{array}
\]

Session types $\ty{\tysend[\cs{o}]{T}{S}}$ and $\ty{\tyrecv[\cs{o}]{T}{S}}$ describe the endpoints of a channel over which we send or receive a value of type $\ty{T}$, and then proceed as $\ty{S}$. Types $\ty{\tyends[\cs{o}]}$ and $\ty{\tyendr[\cs{o}]}$ describe endpoints of a channel whose communication has finished, and over which we must synchronise before closing the channel. Each connective in a session type is annotated with a \emph{priority} $\cs{o}\in\mathbb{N}$.

\subsubsection*{Types}
Types ($\ty{T}$, $\ty{U}$) are defined by the following grammar:
\[
  \begin{array}{lcl}
    \ty{T}, \ty{U}
     & \Coloneqq & \ty{\typrod{T}{U}}
    \sep        \ty{\tyunit}
    \sep        \ty{\tysum{T}{U}}
    \sep        \ty{\tyvoid}
    \sep        \ty{\tylolli[\cs{p},\cs{q}]{T}{U}}
    \sep        \ty{S}
  \end{array}
\]

Types $\ty{\typrod{T}{U}}$, $\ty{\tyunit}$, $\ty{\tysum{T}{U}}$, and $\ty{\tyvoid}$ are the standard linear $\lambda$-calculus product type, unit type, sum type, and empty type.
Type $\ty{\tylolli[\cs{p},\cs{q}]{T}{U}}$ is the standard linear function type, annotated with \emph{priority bounds} $\cs{p},\cs{q}\in\mathbb{N}\cup\{\cs{\pbot},\cs{\ptop}\}$.
Every session type is also a type.
Given a function with type $\ty{\tylolli[\cs{p},\cs{q}]{T}{U}}$, $\cs{p}$ is a \emph{lower bound} on the priorities of the endpoints captured by the body of the function, and $\cs{q}$ is an \emph{upper bound} on the priority of the communications that take place as a result of applying the function. The type of {pure functions} $\ty{\tylolli{T}{U}}$, \ie those which perform no communications, is syntactic sugar for $\ty{\tylolli[\cs{\ptop},\cs{\pbot}]{T}{U}}$. The lower bound for a pure function is $\cs{\ptop}$ as pure functions never start communicating. For similar reasons, the upper bound for a pure function is $\cs{\pbot}$.

We postulate that the only function types---and, consequently, sequents---that are inhabited in PGV are pure functions and functions $\ty{\tylolli[\cs{p},\cs{q}]{T}{U}}$ for which $\cs{p}<\cs{q}$.

\subsubsection*{Typing Environments}
Typing environments $\ty{\Gamma}$, $\ty{\Delta}$, $\ty{\Theta}$ associate types to names. Environments are linear, so two environments can only be combined as $\ty{\Gamma}, \ty{\Delta}$ if their names are distinct, \ie $\fv(\ty{\Gamma})\cap\fv(\ty{\Delta})=\varnothing$.
\[
  \begin{array}{lcl}
    \ty{\Gamma}, \ty{\Delta}
     & \Coloneqq & \ty{\emptyenv}
    \sep        \ty{\Gamma}, \tmty{x}{T}
  \end{array}
\]

\subsubsection*{Type Duality}
Duality plays a crucial role in session types. The two endpoints of a channel are assigned dual types, ensuring that, for instance, whenever one program {sends} a value on a channel, the program on the other end is waiting to {receive}. Each session type $\ty{S}$ has a dual, written $\ty{\co{S}}$. Duality is an involutive function which {preserves priorities}:
\[
  \ty{\co{\tysend[\cs{o}]{T}{S}}} = \ty{\tyrecv[\cs{o}]{T}{\co{S}}}
  \qquad
  \ty{\co{\tyrecv[\cs{o}]{T}{S}}} = \ty{\tysend[\cs{o}]{T}{\co{S}}}
  \qquad
  \ty{\co{\tyends[\cs{o}]}} = \ty{\tyendr[\cs{o}]}
  \qquad
  \ty{\co{\tyendr[\cs{o}]}} = \ty{\tyends[\cs{o}]}
\]

\subsubsection*{Priorities}
Function $\pr(\cdot)$ returns the smallest priority of a session type. The type system guarantees that the top-most connective always holds the smallest priority, so we simply return the priority of the top-most connective:
\[
  \pr(\ty{\tysend[\cs{o}]{T}{S}}) = \cs{o}
  \qquad
  \pr(\ty{\tyrecv[\cs{o}]{T}{S}}) = \cs{o}
  \qquad
  \pr(\ty{\tyends[\cs{o}]})       = \cs{o}
  \qquad
  \pr(\ty{\tyendr[\cs{o}]})       = \cs{o}
\]

We extend the function $\pr(\cdot)$ to types and typing contexts by returning the smallest priority in the type or context, or $\cs{\top}$ if there is no priority. We use $\sqcap$ and $\sqcup$ to denote the minimum and maximum, respectively:
\[
  \begin{array}{lcl}
    \minpr(\ty{\typrod{T}{U}})                 & = & \minpr({\ty{T}})\sqcap\minpr({\ty{U}})  \\
    \minpr(\ty{\tysum{T}{U}})                  & = & \minpr({\ty{T}})\sqcap\minpr({\ty{U}})  \\
    \minpr(\ty{\tylolli[\cs{p},\cs{q}]{T}{U}}) & = & \cs{p}                                  \\
    \minpr(\ty{\Gamma}, \tmty{x}{A})           & = & \minpr(\ty{\Gamma})\sqcap\minpr(\ty{A})
  \end{array}
  \qquad
  \begin{array}{lcl}
    \minpr(\ty{\tyunit})   & = & \ptop       \\
    \minpr(\ty{\tyvoid})   & = & \ptop       \\
    \minpr(\ty{S})         & = & \pr(\ty{S}) \\
    \minpr(\ty{\emptyenv}) & = & \ptop
  \end{array}
\]

\subsubsection*{Terms}
Terms ($\tm{L}$, $\tm{M}$, $\tm{N}$) are defined by the following grammar:
\[
  \begin{array}{lcl}
    \tm{L}, \tm{M}, \tm{N}
     & \Coloneqq & \tm{x}
    \sep        \tm{K}
    \sep        \tm{\lambda x.M}
    \sep        \tm{M\;N}                 \\
     & \sep      & \tm{\unit}
    \sep        \tm{\andthen{M}{N}}
    \sep        \tm{\pair{M}{N}}
    \sep        \tm{\letpair{x}{y}{M}{N}} \\
     & \sep      & \tm{\inl{M}}
    \sep        \tm{\inr{M}}
    \sep        \tm{\casesum{L}{x}{M}{y}{N}}
    \sep        \tm{\absurd{M}}           \\
    \tm{K}
     & \Coloneqq & \tm{\link}
    \sep        \tm{\new}
    \sep        \tm{\spawn}
    \sep        \tm{\send}
    \sep        \tm{\recv}
    \sep        \tm{\close}
    \sep        \tm{\wait}
  \end{array}
\]

Let $\tm{x}$, $\tm{y}$, $\tm{z}$, and $\tm{w}$ range over variable names. Occasionally, we use $\tm{a}$, $\tm{b}$, $\tm{c}$, and $\tm{d}$. The term language is the standard linear $\lambda$-calculus with products, sums, and their units, extended with constants $\tm{K}$ for the communication primitives.

Constants are best understood in conjunction with their typing and reduction rules in~\cref{fig:pgv-typing,fig:pgv-operational-semantics}.

Briefly, $\tm{\link}$ links two endpoints together, forwarding messages from one to the other, $\tm{\new}$ creates a new channel and returns a pair of its endpoints, and $\tm{\spawn}$ spawns off its argument as a new thread.

The $\tm{\send}$ and $\tm{\recv}$ functions send and receive values on a channel.
However, since the typing rules for PGV ensure the linear usage of endpoints, they also return a new copy of the endpoint to continue the session.

The $\tm{\close}$ and $\tm{\wait}$ functions close a channel.

We use syntactic sugar to make terms more readable: we write $\tm{\letbind{x}{M}{N}}$ in place of $\tm{(\lambda x.N)\;M}$, $\tm{\lambda\unit.M}$ in place of $\tm{\lambda z.\andthen{z}{M}}$, and $\tm{\lambda\pair{x}{y}.M}$ in place of $\tm{\lambda z.\letpair{x}{y}{z}{M}}$.
We can recover GV's $\tm{\fork}$ as $\tm{\lambda x.\letpair{y}{z}{\new\;\unit}{\andthen{\spawn\;{(\lambda\unit.x\;y)}}{z}}}$.

\subsubsection*{Internal and External Choice}
Typically, session-typed languages feature constructs for internal and external choice. In GV, these can be defined in terms of the core language, by sending or receiving a value of a sum type~\cite{lindleymorris15}. We use the following syntactic sugar for internal ($\ty{\tyselect[\cs{o}]{S}{S'}}$) and external ($\ty{\tyoffer[\cs{o}]{S}{S'}}$) choice and their units:
\[
  \begin{array}{lcl}
    \ty{\tyselect[\cs{o}]{S}{S'}}
     & \elabarrow & \ty{\tysend[\cs{o}]{(\tysum{\co{S}}{\co{S'}})}{\tyends[\cs{o+1}]}} \\
    \ty{\tyoffer[\cs{o}]{S}{S'}}
     & \elabarrow & \ty{\tyrecv[\cs{o}]{(\tysum{S}{S'})}{\tyendr[\cs{o+1}]}}           \\
  \end{array}
  \qquad
  \begin{array}{lcl}
    \ty{\tyselectemp[\cs{o}]}
     & \elabarrow & \ty{\tysend[\cs{o}]{\tyvoid}{\tyends[\cs{o+1}]}} \\
    \ty{\tyofferemp[\cs{o}]}
     & \elabarrow & \ty{\tyrecv[\cs{o}]{\tyvoid}{\tyendr[\cs{o+1}]}} \\
  \end{array}
\]

As the syntax for units suggests, these are the binary and nullary forms of the more common n-ary choice constructs $\ty{{\oplus}^{\cs{o}}\{l_i:S_i\}_{i\in{I}}}$ and $\ty{{\with}^{\cs{o}}\{l_i:S_i\}_{i\in{I}}}$, which one may obtain generalising the sum types to variant types. For simplicity, we present only the binary and nullary forms.

Similarly, we use syntactic sugar for the term forms of choice, which combine sending and receiving with the introduction and elimination forms for the sum and empty types. There are two constructs for binary internal choice, expressed using the meta-variable $\tm{\ell}$ which ranges over $\{\tm{\labinl},\tm{\labinr}\}$. As there is no introduction for the empty type, there is no construct for nullary internal choice:
\[
  \begin{array}{lcl}
    \tm{\select{\ell}}
     & \elabarrow
     & \tm{\lambda x.\letpair{y}{z}{\new}{\andthen{\close\;(\send\;{\pair{\ell\;y}{x}})}{z}}}
    \\
    \multicolumn{3}{l}{%
      \tm{\offer{L}{x}{M}{y}{N}}\elabarrow}
    \\
    \multicolumn{3}{l}{%
    \qquad\tm{\letpair{z}{w}{\recv\;{L}}{\andthen{\wait\;{w}}{\casesum{z}{x}{M}{y}{N}}}}}
    \\
    \tm{\offeremp{L}}
     & \elabarrow
     & \tm{\letpair{z}{w}{\recv\;L}{\andthen{\wait\;{w}}{\absurd{z}}}}
  \end{array}
\]

\subsection{Operational Semantics}
\subsubsection*{Configurations}
Priority GV terms are evaluated as part of a configuration of processes. Configurations are defined by the following grammar:
\[
  \tm{\phi}
  \Coloneqq \tm{\main}
  \sep      \tm{\child}
  \hfill\qquad\hfill
  \tm{\conf{C}}, \tm{\conf{D}}, \tm{\conf{E}}
  \Coloneqq \tm{\phi\;M}
  \sep      \tm{\ppar{\conf{C}}{\conf{D}}}
  \sep      \tm{\res{x}{x'}{\conf{C}}}
\]

Configurations ($\tm{\conf{C}}$, $\tm{\conf{D}}$, $\tm{\conf{E}}$) consist of threads $\tm{\phi\;M}$, parallel compositions $\tm{\ppar{\conf{C}}{\conf{D}}}$, and name restrictions $\tm{\res{x}{x'}{\conf{C}}}$. To preserve the functional nature of PGV, where programs return a single value, we use flags ($\tm{\phi}$) to differentiate between the main thread, marked $\tm{\main}$, and child threads created by $\tm{\spawn}$, marked $\tm{\child}$. Only the main thread returns a value. We determine the flag of a configuration by combining the flags of all threads in that configuration:
\[
  \tm{\main}  + \tm{\child} = \tm{\main}
  \hfill\qquad\hfill
  \tm{\child} + \tm{\main}  = \tm{\main}
  \hfill\qquad\hfill
  \tm{\child} + \tm{\child} = \tm{\child}
  \hfill\qquad\hfill
  (\tm{\main}  + \tm{\main} \; \text{is undefined})
  \hfill\qquad\hfill
\]

To distinguish between the use of $\tm{\child}$ to mark child threads~\cite{lindleymorris15} and the use of the meta-variable $\cs{o}$ for priorities~\cite{dardhagay18extended}, they are typeset in a different font and colour.

\subsubsection*{Values}
Values ($\tm{V}$, $\tm{W}$), evaluation contexts ($\tm{E}$), thread evaluation contexts ($\tm{\conf{F}}$), and configuration contexts ($\tm{\conf{G}}$) are defined by the following grammars:
\[
  \begin{array}[t]{lcl}
    \tm{V}, \tm{W}
     & \Coloneqq & \tm{x}
    \sep        \tm{K}
    \sep        \tm{\lambda x.M}
    \sep        \tm{\unit}
    \sep        \tm{\pair{V}{W}}
    \sep        \tm{\inl{V}}
    \sep        \tm{\inr{V}}              \\
    \tm{E}
     & \Coloneqq & \tm{\hole}
    \sep        \tm{E\;M}
    \sep        \tm{V\;E}                 \\
     & \sep      & \tm{\andthen{E}{N}}
    \sep        \tm{\pair{E}{M}}
    \sep        \tm{\pair{V}{E}}
    \sep        \tm{\letpair{x}{y}{E}{M}} \\
     & \sep      & \tm{\inl{E}}
    \sep        \tm{\inr{E}}
    \sep        \tm{\casesum{E}{x}{M}{y}{N}}
    \sep        \tm{\absurd{E}}           \\
    \tm{\conf{F}}
     & \Coloneqq & \tm{\phi\;E}
    \\
    \tm{\conf{G}}
     & \Coloneqq & \tm{\hole}
    \sep        \tm{\ppar{\conf{G}}{\conf{C}}}
    \sep        \tm{\res{x}{y}{\conf{G}}}
  \end{array}
\]
Values are the subset of terms which cannot reduce further.
Evaluation contexts are one-hole term contexts, \ie, terms with exactly one hole, written $\tm{\hole}$. We write $\tm{\plug{E}{M}}$ for the evaluation context $\tm{E}$ with its hole replaced by the term $\tm{M}$. Evaluation contexts are specifically those one-hole term contexts under which term reduction can take place.
Thread contexts are a convenient way to lift the notion of evaluation contexts to threads. We write $\tm{\plug{\conf{F}}{M}}$ for the thread context $\tm{\conf{F}}$ with its hole replaced by the term $\tm{M}$.
Configuration contexts are one-hole configuration contexts, \ie, configurations with exactly one hole, written $\tm{\hole}$. Specifically, configuration contexts are those one-hole term contexts under which configuration reduction can take place. The definition for $\tm{\conf{G}}$ only gives the case in which the hole is in the left-most parallel process, \ie, it only defines $\tm{\ppar{\conf{G}}{\conf{C}}}$ and not $\tm{\ppar{\conf{C}}{\conf{G}}}$. The latter is not needed, as $\tm{\parallel}$ is symmetric under structural congruence, though it would be harmless to add. We write $\tm{\plug{\conf{G}}{\conf{C}}}$ for the evaluation context $\tm{\conf{G}}$ with its hole replaced by the term $\tm{\conf{C}}$.

\subsubsection*{Reduction Relation}
\begin{figure}[t!]
  \textbf{Term reduction.}
  \begin{mathpar}
    \begin{array}{llcl}
      \LabTirName{E-Lam}  & \tm{(\lambda x.M) \; V}
                          & \tred & \tm{\subst{M}{V}{x}}
      \\
      \LabTirName{E-Unit} & \tm{\letunit{\unit}{M}}
                          & \tred & \tm{M}
      \\
      \LabTirName{E-Pair} & \tm{\letpair{x}{y}{\pair{V}{W}}{M}}
                          & \tred & \tm{\subst{\subst{M}{V}{x}}{W}{y}}
      \\
      \LabTirName{E-Inl}  & \tm{\casesum{\inl{V}}{x}{M}{y}{N}}
                          & \tred & \tm{\subst{M}{V}{x}}
      \\
      \LabTirName{E-Inr}  & \tm{\casesum{\inr{V}}{x}{M}{y}{N}}
                          & \tred & \tm{\subst{N}{V}{y}}
    \end{array}
    \\
    \inferrule*[lab=E-Lift]{
      \tm{M}\tred\tm{M'}
    }{\tm{\plug{E}{M}}\tred\tm{\plug{E}{M'}}}
  \end{mathpar}

  \textbf{Structural congruence.}
  \begin{mathpar}
    \begin{array}{llcl}
      \LabTirName{SC-LinkSwap}   & \tm{\plug{\conf{F}}{\link\;{\pair{x}{y}}}}
                                 & \equiv & \tm{\plug{\conf{F}}{\link\;{\pair{y}{x}}}}
      \\
      \LabTirName{SC-ResLink}    & \tm{\res{x}{y}{(\phi\;\link\;\pair{x}{y})}}
                                 & \equiv & \tm{\phi\;\unit}
      \\
      \LabTirName{SC-ResSwap}    & \tm{\res{x}{y}{\conf{C}}}
                                 & \equiv & \tm{\res{y}{x}{\conf{C}}}
      \\
      \LabTirName{SC-ResComm}    & \tm{\res{x}{y}{\res{z}{w}{\conf{C}}}}
                                 & \equiv & \tm{\res{z}{w}{\res{x}{y}{\conf{C}}}},
                                            \text{ if }{\{\tm{x},\tm{y}\}\cap\{\tm{z},\tm{w}\}=\varnothing}
      \\
      \LabTirName{SC-ResExt}     & \tm{\res{x}{y}{(\ppar{\conf{C}}{\conf{D}}})}
                                 & \equiv & \tm{\ppar{\conf{C}}{\res{x}{y}{\conf{D}}}},
                                            \text{ if }{\tm{x},\tm{y}\notin\fv(\tm{\conf{C}})}
      \\
      \LabTirName{SC-ParNil}     & \tm{\ppar{\conf{C}}{\child{\unit}}}
                                 & \equiv & \tm{\conf{C}}
      \\
      \LabTirName{SC-ParComm}    & \tm{\ppar{\conf{C}}{\conf{D}}}
                                 & \equiv & \tm{\ppar{\conf{D}}{\conf{C}}}
      \\
      \LabTirName{SC-ParAssoc}   & \tm{\ppar{\conf{C}}{(\ppar{\conf{D}}{\conf{E}})}}
                                 & \equiv & \tm{\ppar{(\ppar{\conf{C}}{\conf{D}})}{\conf{E}}}
    \end{array}
  \end{mathpar}
  \textbf{Configuration reduction.}
  \begin{mathpar}
    \begin{array}{l}
      \begin{array}{llcl}
        \LabTirName{E-Link}  & \tm{\res{x}{y}{(\ppar{\plug{\conf{F}}{\link\;\pair{w}{x}}}{\conf{C}})}}
                             & \cred & \tm{\ppar{\plug{\conf{F}}{\unit}}{\subst{\conf{C}}{w}{y}}}
        \\
        \LabTirName{E-New}   & \tm{\plug{\conf{F}}{\new\;\unit}}
                             & \cred & \tm{\res{x}{y}{(\plug{\conf{F}}{\pair{x}{y}})}},
                                       \text{ if }\tm{x},\tm{y}\notin\fv(\tm{\conf{F}})
        \\
        \LabTirName{E-Spawn} & \tm{\plug{\conf{F}}{(\spawn\;V)}}
                             & \cred & \tm{\ppar{\plug{\conf{F}}{\unit}}{\child\;V\;\unit}}
      \end{array}
      \\
      \begin{array}{llcl}
        \LabTirName{E-Send}  & \tm{\res{x}{y}{(\ppar
                               {\plug{\conf{F}}{\send\;{\pair{V}{x}}}}
                               {\plug{\conf{F'}}{\recv\;{y}}})}}
                             & \cred & \tm{\res{x}{y}{(\ppar
                                       {\plug{\conf{F}}{x}}
                                       {\plug{\conf{F'}}{\pair{V}{y}}})}}
        \\
        \LabTirName{E-Close} & \tm{\res{x}{y}{(\ppar
                               {\plug{\conf{F}}{\wait\;{x}}}
                               {\plug{\conf{F'}}{\close\;{y}}})}}
                             & \cred & \tm{\ppar{\plug{\conf{F}}{\unit}}{\plug{\conf{F'}}{\unit}}}
      \end{array}
    \end{array}
    \\
    \inferrule*[lab=E-LiftC]{
      \tm{\conf{C}}\cred\tm{\conf{C'}}
    }{\tm{\plug{\conf{G}}{\conf{C}}}\cred\tm{\plug{\conf{G}}{\conf{C'}}}}

    \inferrule*[lab=E-LiftM]{
      \tm{M}\tred\tm{M'}
    }{\tm{\plug{\conf{F}}{M}}\tred\tm{\plug{\conf{F}}{M'}}}

    \inferrule*[lab=E-LiftSC]{
      \tm{\conf{C}}\equiv\tm{\conf{C'}}
      \\
      \tm{\conf{C'}}\cred\tm{\conf{D'}}
      \\
      \tm{\conf{D'}}\equiv\tm{\conf{D}}
    }{\tm{\conf{C}}\cred\tm{\conf{D}}}
  \end{mathpar}
  \caption{Operational Semantics for PGV.}%
  \label{fig:pgv-operational-semantics}
\end{figure}

We factor the reduction relation of PGV into a deterministic reduction on terms ($\tred$) and a non-deterministic reduction on configurations ($\cred$), see \cref{fig:pgv-operational-semantics}. We write $\tred^+$ and $\cred^+$ for the transitive closures, and $\tred^\star$ and $\cred^\star$ for the reflexive-transitive closures.

Term reduction is the standard call-by-value, left-to-right evaluation for GV, and only deviates from reduction for the linear $\lambda$-calculus in that it reduces terms to values {or} ready terms waiting to perform a communication action.

Configuration reduction resembles evaluation for a process calculus: \LabTirName{E-Link}, \LabTirName{E-Send}, and \LabTirName{E-Close} perform communications, \LabTirName{E-LiftC} allows reduction under configuration contexts, and \LabTirName{E-LiftSC} embeds a structural congruence $\equiv$. The remaining rules mediate between the process calculus and the functional language: \LabTirName{E-New} and \LabTirName{E-Spawn} evaluate the $\tm{\new}$ and $\tm{\spawn}$ constructs, creating the equivalent configuration constructs, and \LabTirName{E-LiftM} embeds term reduction.

Structural congruence satisfies the following axioms: $\LabTirName{SC-LinkSwap}$ allows swapping channels in the link process. $\LabTirName{SC-ResLink}$ allows restriction to be applied to link which is structurally equivalent to the terminated process, thus allowing elimination of unnecessary restrictions. $\LabTirName{SC-ResSwap}$ allows swapping channels and $\LabTirName{SC-ResComm}$ states that restriction is commutative. $\LabTirName{SC-ResExt}$ is the standard scope extrusion rule. Rules $\LabTirName{SC-ParNil}$, $\LabTirName{SC-ParComm}$ and $\LabTirName{SC-ParAssoc}$ state that parallel composition uses the terminated process as the neutral element; it is commutative and associative.

While our configuration reduction is based on the standard evaluation for GV, the increased expressiveness of PGV allows us to simplify the relation on two counts.
\begin{enumerate}[label=(\roman*)]
  \item
        \emph{We decompose the $\tm{\fork}$ construct}.
        In GV, $\tm{\fork}$ creates a new channel, spawns a child thread, and, when the child thread finishes, it closes the channel to its parent. In PGV, these are three separate operations: $\tm{\new}$, $\tm{\spawn}$, and $\tm{\close}$. We no longer require that every child thread finishes by returning a terminated channel. Consequently, we also simplify the evaluation of the $\tm{\link}$ construct.

        Intuitively, evaluating $\tm{\link}$ causes a substitution: if we have a channel bound as $\tm{\res{x}{y}{}}$, then $\tm{\link\;\pair{w}{x}}$ replaces all occurrences of $\tm{y}$ by $\tm{w}$. However, in GV, $\tm{\link}$ is required to return a terminated channel, which means that the semantics for $\tm{link}$ must {create} a fresh channel of type $\ty{\tyends}/\ty{\tyendr}$. The endpoint of type $\ty{\tyends}$ is returned by the $\tm{link}$ construct, and a $\tm{\wait}$ on the other endpoint guards the {actual} substitution. In PGV, evaluating $\tm{\link}$ simply causes a substitution.
  \item
        \emph{Our structural congruence is type preserving}. Consequently, we can embed it directly into the reduction relation. In GV, this is not the case, and subject reduction relies on proving that if the result of rewriting via $\equiv$ followed by reducing via $\cred$ is an ill-typed configuration, we can rewrite it to a well-typed configuration via $\equiv$.
\end{enumerate}

\subsection{Typing Rules}
\begin{figure}
  \textbf{Static Typing Rules.}
  \begin{mathpar}
    \inferrule*[lab=T-Var]{
    }{\tseq[\cs{\pbot}]{\tmty{x}{T}}{x}{T}}

    \inferrule*[lab=T-Const]{
    }{\tseq[\cs{\pbot}]{\emptyenv}{K}{T}}

    \inferrule*[lab=T-Lam]{
      \tseq[\cs{q}]{\ty{\Gamma},\tmty{x}{T}}{M}{U}
    }{\tseq[\cs{\pbot}]{\ty{\Gamma}}{\lambda x.M}{\tylolli[\cs{\minpr(\ty{\Gamma})},\cs{q}]{T}{U}}}

    \inferrule*[lab=T-App]{
      \tseq[\cs{p}]{\ty{\Gamma}}{M}{\tylolli[\cs{p'},\cs{q'}]{T}{U}}
      \\
      \tseq[\cs{q}]{\ty{\Delta}}{N}{T}
      \\
      \cs{p}<\minpr(\ty{\Delta})
      \\
      \cs{q}<\cs{p'}
    }{\tseq[\cs{p}\sqcup\cs{q}\sqcup\cs{q'}]{\ty{\Gamma},\ty{\Delta}}{M\;N}{U}}
    \\
    \inferrule*[lab=T-Unit]{
    }{\tseq[\cs{\pbot}]{\emptyenv}{\unit}{\tyunit}}

    \inferrule*[lab=T-LetUnit]{
      \tseq[\cs{p}]{\ty{\Gamma}}{M}{\tyunit}
      \\
      \tseq[\cs{q}]{\ty{\Delta}}{N}{T}
      \\
      \cs{p}<\minpr(\ty{\Delta})
    }{\tseq[\cs{p}\sqcup\cs{q}]{\ty{\Gamma},\ty{\Delta}}{\andthen{M}{N}}{T}}

    \inferrule*[lab=T-Pair]{
      \tseq[\cs{p}]{\ty{\Gamma}}{M}{T}
      \\
      \tseq[\cs{q}]{\ty{\Delta}}{N}{U}
      \\
      \cs{p}<\minpr(\ty{\Delta})
    }{\tseq[\cs{p}\sqcup\cs{q}]{\ty{\Gamma},\ty{\Delta}}{\pair{M}{N}}{\typrod{T}{U}}}

    \inferrule*[lab=T-LetPair]{
      \tseq[\cs{p}]{\ty{\Gamma}}{M}{\typrod{T}{T'}}
      \\
      \tseq[\cs{q}]{\ty{\Delta},\tmty{x}{T},\tmty{y}{T'}}{N}{U}
      \\
      \cs{p}<\minpr(\ty{\Delta},\ty{T},\ty{T'})
    }{\tseq[\cs{p}\sqcup\cs{q}]{\ty{\Gamma},\ty{\Delta}}{\letpair{x}{y}{M}{N}}{U}}

    \inferrule*[lab=T-Inl]{
      \tseq[\cs{p}]{\ty{\Gamma}}{M}{T}
      \\
      \minpr(\ty{T})=\minpr(\ty{U})
    }{\tseq[\cs{p}]{\ty{\Gamma}}{\inl{M}}{\tysum{T}{U}}}

    \inferrule*[lab=T-Inr]{
      \tseq[\cs{p}]{\ty{\Gamma}}{M}{U}
      \\
      \minpr(\ty{T})=\minpr(\ty{U})
    }{\tseq[\cs{p}]{\ty{\Gamma}}{\inr{M}}{\tysum{T}{U}}}

    \inferrule*[lab=T-CaseSum]{
      \tseq[\cs{p}]{\ty{\Gamma}}{L}{\tysum{T}{T'}}
      \\
      \tseq[\cs{q}]{\ty{\Delta},\tmty{x}{T}}{M}{U}
      \\
      \tseq[\cs{q}]{\ty{\Delta},\tmty{y}{T'}}{N}{U}
      \\
      \cs{p}<\minpr(\ty{\Delta})
    }{\tseq[\cs{p}\sqcup\cs{q}]{\ty{\Gamma},\ty{\Delta}}{\casesum{L}{x}{M}{y}{N}}{U}}

    \inferrule*[lab=T-Absurd]{
      \tseq[\cs{p}]{\ty{\Gamma}}{M}{\tyvoid}
    }{\tseq[\cs{p}]{\ty{\Gamma},\ty{\Delta}}{\absurd{M}}{T}}
  \end{mathpar}
  \textbf{Type Schemas for Constants.}
  \begin{mathpar}
    \tmty{\link}{\tylolli{\typrod{S}{\co{S}}}{\tyunit}}

    \tmty{\new}{\tylolli{\tyunit}{\typrod{S}{\co{S}}}}

    \tmty{\spawn}{\tylolli{(\tylolli[\cs{p},\cs{q}]{\tyunit}{\tyunit})}{\tyunit}}
    \\
    \tmty{\send}{\tylolli[\cs{\ptop},\cs{o}]{\typrod{T}{\tysend[\cs{o}]{T}{S}}}{S}}

    \tmty{\recv}{\tylolli[\cs{\ptop},\cs{o}]{\tyrecv[\cs{o}]{T}{S}}{\typrod{T}{S}}}
    \\
    \tmty{\close}{\tylolli[\cs{\ptop},\cs{o}]{\tyends[\cs{o}]}{\tyunit}}

    \tmty{\wait}{\tylolli[\cs{\ptop},\cs{o}]{\tyendr[\cs{o}]}{\tyunit}}
  \end{mathpar}
  \textbf{Runtime Typing Rules.}
  \begin{mathpar}
    \inferrule*[lab=T-Main]{
      \tseq[\cs{p}]{\ty{\Gamma}}{M}{T}
    }{\cseq[\main]{\ty{\Gamma}}{\main\;M}}

    \inferrule*[lab=T-Child]{
      \tseq[\cs{p}]{\ty{\Gamma}}{M}{\tyunit}
    }{\cseq[\child]{\ty{\Gamma}}{\child\;M}}

    \inferrule*[lab=T-Res]{
      \cseq[\phi]{\ty{\Gamma},\tmty{x}{S},\tmty{y}{\co{S}}}{\conf{C}}
    }{\cseq[\phi]{\ty{\Gamma}}{\res{x}{y}{\conf{C}}}}

    \inferrule*[lab=T-Par]{
      \cseq[\phi]{\ty{\Gamma}}{\conf{C}}
      \\
      \cseq[\phi']{\ty{\Delta}}{\conf{D}}
    }{\cseq[\phi+\phi']{\ty{\Gamma},\ty{\Delta}}{\ppar{\conf{C}}{\conf{D}}}}
  \end{mathpar}
  \caption{Typing Rules for PGV.}%
  \label{fig:pgv-typing}
\end{figure}

\subsubsection*{Terms Typing}
Typing rules for terms are at the top of \cref{fig:pgv-typing}.
Terms are typed by a judgement $\tseq[\cs{p}]{\ty{\Gamma}}{M}{T}$ stating that ``a term $\tm{M}$ has type $\ty{T}$ and an upper bound on its priority $\cs{p}$ under the typing environment $\ty{\Gamma}$''. Typing for the linear $\lambda$-calculus is standard. Linearity is ensured by splitting environments on branching rules, requiring that the environment in the variable rule consists of just the variable, and the environment in the constant and unit rules are empty. Constants $\tm{K}$ are typed using type schemas, which hold for any concrete assignment of types and priorities to their meta-variables. Instantiated type schemas are embedded into typing derivations using \LabTirName{T-Const} in~\cref{fig:pgv-typing}, \eg, the type schema for $\tm{\send}$ can be instantiated with $\cs{o}=\cs{2}$, $\ty{T}=\ty{\tyunit}$, and $\ty{S}=\ty{\tyvoid}$, and embedded using \LabTirName{T-Const} to give the following typing derivation:
\[
  \inferrule*{
  }{\tmty{\send}{\tylolli[\cs{\ptop},\cs{2}]{\typrod{\tyunit}{\tysend[\cs{2}]{\tyunit}{\tyvoid}}}{\tyvoid}}}
\]

The typing rules treat {all variables} as linear resources, even those of non-linear types such as $\ty{\tyunit}$, though they can easily be extended to allow values with unrestricted usage~\cite{wadler14}.

The only non-standard feature of the typing rules is the priority annotations. Priorities are based on {obligations/capabilities} used by Kobayashi~\cite{kobayashi06}, and simplified to single priorities following Padovani~\cite{padovani14}. The integration of priorities into GV is adapted from Padovani and Novara~\cite{padovaninovara15}. Paraphrasing Dardha and Gay~\cite{dardhagay18extended}, priorities obey the following two laws:
\begin{enumerate}[label= (\roman*) ]
  \item an action with lower priority happens before an action with higher priority; and
  \item communication requires \emph{equal} priorities for dual actions.
\end{enumerate}

In PGV, we keep track of a lower and upper bound on the priorities of a term, \ie, while evaluating the term, when it starts communicating, and when it finishes, respectively. The upper bound is written on the sequent and the lower bound is approximated from the typing environment, \eg, for $\tseq[\cs{p}]{\ty{\Gamma}}{M}{T}$ the upper bound is $\cs{p}$ and the lower bound is at least $\minpr(\ty{\Gamma})$. The latter is {correct} because a term cannot communicate at a priority earlier than the earliest priority amongst the channels it has access to. It is an \emph{approximation} on function terms, as these can ``skip'' communication by returning the corresponding channel unused. However, linearity prevents such functions from being used in well typed configurations: once the unused channel's priority has passed, it can no longer be used.

Typing rules for sequential constructs enforce sequentiality, \eg, the typing for $\tm{\andthen{M}{N}}$ has a side condition which requires that the upper bound of $\tm{M}$ is smaller than the lower bound of $\tm{N}$, \ie, $\tm{M}$ finishes before $\tm{N}$ starts. The typing rule for $\tm{\new}$ ensures that both endpoints of a channel share the same priorities. Together, these two constraints guarantee deadlock freedom.

To illustrate this, let's go back to the deadlocked program in \cref{ex:deadlock}. Crucially, it composes the terms below in parallel. While each of these terms itself is well typed, they impose opposite conditions on the priorities, so connecting $\ty{x}$ to $\ty{x'}$ and $\ty{y}$ to $\ty{y'}$ using \LabTirName{T-Res} is ill-typed, as there is no assignment to $\cs{o}$ and $\cs{o'}$ that can satsify both $\cs{o}<\cs{o'}$ and $\cs{o'}<\cs{o}$. (We omit the priorities on $\ty{\tyends}$ and $\ty{\tyendr}$.)
\begin{mathpar}
  \inferrule*{
  \tseq[\cs{o'}]
  {\tmty{y'}{\tyrecv[\cs{o'}]{\tyunit}{\tyendr}}}
  {\recv\;{y'}}
  {\typrod{\tyunit}{\tyendr}}
  \\
  \tseq[\cs{p}]
  {\tmty{x}{\tysend[\cs{o}]{\tyunit}{\tyends}},\tmty{y'}{\tyendr}}
  {\letbind{x}{\send\;{\pair{\unit}{x}}}{\dots}}
  {\tyunit}
  \and
  \cs{o'}<\cs{o}
  }{%
  \tseq[\cs{p}]
  {%
  \tmty{x}{\tysend[\cs{o}]{\tyunit}{\tyends}},
  \tmty{y'}{\tyrecv[\cs{o'}]{\tyunit}{\tyendr}}
  }
  {\letpair{\unit}{y'}{\recv\;{y'}}{\letbind{x}{\send\;{\pair{\unit}{x}}}{\dots}}}
  {\tyunit}
  }

  \inferrule*{
  \tseq[\cs{o}]
  {\tmty{x'}{\tyrecv[\cs{o}]{\tyunit}{\tyendr}}}
  {\recv\;{x'}}
  {\typrod{\tyunit}{\tyendr}}
  \\
  \tseq[\cs{q}]
  {\tmty{y}{\tysend[\cs{o'}]{\tyunit}{\tyends}},\tmty{x'}{\tyendr}}
  {\letbind{y}{\send\;{\pair{\unit}{y}}}{\dots}}
  {\tyunit}
  \and
  \cs{o}<\cs{o'}
  }{%
  \tseq[\cs{q}]
  {%
  \tmty{y}{\tysend[\cs{o'}]{\tyunit}{\tyends}},
  \tmty{x'}{\tyrecv[\cs{o}]{\tyunit}{\tyendr}}
  }
  {\letpair{\unit}{x'}{\recv\;{x'}}{\letbind{y}{\send\;{\pair{\unit}{y}}}{\dots}}}
  {\tyunit}
  }
\end{mathpar}

Closures suspend communication, so \LabTirName{T-Lam} stores the priority bounds of the function body on the function type, and \LabTirName{T-App} restores them. For instance, $\tm{\lambda{x}.\send\;\pair{x}{y}}$ is assigned the type $\ty{\tylolli[\cs{o},\cs{o}]{A}{S}}$, \ie, a~function which, when applied, starts and finishes communicating at priority $\cs{o}$.
\begin{mathpar}
  \inferrule*{
    \inferrule*{
      \inferrule*{
      }{
        \tmty{\send}{\tylolli[\ptop,\cs{o}]{\typrod{T}{\tysend[\cs{o}]{T}{S}}}{S}}
      }
      \and
      \inferrule*{%
        \inferrule*{
        }{
          \tseq[\pbot]{\tmty{x}{T}}{x}{T}
        }
        \and
        \inferrule*{
        }{
          \tseq[\pbot]
          {\tmty{x}{T},\tmty{y}{\tysend[\cs{o}]{T}{S}}}
          {y}{\tysend[\cs{o}]{T}{S}}
        }
      }{
        \tseq[\pbot]
        {\tmty{x}{T},\tmty{y}{\tysend[\cs{o}]{T}{S}}}
        {\pair{x}{y}}{\typrod{T}{\tysend[\cs{o}]{T}{S}}}
      }
    }{
      \tseq[\cs{o}]
      {\tmty{x}{T},\tmty{y}{\tysend[\cs{o}]{T}{S}}}
      {\send\;\pair{x}{y}}{S}
    }
  }{
    \tseq[\pbot]
    {\tmty{y}{\tysend[\cs{o}]{T}{S}}}
    {\lambda{x}.\send\;\pair{x}{y}}{\tylolli[\cs{o},\cs{o}]{T}{S}}
  }
\end{mathpar}

For simplicity, we assume priority annotations are not inferred, but provided as an input to type checking. However, for any term, priorities can be inferred, \eg, by using the topological ordering of the directed graph where the vertices are the priority meta-variables and the edges are the inequality constraints between the priority meta-variables in the typing derivation.

\subsubsection*{Configurations Typing}
Typing rules for configurations are at the bottom of \cref{fig:pgv-typing}. Configurations are typed by a judgement $\cseq[\phi]{\ty{\Gamma}}{\conf{C}}$ stating that ``a configuration $\tm{\conf{C}}$ with flag $\tm\phi$ is well typed under typing environment $\ty{\Gamma}$''. Configuration typing is based on the standard typing for GV. Terms are embedded either as main or as child threads. The priority bound from the term typing is discarded, as configurations contain no further blocking actions. Main threads are allowed to return a value, whereas child threads are required to return the unit value. Sequents are annotated with a flag $\tm\phi$, which ensures that there is at most one main thread.

While our configuration typing is based on the standard typing for GV, it differs on two counts:
\begin{enumerate}[label= (\roman*) ]
  \item \emph{we require that child threads return the unit value}, as opposed to a terminated channel; and
  \item \emph{we simplify typing for parallel composition}.
\end{enumerate}

In order to guarantee deadlock freedom, in GV each parallel composition must split \emph{exactly one} channel of the channel pseudo-type $\ty{S^\sharp}$ into two endpoints of type $\ty{S}$ and $\ty{\co{S}}$. Consequently, associativity of parallel composition does not preserve typing. In PGV, we guarantee deadlock freedom using priorities, which removes the need for the channel pseudo-type $\ty{S^\sharp}$, and simplifies typing for parallel composition, while restoring type preservation for the structural congruence.

\subsubsection*{Syntactic Sugar Typing}
\begin{figure}
  \begin{mathpar}
    \inferrule*[lab=T-LamUnit]{
      {\tseq[\cs{q}]{\ty{\Gamma}}{M}{T}}
    }{\tseq[\cs{\pbot}]
      {\ty{\Gamma}}
      {\lambda\unit.M}
      {\tylolli[\cs{\minpr(\ty{\Gamma})},\cs{q}]{\tyunit}{T}}}
    \elabarrow
    \inferrule*{
      \inferrule*{
        \inferrule*{
        }{\tseq[\cs{\pbot}]{\tmty{z}{\tyunit}}{z}{\tyunit}}
        \\
        \tseq[\cs{q}]{\ty{\Gamma}}{M}{T}
      }{\tseq[\cs{q}]
        {\ty{\Gamma},\tmty{z}{\tyunit}}
        {\letunit{z}{M}}
        {T}}
    }{\tseq[\cs{\pbot}]
      {\ty{\Gamma}}
      {\lambda z.\letunit{z}{M}}
      {\tylolli[\cs{\minpr(\ty{\Gamma})},\cs{q}]{\tyunit}{T}}}
    \\
    \\
    \inferrule*[lab=T-LamPair]
    {\tseq[\cs{q}]
      {\ty{\Gamma},\tmty{x}{T},\tmty{y}{T'}}
      {M}
      {U}}
    {\tseq[\cs{\pbot}]
      {\ty{\Gamma}}
      {\lambda\pair{x}{y}.M}
      {\tylolli[\cs{\minpr(\ty{\Gamma})},\cs{q}]{\typrod{T}{T'}}{U}}}
    \elabarrow
    \inferrule*{
      \inferrule*{
        \inferrule*{
        }{\tseq[\cs{\pbot}]{\tmty{z}{\typrod{T}{T'}}}{z}{\typrod{T}{T'}}}
        \\
        \tseq[\cs{q}]{\ty{\Gamma},\tmty{x}{T},\tmty{y}{T'}}{M}{U}
      }{\tseq[\cs{q}]
        {\ty{\Gamma},\tmty{z}{\typrod{T}{T'}}}
        {\letpair{x}{y}{z}{M}}
        {T}}
    }{\tseq[\cs{\pbot}]
      {\ty{\Gamma}}
      {\lambda z.\letpair{x}{y}{z}{M}}
      {\tylolli[\cs{\minpr(\ty{\Gamma})},\cs{q}]{\typrod{T}{T'}}{U}}}
    \\
    \\
    \inferrule*[lab=T-Let]{
      \tseq[\cs{p}]{\ty{\Gamma}}{M}{T}
      \\
      \tseq[\cs{q}]{\ty{\Delta},\tmty{x}{T}}{N}{U}
      \\
      \cs{p}<\minpr(\ty{\Delta})
    }{\tseq[\cs{p}\sqcup\cs{q}]{\ty{\Gamma},\ty{\Delta}}{\letbind{x}{M}{N}}{U}}
    \elabarrow
    \inferrule*{
      \inferrule*{
        {\tseq[\cs{q}]{\ty{\Delta},\tmty{x}{T}}{N}{U}}
      }{\tseq[\cs{\pbot}]
        {\ty{\Delta}}
        {\lambda x.N}
        {\tylolli[\cs{\minpr(\ty{\Delta})},\cs{q}]{T}{U}}}
      \\
      \tseq[\cs{p}]{\ty{\Gamma}}{M}{T}
      \\
      \cs{p}<\minpr(\ty{\Delta})
    }{\tseq[\cs{q}\sqcup\cs{p}]
      {\ty{\Gamma},\ty{\Delta}}
      {(\lambda x.N)\;M}
      {U}}
    \\
    \\
    \inferrule*[lab=T-Fork]{
    }{\tseq[\cs{\pbot}]
      {\emptyenv}
      {\fork}
      {\tylolli[]{(\tylolli[\cs{p},\cs{q}]{S}{\tyunit})}{\co{S}}}}
    \elabarrow
    \\
    \inferrule*{
      \inferrule*{
        \inferrule*{
          \LabTirName{(a)}
          \inferrule*{
          }{}
          \\
          \inferrule*{
          }{\tseq[\cs{\pbot}]
            {\emptyenv}
            {\unit}
            {\tyunit}}
        }{\tseq[\cs{\pbot}]
          {\emptyenv}
          {\new\;\unit}
          {\typrod{S}{\co{S}}}}
        \\
        \inferrule*{
          \inferrule*{
            \LabTirName{(b)}
            \\
            \inferrule*{
              \inferrule*{
                \inferrule*{
                }{\tseq[\cs{\pbot}]
                  {\tmty{x}{\tylolli[\cs{p},\cs{q}]{S}{\tyunit}}}
                  {x}{\tylolli[\cs{p},\cs{q}]{S}{\tyunit}}}
                \\
                \inferrule*{
                }{\tseq[\cs{\pbot}]{\tmty{y}{S}}{y}{S}}
              }
              {\tseq[\cs{q}]
                {\tmty{x}{\tylolli[\cs{p},\cs{q}]{S}{\tyunit}},\tmty{y}{S}}
                {x\;y}
                {\tyunit}}
            }
            {\tseq[\cs{\pbot}]
              {\tmty{x}{\tylolli[\cs{p},\cs{q}]{S}{\tyunit}},\tmty{y}{S}}
              {\lambda\unit.x\;y}
              {\tylolli[\cs{p},\cs{q}]{\tyunit}{\tyunit}}}
          }
          {\tseq[\cs{\pbot}]
            {\tmty{x}{\tylolli[\cs{p},\cs{q}]{S}{\tyunit}},\tmty{y}{S}}
            {\spawn\;{(\lambda\unit.x\;y)}}
            {\tyunit}}
          \\
          \inferrule*{
          }{\tseq[\cs{\pbot}]{\tmty{z}{\co{S}}}{z}{\co{S}}}
        }
        {\tseq[\cs{\pbot}]
          {\tmty{x}{\tylolli[\cs{p},\cs{q}]{S}{\tyunit}},\tmty{y}{S},\tmty{z}{\co{S}}}
          {\andthen{\spawn\;{(\lambda\unit.x\;y)}}{z}}
          {\co{S}}}
      }
      {\tseq[\cs{\pbot}]
        {\tmty{x}{\tylolli[\cs{p},\cs{q}]{S}{\tyunit}}}
        {\letpair{y}{z}{\new\;\unit}{\andthen{\spawn\;{(\lambda\unit.x\;y)}}{z}}}
        {\co{S}}}
    }
    {\tseq[\cs{\pbot}]
      {\emptyenv}
      {\lambda x.\letpair{y}{z}{\new\;\unit}{\andthen{\spawn\;{(\lambda\unit.x\;y)}}{z}}}
      {\tylolli[]{(\tylolli[\cs{p},\cs{q}]{S}{\tyunit})}{\co{S}}}}
    \\
    \LabTirName{(a)} =
    \tmty{\new}
    {\tylolli{\tyunit}{{\typrod{S}{\co{S}}}}}
  
    \LabTirName{(b)} =
    \tmty{\spawn}
    {\tylolli[]{(\tylolli[\cs{p},\cs{q}]{\tyunit}{\tyunit})}{\tyunit}}
  \end{mathpar}
  \caption{Typing Rules for Syntactic Sugar for PGV (\LabTirName{T-LamUnit}, \LabTirName{T-LamPair}, \LabTirName{T-Let}, and \LabTirName{T-Fork}).}
  \label{fig:pgv-typing-sugar}
\end{figure}

\begin{figure}
  \begin{mathpar}
    \inferrule*[lab=T-Select-Inl]{
      \minpr(\ty{S})=\minpr(\ty{S'})
    }{\tseq[\cs{\pbot}]{\emptyenv}{\select{\labinl}}{\tylolli[\cs{\ptop},\cs{o}]{\tyselect[\cs{o}]{S}{S'}}{S}}}
    \elabarrow
    \\
    \inferrule*{
      \inferrule*{
        \inferrule*{
          \LabTirName{(a)}
          \\
          \inferrule*{
          }{\tseq
            {\emptyenv}
            {\unit}
            {\tyunit}}
        }{\tseq
          {\emptyenv}
          {\new\;\unit}
          {\typrod{\co{S}}{S}}}
        \\
        \inferrule*{
          \inferrule*{
            \LabTirName{(b)}
            \\
            \inferrule*{
              \LabTirName{(c)}
              \\
              \inferrule*{
                \inferrule*{
                  \inferrule*{
                  }{\tseq
                    {\tmty{y}{\co{S}}}
                    {y}
                    {\co{S}}}
                }{\tseq
                  {\tmty{y}{\co{S}}}
                  {\inl{y}}
                  {\tysum{\co{S}}{\co{S'}}}}
                \\
                \inferrule*{
                }{\tseq
                  {\tmty{x}{\tyselect[\cs{o}]{S}{S'}}}
                  {x}
                  {\tyselect[\cs{o}]{S}{S'}}}
              }{\tseq
                {\tmty{x}{\tyselect[\cs{o}]{S}{S'}},\tmty{y}{\co{S}}}
                {\pair{\inl{y}}{x}}
                {\typrod{(\tysum{\co{S}}{\co{S'}})}{(\tyselect[\cs{o}]{S}{S'})}}}
            }{\tseq
              {\tmty{x}{\tyselect[\cs{o}]{S}{S'}},\tmty{y}{\co{S}}}
              {\send\;{\pair{\inl{y}}{x}}}
              {\tyends[\cs{o+1}]}}
          }{\tseq
            {\tmty{x}{\tyselect[\cs{o}]{S}{S'}},\tmty{y}{\co{S}}}
            {\close\;(\send\;{\pair{\inl{y}}{x}})}
            {\tyunit}}
          \\
          \inferrule*{
          }{\tseq
            {\tmty{z}{S}}
            {z}
            {S}}
        }{\tseq
          {\tmty{x}{\tyselect[\cs{o}]{S}{S'}},\tmty{y}{\co{S}},\tmty{z}{S}}
          {\andthen{\close\;(\send\;{\pair{\inl{y}}{x}})}{z}}
          {S}}
      }{\tseq
        {\tmty{x}{\tyselect[\cs{o}]{S}{S'}}}
        {\letpair{y}{z}{\new\;\unit}{\andthen{\close\;(\send\;{\pair{\inl{y}}{x}})}{z}}}
        {S}}
    }{\tseq
      {\emptyenv}
      {\lambda x.\letpair{y}{z}{\new\;\unit}{\andthen{\close\;(\send\;{\pair{\inl{y}}{x}})}{z}}}
      {\tylolli[\cs{\ptop},\cs{o}]{\tyselect[\cs{o}]{S}{S'}}{S}}}
    \\
    \LabTirName{(a)} = 
    \tmty{\new}
    {\tylolli[\cs{\ptop},\cs{o}]{\tyunit}{\typrod{\co{S}}{S}}}
  
    \LabTirName{(b)} =
    \tmty{\close}
    {\tylolli[\cs{\ptop},\cs{o+1}]{\tyends[\cs{o+1}]}{\tyunit}}
  
    \LabTirName{(c)} =
    \tmty{\send}
    {\tylolli[\cs{\ptop},\cs{o}]
      {\typrod{(\tysum{\co{S}}{\co{S'}})}{(\tyselect[\cs{o}]{S}{S'})}}{\tyends[\cs{o+1}]}}
    \\
    \\
    \inferrule*[lab=T-Select-Inr]{
      \minpr(\ty{S})=\minpr(\ty{S'})
    }{\tseq[\cs{\pbot}]{\emptyenv}{\select{\labinr}}{\tylolli[\cs{\ptop},\cs{o}]{\tyselect[\cs{o}]{S}{S'}}{S'}}}
    \elabarrow
    \\
    \inferrule*{
      \inferrule*{
        \inferrule*{
          \LabTirName{(a)}
          \\
          \inferrule*{
          }{\tseq
            {\emptyenv}
            {\unit}
            {\tyunit}}
        }{\tseq
          {\emptyenv}
          {\new\;\unit}
          {\typrod{\co{S'}}{S'}}}
        \\
        \inferrule*{
          \inferrule*{
            \LabTirName{(b)}
            \\
            \inferrule*{
              \LabTirName{(c)}
              \\
              \inferrule*{
                \inferrule*{
                  \inferrule*{
                  }{\tseq
                    {\tmty{y}{\co{S'}}}
                    {y}
                    {\co{S'}}}
                }{\tseq
                  {\tmty{y}{\co{S'}}}
                  {\inr{y}}
                  {\tysum{\co{S}}{\co{S'}}}}
                \\
                \inferrule*{
                }{\tseq
                  {\tmty{x}{\tyselect[\cs{o}]{S}{S'}}}
                  {x}
                  {\tyselect[\cs{o}]{S}{S'}}}
              }{\tseq
                {\tmty{x}{\tyselect[\cs{o}]{S}{S'}},\tmty{y}{\co{S'}}}
                {\pair{\inr{y}}{x}}
                {\typrod{(\tysum{\co{S}}{\co{S'}})}{(\tyselect[\cs{o}]{S}{S'})}}}
            }{\tseq
              {\tmty{x}{\tyselect[\cs{o}]{S}{S'}},\tmty{y}{\co{S'}}}
              {\send\;{\pair{\inr{y}}{x}}}
              {\tyends[\cs{o+1}]}}
          }{\tseq
            {\tmty{x}{\tyselect[\cs{o}]{S}{S'}},\tmty{y}{\co{S'}}}
            {\close\;(\send\;{\pair{\inr{y}}{x}})}
            {\tyunit}}
          \\
          \inferrule*{
          }{\tseq
            {\tmty{z}{S'}}
            {z}
            {S'}}
        }{\tseq
          {\tmty{x}{\tyselect[\cs{o}]{S}{S'}},\tmty{y}{\co{S'}},\tmty{z}{S'}}
          {\andthen{\close\;(\send\;{\pair{\inr{y}}{x}})}{z}}
          {S'}}
      }{\tseq
        {\tmty{x}{\tyselect[\cs{o}]{S}{S'}}}
        {\letpair{y}{z}{\new\;\unit}{\andthen{\close\;(\send\;{\pair{\inr{y}}{x}})}{z}}}
        {S'}}
    }{\tseq
      {\emptyenv}
      {\lambda x.\letpair{y}{z}{\new\;\unit}{\andthen{\close\;(\send\;{\pair{\inr{y}}{x}})}{z}}}
      {\tylolli[\cs{\ptop},\cs{o}]{\tyselect[\cs{o}]{S}{S'}}{S'}}}
    \\
    \LabTirName{(a)} = 
    \tmty{\new}
    {\tylolli[\cs{\ptop},\cs{o}]{\tyunit}{\typrod{\co{S'}}{S'}}}
  
    \LabTirName{(b)} =
    \tmty{\close}
    {\tylolli[\cs{\ptop},\cs{o+1}]{\tyends[\cs{o+1}]}{\tyunit}}
  
    \LabTirName{(c)} =
    \tmty{\send}
    {\tylolli[\cs{\ptop},\cs{o}]
      {\typrod{(\tysum{\co{S}}{\co{S'}})}{(\tyselect[\cs{o}]{S}{S'})}}{\tyends[\cs{o+1}]}}
  \end{mathpar}
  \caption{Typing Rules for Syntactic Sugar for PGV (\LabTirName{T-Select-Inl} and \LabTirName{T-Select-Inr}).
  \label{fig:pgv-typing-sugar-select}}
\end{figure}

\begin{figure}
  \begin{mathpar}
    \mprset{sep=0.5em}
    \inferrule*[lab=T-Offer]{
      {\tseq[\cs{p}]
        {\ty{\Gamma}}
        {L}
        {\tyoffer[\cs{o}]{S}{S'}}}
      \\
      {\tseq[\cs{q}]
        {\ty{\Delta},\tmty{x}{S}}
        {M}
        {T}}
      \\
      {\tseq[\cs{q}]
        {\ty{\Delta},\tmty{y}{S'}}
        {N}
        {T}}
      \\
      \cs{o}\sqcup\cs{p}<\minpr(\ty{\Delta},\ty{S},\ty{S'})
    }{\tseq[\cs{o}\sqcup\cs{p}\sqcup\cs{q}]
      {\ty{\Gamma},\ty{\Delta}}
      {\offer{L}{x}{M}{y}{N}}
      {T}}
    \elabarrow
    \\
    \inferrule*[lab=(c)]{
      \inferrule*{
        \LabTirName{(b)}
        \\
        \inferrule*{
        }{\tseq[\cs{\pbot}]
          {\tmty{w}{\tyendr[\cs{o+1}]}}
          {w}
          {\tyendr[\cs{o+1}]}}
      }{\tseq[\cs{o}]
        {\tmty{w}{\tyendr[\cs{o+1}]}}
        {\wait\;{w}}
        {\tyunit}}
      \\
      \inferrule*{
        \inferrule*{
        }{\tseq[\cs{\pbot}]
          {\tmty{z}{\tysum{S}{S'}}}
          {z}
          {\tysum{S}{S'}}}
        \\
        {\tseq[\cs{q}]
          {\ty{\Delta},\tmty{x}{S}}
          {M}
          {T}}
        \\
        {\tseq[\cs{q}]
          {\ty{\Delta},\tmty{y}{S'}}
          {N}
          {T}}
      }{\tseq[\cs{q}]
        {\ty{\Delta},\tmty{z}{\tysum{S}{S'}}}
        {\casesum{z}{x}{M}{y}{N}}
        {T}}
      \\
      \cs{o}<\minpr(\ty{\Delta},\ty{\tysum{S}{S'}})
    }{
      \tseq[\cs{o}\sqcup\cs{q}]
      {\ty{\Delta},\tmty{z}{\tysum{S}{S'}},\tmty{w}{\tyendr[\cs{o+1}]}}
      {\andthen{\wait\;{w}}{\casesum{z}{x}{M}{y}{N}}}
      {T}}
    \\
    \inferrule*{
      \inferrule*{
        \LabTirName{(a)}
        \inferrule*{
        }{}
        \\
        {\tseq[\cs{p}]
          {\ty{\Gamma}}
          {L}
          {\tyrecv[\cs{o}]{(\tysum{S}{S'})}{\tyendr[\cs{o+1}]}}}
      }{\tseq[\cs{o}\sqcup\cs{p}]
        {\ty{\Gamma}}
        {\recv\;{L}}
        {\typrod{(\tysum{S}{S'})}{\tyendr[\cs{o+1}]}}}
      \\
      \LabTirName{(c)}
      \\
     \cs{o}\sqcup\cs{p}<\minpr(\ty{\Delta})
    }{\tseq[\cs{o}\sqcup\cs{p}\sqcup\cs{q}]
      {\ty{\Gamma},\ty{\Delta}}
      {\letpair{z}{w}{\recv\;{L}}{\andthen{\wait\;{w}}{\casesum{z}{x}{M}{y}{N}}}}
      {T}}
    \\
    \LabTirName{(a)} =
    \tmty{\recv}
    {\tylolli[\cs{\ptop},\cs{o}]
      {\tyrecv[\cs{o}]{(\tysum{S}{S'})}{\tyendr[\cs{o+1}]}}
      {\typrod{(\tysum{S}{S'})}{\tyendr[\cs{o+1}]}}}

    \LabTirName{(b)} =
    \tmty{\wait}
    {\tylolli[\cs{\ptop},\cs{o}]{\tyendr[\cs{o+1}]}{\tyunit}} 
    \\
    \\
    \inferrule*[lab=T-Offer-Absurd]{
      \tseq[\cs{p}]
      {\ty{\Gamma}}
      {L}
      {\tyofferemp[\cs{o}]}
      \\
      \cs{o}\sqcup\cs{p}<\minpr(\ty{\Delta})
    }{\tseq[\cs{o}\sqcup\cs{p}]
      {\ty{\Gamma},\ty{\Delta}}
      {\offeremp{L}}
      {T}}
    \elabarrow
    \inferrule*{
      \inferrule*{
        \LabTirName{(a)}
        \\
        {\tseq[\cs{p}]
          {\ty{\Gamma}}
          {L}
          {\tyrecv[\cs{o}]{\tyvoid}{\tyendr[\cs{o+1}]}}}
      }{\tseq[\cs{o}\sqcup\cs{p}]
        {\ty{\Gamma}}
        {\recv\;{L}}
        {\typrod{\tyvoid}{\tyendr[\cs{o+1}]}}}
      \\
      \inferrule*{
        \inferrule*{
          \LabTirName{(b)}
          \\
          \inferrule*{
          }{\tseq[\cs{\pbot}]
            {\tmty{w}{\tyendr[\cs{o+1}]}}
            {w}
            {\tyendr[\cs{o+1}]}}
        }{\tseq[\cs{o}]
          {\tmty{w}{\tyendr[\cs{o+1}]}}
          {\wait\;{w}}
          {\tyunit}}
        \\
        \inferrule*{
          \inferrule*{
          }{\tseq[\cs{\pbot}]{\tmty{z}{\tyvoid}}{z}{\tyvoid}}
        }{\tseq[\cs{\pbot}]
          {\ty{\Delta},\tmty{z}{\tyvoid}}
          {\absurd{z}}
          {T}}
        \\
        \cs{o}<\minpr(\ty{\Delta})
      }{\tseq[\cs{o}]
        {\ty{\Delta},\tmty{z}{\tyvoid},\tmty{w}{\tyendr[\cs{o+1}]}}
        {\andthen{\wait\;{w}}{\absurd{z}}}
        {T}}
      \\
      \cs{o}\sqcup\cs{p}<\minpr(\ty{\Delta})
    }{\tseq[\cs{o}\sqcup\cs{p}]
      {\ty{\Gamma},\ty{\Delta}}
      {\letpair{z}{w}{\recv\;{L}}{\andthen{\wait\;{w}}{\absurd{z}}}}
      {T}}
    \\
    \LabTirName{(a)} =
    \tmty{\recv}
    {\tylolli[\cs{\ptop},\cs{o}]
      {\tyrecv[\cs{o}]{\tyvoid}{\tyendr[\cs{o+1}]}}
      {\typrod{\tyvoid}{\tyendr[\cs{o+1}]}}}
  
    \LabTirName{(b)} =
    \tmty{\wait}
    {\tylolli[\cs{\ptop},\cs{o}]{\tyendr[\cs{o+1}]}{\tyunit}}
  \end{mathpar}
  \caption{Typing Rules for Syntactic Sugar for PGV (\LabTirName{T-Offer} and \LabTirName{T-Offer-Absurd}).}
  \label{fig:pgv-typing-sugar-offer}
\end{figure}


The following typing rules given in \cref{{fig:pgv-typing-sugar},fig:pgv-typing-sugar-select,fig:pgv-typing-sugar-offer}, cover syntactic sugar typing for PGV.

\section{Technical Developments}
\subsection{Subject Reduction}
Unlike with previous versions of GV, structural congruence, term reduction, and configuration reduction are all type preserving.

We must show that substitution preserves priority constraints. For this, we prove~\cref{lem:pgv-value-done}, which shows that values have finished all their communication, and that any priorities in the type of the value come from the typing environment.
\begin{lem}\label{lem:pgv-value-done}
  If $\tseq[\cs{p}]{\ty{\Gamma}}{V}{T}$, then $\cs{p}=\cs{\pbot}$, and $\minpr(\ty{\Gamma})=\minpr(\ty{T})$.
\end{lem}
\proof
\label{prf:lem-pgv-value-done}
By induction on the derivation of $\tseq[\cs{o}]{\ty{\Gamma}}{V}{T}$.

\begin{case*}[\LabTirName{T-Var}]
  Immediately.
  \begin{mathpar}
    \inferrule*{
    }{\tseq[\cs{\pbot}]{\tmty{x}{T}}{x}{T}}
  \end{mathpar}
\end{case*}
\begin{case*}[\LabTirName{T-Const}]
  Immediately.
  \begin{mathpar}
    \inferrule*{
    }{\tseq[\cs{\pbot}]{\emptyenv}{K}{T}}
  \end{mathpar}
\end{case*}
\begin{case*}[\LabTirName{T-Lam}]
  Immediately.
  \begin{mathpar}
    \inferrule*{
      \tseq[\cs{q}]{\ty{\Gamma},\tmty{x}{T}}{M}{U}
    }{\tseq[\cs{\pbot}]{\ty{\Gamma}}{\lambda x.M}{\tylolli[\cs{\pr(\ty{\Gamma})},\cs{q}]{T}{U}}}
  \end{mathpar}
\end{case*}
\begin{case*}[\LabTirName{T-Unit}]
  Immediately.
  \begin{mathpar}
    \inferrule*{
    }{\tseq[\cs{\pbot}]{\emptyenv}{\unit}{\tyunit}}
  \end{mathpar}
\end{case*}
\begin{case*}[\LabTirName{T-Pair}]
  The induction hypotheses give us $\cs{p}=\cs{q}=\cs{\pbot}$, hence $\cs{p}\sqcup\cs{q}=\cs{\pbot}$, and $\pr(\ty{\Gamma})=\pr(\ty{T})$ and $\pr(\ty{\Delta})=\pr(\ty{U})$, hence $\pr(\ty{\Gamma},\ty{\Delta})=\pr(\ty{\Gamma})\sqcap\pr(\ty{\Delta})=\pr(\ty{T})\sqcap\pr(\ty{U})=\pr(\ty{\typrod{T}{U}})$.
  \begin{mathpar}
    \inferrule*{
      \tseq[\cs{p}]{\ty{\Gamma}}{V}{T}
      \\
      \tseq[\cs{q}]{\ty{\Delta}}{W}{U}
      \\
      \cs{p}<\pr(\ty{\Delta})
    }{\tseq[\cs{p}\sqcup\cs{q}]{\ty{\Gamma},\ty{\Delta}}{\pair{V}{W}}{\typrod{T}{U}}}
  \end{mathpar}
\end{case*}
\begin{case*}[\LabTirName{T-Inl}]
  The induction hypothesis gives us $\cs{p}=\cs{\pbot}$, and $\pr(\ty{\Gamma})=\pr{(\ty{T})}$. We know $\pr(\ty{T})=\pr({\ty{U}})$, hence $\pr(\ty{\Gamma})=\pr(\ty{\tysum{T}{U}})$.
  \begin{mathpar}
    \inferrule*{
      \tseq[\cs{p}]{\ty{\Gamma}}{V}{T}
      \\
      \pr(\ty{T})=\pr(\ty{U})
    }{\tseq[\cs{p}]{\ty{\Gamma}}{\inl{V}}{\tysum{T}{U}}}
  \end{mathpar}
\end{case*}
\begin{case*}[\LabTirName{T-Inr}]
  The induction hypothesis gives us $\cs{p}=\cs{\pbot}$, and $\pr(\ty{\Gamma})=\pr{(\ty{U})}$. We know $\pr(\ty{T})=\pr({\ty{U}})$, hence $\pr(\ty{\Gamma})=\pr(\ty{\tysum{T}{U}})$.
  \begin{mathpar}
    \inferrule*{
      \tseq[\cs{p}]{\ty{\Gamma}}{V}{U}
      \\
      \pr(\ty{T})=\pr(\ty{U})
    }{\tseq[\cs{p}]{\ty{\Gamma}}{\inr{V}}{\tysum{T}{U}}}
  \end{mathpar}
\end{case*}
\qed


\begin{lem}\label{lem:pgv-substitution}
  If $\tseq[\cs{p}]{\ty{\Gamma},\tmty{x}{U'}}{M}{T}$ and $\tseq[\cs{q}]{\ty{\Theta}}{V}{U'}$, then $\tseq[\cs{p}]{\ty{\Gamma},\ty{\Theta}}{\subst{M}{V}{x}}{T}$.
\end{lem}
\proof
\label{prf:lem-pgv-substitution}
By induction on the derivation of $\tseq[\cs{p}]{\ty{\Gamma},\tmty{x}{U'}}{M}{T}$.

\begin{case*}[\LabTirName{T-Var}]
  By \cref{lem:pgv-value-done}, $\cs{q}=\cs{\pbot}$.
  \begin{mathpar}
    \inferrule*{
    }{\tseq[\cs{\pbot}]{\tmty{x}{U'}}{x}{U'}}
    \substarrow{V}{x}
    \tseq[\cs{\pbot}]{\ty{\Theta}}{V}{U'}
  \end{mathpar}
\end{case*}
\begin{case*}[\LabTirName{T-Lam}]
  By \cref{lem:pgv-value-done}, $\minpr(\ty{\Theta})=\minpr(\ty{U'})$, hence $\minpr(\ty{\Gamma},\ty{\Theta})=\minpr(\ty{\Gamma},\ty{U'})$.
  \begin{mathpar}
    \inferrule*{
      \tseq[\cs{q}]{\ty{\Gamma},\tmty{x}{U'},\tmty{y}{T}}{M}{U}
    }{\tseq[\cs{\pbot}]{\ty{\Gamma},\tmty{x}{U'}}
      {\lambda y.M}
      {\tylolli[\cs{\minpr(\ty{\Gamma},\ty{U'})},\cs{q}]{T}{U}}}
    \substarrow{V}{x}
    \inferrule*{
      \tseq[\cs{q}]{\ty{\Gamma},\ty{\Theta},\tmty{y}{T}}{\subst{M}{V}{x}}{U}
    }{\tseq[\cs{\pbot}]{\ty{\Gamma},\ty{\Theta}}
      {\lambda y.\subst{M}{V}{x}}
      {\tylolli[\cs{\minpr(\ty{\Gamma},\ty{\Theta})},\cs{q}]{T}{U}}}
  \end{mathpar}
\end{case*}
\begin{case*}[\LabTirName{T-App}]
  There are two subcases:
  \begin{subcase*}[$\tm{x}\in\tm{M}$]
    Immediately, from the induction hypothesis.
    \begin{mathpar}
      \inferrule*{
        \tseq[\cs{p}]{\ty{\Gamma},\tmty{x}{U'}}{M}{\tylolli[\cs{p'},\cs{q'}]{T}{U}}
        \\
        \tseq[\cs{q}]{\ty{\Delta}}{N}{T}
        \\
        \cs{p}<\minpr(\ty{\Delta})
        \\
        \cs{q}<\cs{p'}
      }{\tseq[\cs{p}\sqcup\cs{q}\sqcup\cs{q'}]{\ty{\Gamma},\ty{\Delta},\tmty{x}{U'}}{M\;N}{U}}
      \substarrow{V}{x}
      \inferrule*{
        \tseq[\cs{p}]{\ty{\Gamma},\ty{\Theta}}{\subst{M}{V}{x}}{\tylolli[\cs{p'},\cs{q'}]{T}{U}}
        \\
        \tseq[\cs{q}]{\ty{\Delta}}{N}{T}
        \\
        \cs{p}<\minpr(\ty{\Delta})
        \\
        \cs{q}<\cs{p'}
      }{\tseq[\cs{p}\sqcup\cs{q}\sqcup\cs{q'}]{\ty{\Gamma},\ty{\Delta},\ty{\Theta}}{(\subst{M}{V}{x})\;N}{U}}
    \end{mathpar}
  \end{subcase*}
  \begin{subcase*}[$\tm{x}\in\tm{N}$]
    By \cref{lem:pgv-value-done}, $\minpr(\ty{\Theta})=\minpr(\ty{U'})$, hence $\minpr(\ty{\Delta},\ty{\Theta})=\minpr(\ty{\Delta},\ty{U'})$.
    \begin{mathpar}
      \inferrule*{
        \tseq[\cs{p}]{\ty{\Gamma}}{M}{\tylolli[\cs{p'},\cs{q'}]{T}{U}}
        \\
        \tseq[\cs{q}]{\ty{\Delta},\tmty{x}{U'}}{N}{T}
        \\
        \cs{p}<\minpr(\ty{\Delta},\ty{U'})
        \\
        \cs{q}<\cs{p'}
      }{\tseq[\cs{p}\sqcup\cs{q}\sqcup\cs{q'}]{\ty{\Gamma},\ty{\Delta},\tmty{x}{U'}}{M\;N}{U}}
      \substarrow{V}{x}
      \inferrule*{
        \tseq[\cs{p}]{\ty{\Gamma}}{M}{\tylolli[\cs{p'},\cs{q'}]{T}{U}}
        \\
        \tseq[\cs{q}]{\ty{\Delta},\ty{\Theta}}{\subst{N}{V}{x}}{T}
        \\
        \cs{p}<\minpr(\ty{\Delta},\ty{\Theta})
        \\
        \cs{q}<\cs{p'}
      }{\tseq[\cs{p}\sqcup\cs{q}\sqcup\cs{q'}]{\ty{\Gamma},\ty{\Delta},\ty{\Theta}}{M\;(\subst{N}{V}{x})}{U}}
    \end{mathpar}
  \end{subcase*}
\end{case*}
\begin{case*}[\LabTirName{T-LetUnit}]
  There are two subcases:
  \begin{subcase*}[$\tm{x}\in\tm{M}$]
    Immediately, from the induction hypothesis.
    \begin{mathpar}
      \inferrule*{
        \tseq[\cs{p}]{\ty{\Gamma},\tmty{x}{U'}}{M}{\tyunit}
        \\
        \tseq[\cs{q}]{\ty{\Delta}}{N}{T}
        \\
        \cs{p}<\minpr(\ty{\Delta})
      }{\tseq[\cs{p}\sqcup\cs{q}]{\ty{\Gamma},\ty{\Delta},\tmty{x}{U'}}{\letunit{M}{N}}{T}}
      \substarrow{V}{x}
      \inferrule*{
        \tseq[\cs{p}]{\ty{\Gamma},\ty{\Theta}}{\subst{M}{V}{x}}{\tyunit}
        \\
        \tseq[\cs{q}]{\ty{\Delta}}{N}{T}
        \\
        \cs{p}<\minpr(\ty{\Delta})
      }{\tseq[\cs{p}\sqcup\cs{q}]{\ty{\Gamma},\ty{\Delta},\ty{\Theta}}{\letunit{\subst{M}{V}{x}}{N}}{T}}
    \end{mathpar}
  \end{subcase*}
  \begin{subcase*}[$\tm{x}\in\tm{N}$]
    By \cref{lem:pgv-value-done}, $\minpr(\ty{\Theta})=\minpr(\ty{U'})$, hence $\minpr(\ty{\Delta},\ty{\Theta})=\minpr(\ty{\Delta},\ty{U'})$.
    \begin{mathpar}
      \inferrule*{
        \tseq[\cs{p}]{\ty{\Gamma}}{M}{\tyunit}
        \\
        \tseq[\cs{q}]{\ty{\Delta},\tmty{x}{U'}}{N}{T}
        \\
        \cs{p}<\minpr(\ty{\Delta},\ty{U'})
      }{\tseq[\cs{p}\sqcup\cs{q}]{\ty{\Gamma},\ty{\Delta},\tmty{x}{U'}}{\letunit{M}{N}}{T}}
      \substarrow{V}{x}
      \inferrule*{
        \tseq[\cs{p}]{\ty{\Gamma}}{M}{\tyunit}
        \\
        \tseq[\cs{q}]{\ty{\Delta},\ty{\Theta}}{\subst{N}{V}{x}}{T}
        \\
        \cs{p}<\minpr(\ty{\Delta},\ty{\Theta})
      }{\tseq[\cs{p}\sqcup\cs{q}]{\ty{\Gamma},\ty{\Delta},\ty{\Theta}}{\letunit{M}{\subst{N}{V}{x}}}{T}}
    \end{mathpar}
  \end{subcase*}
\end{case*}
\begin{case*}[\LabTirName{T-Pair}]
  There are two subcases:
  \begin{subcase*}[$\tm{x}\in\tm{M}$]
    Immediately, from the induction hypothesis.
    \begin{mathpar}
      \inferrule*{
        \tseq[\cs{p}]{\ty{\Gamma},\tmty{x}{U'}}{M}{T}
        \\
        \tseq[\cs{q}]{\ty{\Delta}}{N}{U}
        \\
        \cs{p}<\minpr(\ty{\Delta},\ty{U'})
      }{\tseq[\cs{p}\sqcup\cs{q}]{\ty{\Gamma},\ty{\Delta},\tmty{x}{U'}}{\pair{M}{N}}{\typrod{T}{U}}}
      \substarrow{V}{x}
      \inferrule*{
        \tseq[\cs{p}]{\ty{\Gamma},\ty{\Theta}}{\subst{M}{V}{x}}{T}
        \\
        \tseq[\cs{q}]{\ty{\Delta}}{N}{U}
        \\
        \cs{p}<\minpr(\ty{\Delta},\ty{\Theta})
      }{\tseq[\cs{p}\sqcup\cs{q}]{\ty{\Gamma},\ty{\Delta},\ty{\Theta}}{\pair{\subst{M}{V}{x}}{N}}{\typrod{T}{U}}}
    \end{mathpar}
  \end{subcase*}
  \begin{subcase*}[$\tm{x}\in\tm{N}$]
    By \cref{lem:pgv-value-done}, $\minpr(\ty{\Theta})=\minpr(\ty{U'})$, hence $\minpr(\ty{\Delta},\ty{\Theta})=\minpr(\ty{\Delta},\ty{U'})$.
    \begin{mathpar}
      \inferrule*{
        \tseq[\cs{p}]{\ty{\Gamma}}{M}{T}
        \\
        \tseq[\cs{q}]{\ty{\Delta},\tmty{x}{U'}}{N}{U}
        \\
        \cs{p}<\minpr(\ty{\Delta},\ty{U'})
      }{\tseq[\cs{p}\sqcup\cs{q}]{\ty{\Gamma},\ty{\Delta},\tmty{x}{U'}}{\pair{M}{N}}{\typrod{T}{U}}}
      \substarrow{V}{x}
      \inferrule*{
        \tseq[\cs{p}]{\ty{\Gamma}}{M}{T}
        \\
        \tseq[\cs{q}]{\ty{\Delta},\ty{\Theta}}{\subst{N}{V}{x}}{U}
        \\
        \cs{p}<\minpr(\ty{\Delta},\ty{\Theta})
      }{\tseq[\cs{p}\sqcup\cs{q}]{\ty{\Gamma},\ty{\Delta},\ty{\Theta}}{\pair{M}{\subst{N}{V}{x}}}{\typrod{T}{U}}}
    \end{mathpar}
  \end{subcase*}
\end{case*}
\begin{case*}[\LabTirName{T-LetPair}]
  There are two subcases:
  \begin{subcase*}[$\tm{x}\in\tm{M}$]
    Immediately, from the induction hypothesis.
    \begin{mathpar}
      \inferrule*{
        \tseq[\cs{p}]{\ty{\Gamma},\tmty{x}{U'}}{M}{\typrod{T}{T'}}
        \\
        \tseq[\cs{q}]{\ty{\Delta},\tmty{y}{T},\tmty{z}{T'}}{N}{U}
        \\
        \cs{p}<\minpr(\ty{\Delta},\ty{T},\ty{T'})
      }{\tseq[\cs{p}\sqcup\cs{q}]{\ty{\Gamma},\ty{\Delta},\tmty{x}{U'}}{\letpair{y}{z}{M}{N}}{U}}
      \substarrow{V}{x}
      \inferrule*{
        \tseq[\cs{p}]{\ty{\Gamma},\ty{\Theta}}{\subst{M}{V}{x}}{\typrod{T}{T'}}
        \\
        \tseq[\cs{q}]{\ty{\Delta},\tmty{y}{T},\tmty{z}{T'}}{N}{U}
        \\
        \cs{p}<\minpr(\ty{\Delta},\ty{T},\ty{T'})
      }{\tseq[\cs{p}\sqcup\cs{q}]{\ty{\Gamma},\ty{\Delta},\ty{\Theta}}{\letpair{y}{z}{\subst{M}{V}{x}}{N}}{U}}
    \end{mathpar}
  \end{subcase*}
  \begin{subcase*}[$\tm{x}\in\tm{N}$]
    By \cref{lem:pgv-value-done}, $\minpr(\ty{\Theta})=\minpr(\ty{U'})$, hence $\minpr(\ty{\Delta},\ty{\Theta},\ty{T},\ty{T'})=\minpr(\ty{\Delta},\ty{U'},\ty{T},\ty{T'})$.
    \begin{mathpar}
      \inferrule*{
        \tseq[\cs{p}]{\ty{\Gamma}}{M}{\typrod{T}{T'}}
        \\
        \tseq[\cs{q}]{\ty{\Delta},\tmty{x}{U'},\tmty{y}{T},\tmty{z}{T'}}{N}{U}
        \\
        \cs{p}<\minpr(\ty{\Delta},\ty{U'},\ty{T},\ty{T'})
      }{\tseq[\cs{p}\sqcup\cs{q}]{\ty{\Gamma},\ty{\Delta},\tmty{x}{U'}}{\letpair{y}{z}{M}{N}}{U}}
      \substarrow{V}{x}
      \inferrule*{
        \tseq[\cs{p}]{\ty{\Gamma}}{M}{\typrod{T}{T'}}
        \\
        \tseq[\cs{q}]{\ty{\Delta},\ty{\Theta},\tmty{y}{T},\tmty{z}{T'}}{\subst{N}{V}{x}}{U}
        \\
        \cs{p}<\minpr(\ty{\Delta},\ty{\Theta},\ty{T},\ty{T'})
      }{\tseq[\cs{p}\sqcup\cs{q}]{\ty{\Gamma},\ty{\Delta},\ty{\Theta}}{\letpair{y}{z}{M}{\subst{N}{V}{x}}}{U}}
    \end{mathpar}
  \end{subcase*}
\end{case*}
\begin{case*}[\LabTirName{T-Absurd}]
  \begin{mathpar}
    \inferrule*{
      \tseq[\cs{p}]{\ty{\Gamma},\tmty{x}{U'}}{M}{\tyvoid}
    }{\tseq[\cs{p}]{\ty{\Gamma},\ty{\Delta},\tmty{x}{U'}}{\absurd{M}}{T}}
    \substarrow{V}{x}
    \inferrule*{
      \tseq[\cs{p}]{\ty{\Gamma},\ty{\Theta}}{\subst{M}{V}{x}}{\tyvoid}
    }{\tseq[\cs{p}]{\ty{\Gamma},\ty{\Delta},\ty{\Theta}}{\absurd{\subst{M}{V}{x}}}{T}}
  \end{mathpar}
\end{case*}
\begin{case*}[\LabTirName{T-Inl}]
  \begin{mathpar}
    \inferrule*{
      \tseq[\cs{p}]{\ty{\Gamma},\tmty{x}{U'}}{M}{T}
      \\
      \minpr(\ty{T})=\minpr(\ty{U})
    }{\tseq[\cs{p}]{\ty{\Gamma},\tmty{x}{U'}}{\inl{M}}{\tysum{T}{U}}}
    \substarrow{V}{x}
    \inferrule*{
      \tseq[\cs{p}]{\ty{\Gamma},\ty{\Theta}}{\subst{M}{V}{x}}{T}
      \\
      \minpr(\ty{T})=\minpr(\ty{U})
    }{\tseq[\cs{p}]{\ty{\Gamma},\ty{\Theta}}{\inl{\subst{M}{V}{x}}}{\tysum{T}{U}}}
  \end{mathpar}
\end{case*}
\begin{case*}[\LabTirName{T-Inr}]
  \begin{mathpar}
    \inferrule*{
      \tseq[\cs{p}]{\ty{\Gamma},\tmty{x}{U'}}{M}{U}
      \\
      \minpr(\ty{T})=\minpr(\ty{U})
    }{\tseq[\cs{p}]{\ty{\Gamma},\tmty{x}{U'}}{\inr{M}}{\tysum{T}{U}}}
    \substarrow{V}{x}
    \inferrule*{
      \tseq[\cs{p}]{\ty{\Gamma},\ty{\Theta}}{\subst{M}{V}{x}}{U}
      \\
      \minpr(\ty{T})=\minpr(\ty{U})
    }{\tseq[\cs{p}]{\ty{\Gamma},\ty{\Theta}}{\inr{\subst{M}{V}{x}}}{\tysum{T}{U}}}
  \end{mathpar}
\end{case*}
\begin{case*}[\LabTirName{T-CaseSum}]
  There are two subcases:
  \begin{subcase*}[$\tm{x}\in\tm{L}$]
    Immediately, from the induction hypothesis.
    \begin{mathpar}
      \inferrule*{
        \tseq[\cs{p}]{\ty{\Gamma},\tmty{x}{U'}}{L}{\tysum{T}{T'}}
        \\
        \tseq[\cs{q}]{\ty{\Delta},\tmty{y}{T}}{M}{U}
        \\
        \tseq[\cs{q}]{\ty{\Delta},\tmty{z}{T'}}{N}{U}
        \\
        \cs{p}<\minpr(\ty{\Delta})
      }{\tseq[\cs{p}\sqcup\cs{q}]{\ty{\Gamma},\ty{\Delta},\tmty{x}{U'}}{\casesum{L}{y}{M}{z}{N}}{U}}
      \substarrow{V}{x}
      \inferrule*{
        \tseq[\cs{p}]{\ty{\Gamma},\ty{\Theta}}{\subst{L}{V}{x}}{\tysum{T}{T'}}
        \\
        \tseq[\cs{q}]{\ty{\Delta},\tmty{y}{T}}{M}{U}
        \\
        \tseq[\cs{q}]{\ty{\Delta},\tmty{z}{T'}}{N}{U}
        \\
        \cs{p}<\minpr(\ty{\Delta})
      }{\tseq[\cs{p}\sqcup\cs{q}]{\ty{\Gamma},\ty{\Delta},\ty{\Theta}}{\casesum{\subst{L}{V}{x}}{y}{M}{z}{N}}{U}}
    \end{mathpar}
  \end{subcase*}
  \begin{subcase*}[$\tm{x}\in\tm{M}$ and $\tm{x}\in\tm{N}$]
    By \cref{lem:pgv-value-done}, $\minpr(\ty{\Theta})=\minpr(\ty{U'})$, hence $\minpr(\ty{\Delta},\ty{\Theta},\ty{T})=\minpr(\ty{\Delta},\ty{U'},\ty{T})$ and $\minpr(\ty{\Delta},\ty{\Theta},\ty{T'})=\minpr(\ty{\Delta},\ty{U'},\ty{T'})$.
    \begin{mathpar}
      \inferrule*{
        \tseq[\cs{p}]{\ty{\Gamma}}{L}{\tysum{T}{T'}}
        \\
        \tseq[\cs{q}]{\ty{\Delta},\tmty{x}{U'},\tmty{y}{T}}{M}{U}
        \\
        \tseq[\cs{q}]{\ty{\Delta},\tmty{x}{U'},\tmty{z}{T'}}{N}{U}
        \\
        \cs{p}<\minpr(\ty{\Delta},\ty{U'})
      }{\tseq[\cs{p}\sqcup\cs{q}]{\ty{\Gamma},\ty{\Delta},\tmty{x}{U'}}{\casesum{L}{y}{M}{z}{N}}{U}}
      \substarrow{V}{x}
      \inferrule*{
        \tseq[\cs{p}]{\ty{\Gamma}}{L}{\tysum{T}{T'}}
        \\
        \tseq[\cs{q}]{\ty{\Delta},\ty{\Theta},\tmty{y}{T}}{\subst{M}{V}{x}}{U}
        \\
        \tseq[\cs{q}]{\ty{\Delta},\ty{\Theta},\tmty{z}{T'}}{\subst{N}{V}{x}}{U}
        \\
        \cs{p}<\minpr(\ty{\Delta},\ty{\Theta})
      }{\tseq[\cs{p}\sqcup\cs{q}]{\ty{\Gamma},\ty{\Delta},\ty{\Theta}}{\casesum{L}{y}{\subst{M}{V}{x}}{z}{\subst{N}{V}{x}}}{U}}
    \end{mathpar}
  \end{subcase*}
\end{case*}
\noindent
We omit the cases where $\tm{x}\not\in\tm{M}$, as they are straightforward.
\qed


\begin{lem}\label{lem:pgv-subject-reduction-terms}
  If $\tseq[\cs{p}]{\ty{\Gamma}}{M}{T}$ and $\tm{M}\tred\tm{M'}$,
  then $\tseq[\cs{p}]{\ty{\Gamma}}{M'}{T}$.
\end{lem}
\proof
\label{prf:lem-pgv-subject-reduction-terms}
The proof closely follows the standard proof of subject reduction for the simply-typed linear $\lambda$-calculus, as the constants are uninterpreted by the term reduction ($\tred$) and priority constraints are maintained consequence of~\cref{prf:lem-pgv-substitution}.
\\
By induction on the derivation of $\tm{M}\tred\tm{M'}$.

\begin{case*}[\LabTirName{E-Lam}]
  By \cref{lem:pgv-substitution}.
  \begin{mathpar}
    \inferrule*{
      \inferrule*{
        \tseq[\cs{p}]{\ty{\Gamma},\tmty{x}{T}}{M}{U}
      }{\tseq[\cs{\pbot}]{\ty{\Gamma}}{\lambda x.M}{\tylolli[\cs{\pr(\ty{\Gamma})},\cs{p}]{T}{U}}}
      \\
      \tseq[\cs{\pbot}]{\ty{\Delta}}{V}{T}
    }{\tseq[\cs{p}]{\ty{\Gamma},\ty{\Delta}}{(\lambda x.M)\;V}{U}}
    \tred
    \tseq[\cs{p}]{\ty{\Gamma},\ty{\Delta}}{\subst{M}{V}{x}}{U}
  \end{mathpar}
\end{case*}
\begin{case*}[\LabTirName{E-Unit}]
  By \cref{lem:pgv-substitution}.
  \begin{mathpar}
    \inferrule*{
      \inferrule*{
      }{\tseq[\cs{\pbot}]{\emptyenv}{\unit}{\tyunit}}
      \\
      \tseq[\cs{p}]{\ty{\Gamma}}{M}{T}
    }{\tseq[\cs{p}]{\ty{\Gamma}}{\letunit{\unit}{M}}{T}}
    \tred
    \tseq[\cs{p}]{\ty{\Gamma}}{M}{T}
  \end{mathpar}
\end{case*}
\begin{case*}[\LabTirName{E-Pair}]
  By \cref{lem:pgv-substitution}.
  \begin{mathpar}
    \inferrule*{
      \inferrule*{
        \tseq[\cs{\pbot}]{\ty{\Gamma}}{V}{T}
        \\
        \tseq[\cs{\pbot}]{\ty{\Delta}}{W}{T'}
      }{\tseq[\cs{\pbot}]{\ty{\Gamma},\ty{\Delta}}{\pair{V}{W}}{\typrod{T}{T'}}}
      \\
      \tseq[\cs{p}]{\ty{\Theta},\tmty{x}{T},\tmty{y}{T'}}{M}{U}
    }{\tseq[]{\ty{\Gamma},\ty{\Delta},\ty{\Theta}}{\letpair{x}{y}{\pair{V}{W}}{M}}{U}}
    \\
    \begin{turn}{270}
      \tred
    \end{turn}
    \\
    \tseq[\cs{p}]{\ty{\Gamma},\ty{\Delta},\ty{\Theta}}{\subst{\subst{M}{V}{x}}{W}{y}}{U}
  \end{mathpar}
\end{case*}
\begin{case*}[\LabTirName{E-Inl}]
  By \cref{lem:pgv-substitution}.
  \begin{mathpar}
    \inferrule*{
      \inferrule*{
        \tseq[\cs{\pbot}]{\ty{\Gamma}}{V}{T}
      }{\tseq[\cs{\pbot}]{\ty{\Gamma}}{\inl{V}}{\tysum{T}{T'}}}
      \\
      \tseq[\cs{p}]{\ty{\Delta},\tmty{x}{T}}{M}{U}
      \\
      \tseq[\cs{p}]{\ty{\Delta},\tmty{y}{T'}}{N}{U}
    }{\tseq[\cs{p}]{\ty{\Gamma},\ty{\Delta}}{\casesum{\inl{V}}{x}{M}{y}{N}}{U}}
    \\
    \begin{turn}{270}
      \tred
    \end{turn}
    \\
    \tseq[\cs{p}]{\ty{\Gamma},\ty{\Delta}}{\subst{M}{V}{x}}{U}
  \end{mathpar}
\end{case*}
\begin{case*}[\LabTirName{E-Inr}]
  By \cref{lem:pgv-substitution}.
  \begin{mathpar}
    \inferrule*{
      \inferrule*{
        \tseq[\cs{\pbot}]{\ty{\Gamma}}{V}{T'}
      }{\tseq[\cs{\pbot}]{\ty{\Gamma}}{\inr{V}}{\tysum{T}{T'}}}
      \\
      \tseq[\cs{p}]{\ty{\Delta},\tmty{x}{T}}{M}{U}
      \\
      \tseq[\cs{p}]{\ty{\Delta},\tmty{y}{T'}}{N}{U}
    }{\tseq[\cs{p}]{\ty{\Gamma},\ty{\Delta}}{\casesum{\inr{V}}{x}{M}{y}{N}}{U}}
    \\
    \begin{turn}{270}
      \tred
    \end{turn}
    \\
    \tseq[\cs{p}]{\ty{\Gamma},\ty{\Delta}}{\subst{N}{V}{y}}{U}
  \end{mathpar}
\end{case*}
\begin{case*}[\LabTirName{E-Lift}]
  Immediately by induction on the evaluation context $\tm{E}$.
\end{case*}
\qed


\begin{lem}\label{lem:pgv-subject-congruence}
  If $\cseq[\phi]{\ty{\Gamma}}{\conf{C}}$ and $\tm{\conf{C}}\equiv\tm{\conf{C'}}$,
  then $\cseq[\phi]{\ty{\Gamma}}{\conf{C'}}$.
\end{lem}
\proof
\label{prf:lem-pgv-subject-congruence}
By induction on the derivation of $\tm{\conf{C}}\equiv\tm{\conf{C'}}$.

\begin{case*}[\LabTirName{SC-LinkSwap}]
  \begin{mathpar}
    \inferrule*{
    \inferrule*[vdots=1.5em]{
    \inferrule*{
    }{\tmty{\link}{\tylolli{\typrod{S}{\co{S}}}{\tyunit}}}
    \\
    \inferrule*{
      \inferrule*{
      }{\tseq[\cs{\pbot}]{\tmty{x}{S}}{x}{S}}
      \\
      \inferrule*{
      }{\tseq[\cs{\pbot}]{\tmty{y}{\co{S}}}{y}{\co{S}}}
    }{\tseq[\cs{\pbot}]{\tmty{x}{S},\tmty{y}{\co{S}}}{\pair{x}{y}}{\typrod{S}{\co{S}}}}
    }{\tseq[\cs{\pbot}]{\tmty{x}{S},\tmty{y}{\co{S}}}{\link\;{\pair{x}{y}}}{\tyunit}}
    }{\cseq[\phi]{\ty{\Gamma},\tmty{x}{S},\tmty{y}{\co{S}}}{\plug{\conf{F}}{\link\;{\pair{x}{y}}}}}
    \\
    \begin{turn}{270}
      \ensuremath{\equiv}
    \end{turn}
    \\
    \inferrule*{
    \inferrule*[vdots=1.5em]{
    \inferrule*{
    }{\tmty{\link}{\tylolli{\typrod{\co{S}}{S}}{\tyunit}}}
    \\
    \inferrule*{
      \inferrule*{
      }{\tseq[\cs{\pbot}]{\tmty{y}{\co{S}}}{y}{\co{S}}}
      \\
      \inferrule*{
      }{\tseq[\cs{\pbot}]{\tmty{x}{S}}{x}{S}}
    }{\tseq[\cs{\pbot}]{\tmty{x}{S},\tmty{y}{\co{S}}}{\pair{y}{x}}{\typrod{S}{\co{S}}}}
    }{\tseq[\cs{\pbot}]{\tmty{x}{S},\tmty{y}{\co{S}}}{\link\;{\pair{y}{x}}}{\tyunit}}
    }{\cseq[\phi]{\ty{\Gamma},\tmty{x}{S},\tmty{y}{\co{S}}}{\plug{\conf{F}}{\link\;{\pair{y}{x}}}}}
  \end{mathpar}
\end{case*}

\begin{case*}[\LabTirName{SC-ResLink}]
  \begin{mathpar}
    \inferrule*{
      \inferrule*{
        \inferrule*{
          \inferrule*{
          }{\tmty
            {\link}
            {\tylolli{\typrod{S}{\co{S}}}{\tyunit}}}
          \\
          \inferrule*{
            \inferrule*{
            }{\tseq[\cs{\pbot}]
              {\tmty{x}{S}}
              {x}
              {S}}
            \\
            \inferrule*{
            }{\tseq[\cs{\pbot}]
              {\tmty{y}{\co{S}}}
              {y}
              {\co{S}}}
          }{\tseq[\cs{\pbot}]
            {\tmty{x}{S},\tmty{y}{\co{S}}}
            {\pair{x}{y}}
            {\typrod{S}{\co{S}}}}
        }{\tseq[\cs{\pbot}]
          {\tmty{x}{S},\tmty{y}{\co{S}}}
          {\link\;\pair{x}{y}}
          {\tyunit}}
      }{\cseq[\phi]
        {\tmty{x}{S},\tmty{y}{\co{S}}}
        {\phi\;\link\;\pair{x}{y}}}
    }{\cseq[\phi]
      {\ty{\emptyenv}}
      {\res{x}{y}{(\phi\;\link\;\pair{x}{y})}}}
    \equiv
    \inferrule*{
      \inferrule*{
      }{\tseq[\phi]
        {\ty{\emptyenv}}
        {\unit}
        {\tyunit}}
    }{\cseq[\phi]
      {\ty{\emptyenv}}
      {\phi\;\unit}}
  \end{mathpar}
\end{case*}

\begin{case*}[\LabTirName{SC-ResSwap}]
  \begin{mathpar}
    \inferrule*{
      \cseq[\phi]
      {\ty{\Gamma},\tmty{x}{S},\tmty{y}{\co{S}}}
      {\conf{C}}
    }{\cseq[\phi]
      {\ty{\Gamma}}
      {\res{x}{y}{\conf{C}}}}
    \equiv
    \inferrule*{
      \cseq[\phi]
      {\ty{\Gamma},\tmty{x}{S},\tmty{y}{\co{S}}}
      {\conf{C}}
    }{\cseq[\phi]
      {\ty{\Gamma}}
      {\res{y}{x}{\conf{C}}}}
  \end{mathpar}
\end{case*}

\begin{case*}[\LabTirName{SC-ResComm}]
  \begin{mathpar}
    \inferrule*{
      \inferrule*{
        \cseq[\phi]{\ty{\Gamma},\tmty{x}{S},\tmty{y}{\co{S}},\tmty{z}{S'},\tmty{w}{\co{S'}}}{\conf{C}}
      }{\cseq[\phi]{\ty{\Gamma},\tmty{x}{S},\tmty{y}{\co{S}}}{\res{z}{w}{\conf{C}}}}
    }{\cseq[\phi]{\ty{\Gamma}}{\res{x}{y}{\res{z}{w}{\conf{C}}}}}
    \equiv
    \inferrule*{
      \inferrule*{
        \cseq[\phi]{\ty{\Gamma},\tmty{x}{S},\tmty{y}{\co{S}},\tmty{z}{S'},\tmty{w}{\co{S'}}}{\conf{C}}
      }{\cseq[\phi]{\ty{\Gamma},\tmty{z}{S'},\tmty{w}{\co{S'}}}{\res{x}{y}{\conf{C}}}}
    }{\cseq[\phi]{\ty{\Gamma}}{\cseq{}{\res{z}{w}{\res{x}{y}{\conf{C}}}}}}
  \end{mathpar}
\end{case*}

\begin{case*}[\LabTirName{SC-ResExt}]
  \begin{mathpar}
    \inferrule*{
      \inferrule*{
        \cseq[\phi]{\ty{\Gamma}}{\conf{C}}
        \\
        \cseq[\phi]{\ty{\Delta},\tmty{x}{S},\tmty{y}{\co{S}}}{\conf{D}}
      }{\cseq[\phi]{\ty{\Gamma},\ty{\Delta},\tmty{x}{S},\tmty{y}{\co{S}}}{(\ppar{\conf{C}}{\conf{D}})}}
    }{\cseq[\phi]{\ty{\Gamma},\ty{\Delta}}{\res{x}{y}{(\ppar{\conf{C}}{\conf{D}}})}}
    \equiv
    \inferrule*{
      \cseq[\phi]{\ty{\Gamma}}{\conf{C}}
      \\
      \inferrule*{
        \cseq[\phi]{\ty{\Delta},\tmty{x}{S},\tmty{y}{\co{S}}}{\conf{D}}
      }{\cseq[\phi]{\ty{\Delta}}{\res{x}{y}{\conf{D}}}}
    }{\cseq[\phi]{\ty{\Gamma},\ty{\Delta}}{\ppar{\conf{C}}{\res{x}{y}{\conf{D}}}}}
  \end{mathpar}
\end{case*}

\begin{case*}[\LabTirName{SC-ParNil}]
  \begin{mathpar}
    \inferrule*{
      \cseq[\phi]{\ty{\Gamma}}{\conf{C}}
      \\
      \inferrule*{
        \inferrule*{
        }{\tseq[\cs{\pbot}]{\ty{\emptyenv}}{\unit}{\tyunit}}
      }{\cseq[\child]{\ty{\emptyenv}}{\child{\unit}}}
    }{\cseq[\phi]{\ty{\Gamma}}{\ppar{\conf{C}}{\child{\unit}}}}
    \equiv
    \cseq[\phi]{\ty{\Gamma}}{\conf{C}}
  \end{mathpar}
\end{case*}

\begin{case*}[\LabTirName{SC-ParComm}]
  \begin{mathpar}
    \inferrule*{
      \cseq[\phi]{\ty{\Gamma}}{\conf{C}}
      \\
      \cseq[\phi']{\ty{\Delta}}{\conf{D}}
    }{\cseq[\phi+\phi']{\ty{\Gamma},\ty{\Delta}}{(\ppar{\conf{C}}{\conf{D}})}}
    \equiv
    \inferrule*{
      \cseq[\phi']{\ty{\Delta}}{\conf{D}}
      \\
      \cseq[\phi]{\ty{\Gamma}}{\conf{C}}
    }{\cseq[\phi'+\phi]{\ty{\Gamma},\ty{\Delta}}{(\ppar{\conf{D}}{\conf{C}})}}
  \end{mathpar}
\end{case*}

\begin{case*}[\LabTirName{SC-ParAssoc}]
  \begin{mathpar}
    \inferrule*{
      \cseq[\phi]{\ty{\Gamma}}{\conf{C}}
      \\
      \inferrule*{
        \cseq[\phi']{\ty{\Delta}}{\conf{D}}
        \\
        \cseq[\phi'']{\ty{\Theta}}{\conf{E}}
      }{\cseq[\phi'+\phi'']{\ty{\Delta},\ty{\Theta}}{(\ppar{\conf{D}}{\conf{E}})}}
    }{\cseq[\phi+\phi'+\phi'']{\ty{\Gamma},\ty{\Delta},\ty{\Theta}}{\ppar{\conf{C}}{(\ppar{\conf{D}}{\conf{E}})}}}
    \equiv
    \inferrule*{
      \inferrule*{
        \cseq[\phi]{\ty{\Gamma}}{\conf{C}}
        \\
        \cseq[\phi']{\ty{\Delta}}{\conf{D}}
      }{\cseq[\phi+\phi']{\ty{\Gamma},\ty{\Delta}}{(\ppar{\conf{C}}{\conf{D}})}}
      \\
      \cseq[\phi'']{\ty{\Theta}}{\conf{E}}
    }{\cseq[\phi+\phi'+\phi'']{\ty{\Gamma},\ty{\Delta},\ty{\Theta}}{\ppar{(\ppar{\conf{C}}{\conf{D}})}{\conf{E}}}}
  \end{mathpar}
\end{case*}
\qed


\begin{thm}\label{thm:pgv-subject-reduction-confs}
  If $\cseq[\phi]{\ty{\Gamma}}{\conf{C}}$ and $\tm{\conf{C}}\cred\tm{\conf{C'}}$,
  then $\cseq[\phi]{\ty{\Gamma}}{\conf{C'}}$.
\end{thm}
\proof
\label{prf:thm-pgv-subject-reduction-confs}
By induction on the derivation of $\tm{\conf{C}}\cred\tm{\conf{C'}}$.

\begin{case*}[\LabTirName{E-New}]
  \begin{mathpar}
    \inferrule*{
      \inferrule*[vdots=1.5em]{
        \inferrule*{
        }{\tmty{\new}{\tylolli{\tyunit}{\typrod{S}{\co{S}}}}}
        \\
        \inferrule*{
        }{\tseq[\cs{\pbot}]{\emptyenv}{\unit}{\tyunit}}
      }{\tseq[\cs{\pbot}]{\emptyenv}{\new\;\unit}{\typrod{S}{\co{S}}}}
    }{\cseq[\phi]{\ty{\Gamma}}{\plug{\conf{F}}{\new\;\unit}}}
    \cred
    \inferrule*{
      \inferrule*{
        \inferrule*[vdots=1.5em]{
          \inferrule*{
          }{\tseq[\cs{\pbot}]{\tmty{x}{S}}{x}{S}}
          \\
          \inferrule*{
          }{\tseq[\cs{\pbot}]{\tmty{y}{\co{S}}}{y}{\co{S}}}
        }{\tseq[\cs{\pbot}]{\tmty{x}{S},\tmty{y}{\co{S}}}{\pair{x}{y}}{\typrod{S}{\co{S}}}}
      }{\cseq[\phi]{\ty{\Gamma},\tmty{x}{S},\tmty{y}{\co{S}}}{\plug{\conf{F}}{\pair{x}{y}}}}
    }{\cseq[\phi]{\ty{\Gamma}}{\res{x}{y}{\plug{\conf{F}}{\pair{x}{y}}}}}
  \end{mathpar}
\end{case*}
\begin{case*}[\LabTirName{E-Spawn}]
  \begin{mathpar}
    \inferrule*{
      \inferrule*[vdots=1.5em]{
        \inferrule*{
        }{\tmty{\spawn}{\tylolli{(\tylolli[\cs{p},\cs{q}]{\tyunit}{\tyunit})}{\tyunit}}}
        \tseq[\cs{\pbot}]{\ty{\Delta}}{V}{\tylolli[\cs{p},\cs{q}]{\tyunit}{\tyunit}}
      }{\tseq[\cs{\pbot}]{\ty{\Delta}}{\spawn\;V}{\tyunit}}
    }{\cseq[\phi]{\ty{\Gamma},\ty{\Delta}}{\plug{\conf{F}}{\spawn\;V}}}
    \\
    \begin{turn}{270}
      \cred
    \end{turn}
    \\
    \inferrule*{
      \inferrule*{
        \inferrule*[vdots=1.5em]{
        }{\tseq[\cs{\pbot}]{\emptyenv}{\unit}{\tyunit}}
      }{\cseq[\phi]{\ty{\Gamma}}{\plug{\conf{F}}{\unit}}}
      \\
      \inferrule*{
        \inferrule*{
          \tseq[\cs{\pbot}]{\ty{\Delta}}{V}{\tylolli[\cs{p},\cs{q}]{\tyunit}{\tyunit}}
          \\
          \inferrule*{
          }{\tseq[\cs{\pbot}]{\emptyenv}{\unit}{\tyunit}}
        }{\tseq[\cs{q}]{\ty{\Delta}}{V\;\unit}{\tyunit}}
      }{\cseq[\child]{\ty{\Delta}}{\child\;(V\;\unit)}}
    }{\cseq[\phi]{\ty{\Gamma},\ty{\Delta}}{\ppar{\plug{\conf{F}}{\unit}}{\child\;(V\;\unit)}}}
  \end{mathpar}
\end{case*}
\begin{case*}[\LabTirName{E-Send}]
  See \cref{fig:pgv-subject-reduction-cred-send}.
\end{case*}
\begin{case*}[\LabTirName{E-Close}]
  \begin{mathpar}
    \footnotesize
    \inferrule*{
    \inferrule*{
    \inferrule*{
    \inferrule*[vdots=1.5em]{
    \inferrule*{
    }{\tmty{\close}{\tylolli[\cs{\ptop},\cs{o}]{\tyends[\cs{o}]}{\tyunit}}}
    \\
    \inferrule*{
    }{\tseq[\cs{\pbot}]{\tmty{x}{\tyends[\cs{o}]}}{x}{\tyends[\cs{o}]}}
    }{\tseq[\cs{o}]{\tmty{x}{\tyends[\cs{o}]}}{\close\;{x}}{\tyunit}}
    }{\cseq[\phi]
    {\ty{\Gamma},\tmty{x}{\tyends[\cs{o}]}}
    {\plug{\conf{F}}{\close\;{x}}}}
    \and
    \inferrule*{
    \inferrule*[vdots=1.5em]{
    \inferrule*{
    }{\tmty{\wait}{\tylolli[\cs{\ptop},\cs{o}]{\tyendr[\cs{o}]}{\tyunit}}}
    \\
    \inferrule*{
    }{\tseq[\cs{\pbot}]{\tmty{y}{\tyendr[\cs{o}]}}{y}{\tyendr[\cs{o}]}}
    }{\tseq[\cs{o}]{\tmty{y}{\tyendr[\cs{o}]}}{\wait\;{y}}{\tyunit}}
    }{\cseq[\phi']
    {\ty{\Delta},\tmty{y}{\tyendr[\cs{o}]}}
    {\plug{\conf{F'}}{\wait\;{y}}}}
    }{\cseq[\phi+\phi']
    {\ty{\Gamma},\ty{\Delta},
    \tmty{x}{\tyends[\cs{o}]},\tmty{y}{\tyendr[\cs{o}]}}
    {\ppar{\plug{\conf{F}}{\close\;{x}}}{\plug{\conf{F'}}{\wait\;{y}}}}}
    }{\cseq[\phi+\phi']
    {\ty{\Gamma},\ty{\Delta}}
    {\res{x}{y}{(\ppar
    {\plug{\conf{F}}{\close\;{x}}}
    {\plug{\conf{F'}}{\wait\;{y}}})}}}
    \\
    \begin{turn}{270}
      \cred
    \end{turn}
    \\
    \inferrule*{
      \inferrule*{
        \inferrule*[vdots=1.5em]{
        }{\tseq[\cs{\pbot}]{\emptyenv}{\unit}{\tyunit}}
      }{\cseq[\phi]{\ty{\Gamma}}{\plug{\conf{F}}{\unit}}}
      \\
      \inferrule*{
        \inferrule*[vdots=1.5em]{
        }{\tseq[\cs{\pbot}]{\emptyenv}{\unit}{\tyunit}}
      }{\cseq[\phi']
        {\ty{\Delta}}
        {\plug{\conf{F'}}{\unit}}}
    }{\cseq[\phi+\phi']
      {\ty{\Gamma},\ty{\Delta}}
      {\ppar{\plug{\conf{F}}{\unit}}{\plug{\conf{F'}}{\unit}}}}
  \end{mathpar}
\end{case*}
\begin{case*}[\LabTirName{E-LiftC}]
  By induction on the evaluation context $\ty{\conf{G}}$.
\end{case*}
\begin{case*}[\LabTirName{E-LiftM}]
  By \cref{lem:pgv-subject-reduction-terms}.
\end{case*}
\begin{case*}[\LabTirName{E-LiftSC}]
  By \cref{lem:pgv-subject-congruence}.
\end{case*}
\qed
\begin{figure}[ht!]
  \begin{mathpar}
    \inferrule*{
    \inferrule*{
    \LabTirName{(a)}
    \\
    \LabTirName{(b)}
    }{\cseq[\phi+\phi']
    {\ty{\Gamma},\ty{\Delta},\ty{\Theta},
    \tmty{x}{\tysend[\cs{o}]{T}{S}},\tmty{y}{\tyrecv[\cs{o}]{T}{\co{S}}}}
    {\ppar{\plug{\conf{F}}{\send\;{\pair{V}{x}}}}{\plug{\conf{F'}}{\recv\;{y}}}}}
    }{\cseq[\phi+\phi']
    {\ty{\Gamma},\ty{\Delta},\ty{\Theta}}
    {\res{x}{y}{(\ppar
    {\plug{\conf{F}}{\send\;{\pair{V}{x}}}}
    {\plug{\conf{F'}}{\recv\;{y}}})}}}

    \LabTirName{(a)}
    \elabarrow
    \inferrule*{
    \inferrule*[vdots=1.5em]{
    \inferrule*{
    }{\tmty{\send}{\tylolli[\cs{\ptop},\cs{o}]{\typrod{T}{\tysend[\cs{o}]{T}{S}}}{S}}}
    \inferrule*{
      \tseq[\cs{p}]{\ty{\Delta}}{V}{T}
      \\
      \inferrule*{
      }{\tseq[\cs{\pbot}]{\tmty{x}{\tysend[\cs{o}]{T}{S}}}{x}{\tysend[\cs{o}]{T}{S}}}
    }{\tseq[\cs{p}]{\ty{\Delta},\tmty{x}{\tysend[\cs{o}]{T}{S}}}
      {\pair{V}{x}}{\typrod{T}{\tysend[\cs{o}]{T}{S}}}}
    }{\tseq[\cs{p}\sqcup\cs{o}]{\ty{\Delta},\tmty{x}{\tysend[\cs{o}]{T}{S}}}{\send\;{\pair{V}{x}}}{S}}
    }{\cseq[\phi]
    {\ty{\Gamma},\ty{\Delta},\tmty{x}{\tysend[\cs{o}]{T}{S}}}
    {\plug{\conf{F}}{\send\;{\pair{V}{x}}}}}

    \LabTirName{(b)}
    \elabarrow
    \inferrule*{
    \inferrule*[vdots=1.5em]{
      \inferrule*{
      }{\tmty{\recv}{\tylolli[\cs{\ptop},\cs{o}]{\tyrecv[\cs{o}]{T}{\co{S}}}{\typrod{T}{\co{S}}}}}
      \\
      \inferrule*{
      }{\tseq[\cs{\pbot}]{\tmty{y}{\tyrecv[\cs{o}]{T}{\co{S}}}}{y}{\tyrecv[\cs{o}]{T}{\co{S}}}}
    }{\tseq[\cs{o}]
      {\tmty{y}{\tyrecv[\cs{o}]{T}{\co{S}}}}
      {\recv\;y}
      {\typrod{T}{\co{S}}}}
    }{\cseq[\phi']
    {\ty{\Theta},\tmty{y}{\tyrecv[\cs{o}]{T}{\co{S}}}}
    {\plug{\conf{F'}}{\recv\;{y}}}}

    \\
    \begin{turn}{270}
      \cred
    \end{turn}
    \\
    \inferrule*{
      \inferrule*{
        \inferrule*{
          \inferrule*[vdots=1.5em]{
          }{\tseq[\cs{\pbot}]{\tmty{x}{S}}{x}{S}}
        }{\cseq[\phi]{\ty{\Gamma},\tmty{x}{S}}{\plug{\conf{F}}{x}}}
        \\
        \inferrule*{
          \inferrule*[vdots=1.5em]{
            \tseq[\cs{p}]{\ty{\Delta}}{V}{T}
            \\
            \inferrule*{
            }{\tseq[\cs{\pbot}]{\ty{\Delta},\tmty{y}{\co{S}}}{y}{\co{S}}}
          }{\tseq[\cs{p}]{\ty{\Delta},\tmty{y}{\co{S}}}{\pair{V}{y}}{\typrod{T}{\co{S}}}}
        }{\cseq[\phi']
          {\ty{\Delta},\ty{\Theta},\tmty{y}{\co{S}}}
          {\plug{\conf{F'}}{\pair{V}{y}}}}
      }{\cseq[\phi+\phi']
        {\ty{\Gamma},\ty{\Delta},\ty{\Theta},
          \tmty{x}{S},\tmty{y}{\co{S}}}
        {\ppar
          {\plug{\conf{F}}{x}}
          {\plug{\conf{F'}}{\pair{V}{y}}}}}
    }{\cseq[\phi+\phi']
      {\ty{\Gamma},\ty{\Delta},\ty{\Theta}}
      {\res{x}{y}{(\ppar
          {\plug{\conf{F}}{x}}
          {\plug{\conf{F'}}{\pair{V}{y}}})}}}
  \end{mathpar}
  \caption{Subject Reduction (\LabTirName{E-Send})}
  \label{fig:pgv-subject-reduction-cred-send}
\end{figure}


\subsection{Progress and Deadlock Freedom}
PGV satisfies progress, as PGV configurations either reduce or are in normal form. However, the normal forms may seem surprising at first, as evaluating a well-typed PGV term does not necessarily produce {just} a value. If a term returns an endpoint, then its normal form contains a thread which is ready to communicate on the dual of that endpoint. This behaviour is not new to PGV.

Let us consider an example, adapted from Lindley and Morris~\cite{lindleymorris15}, in which a term returns an endpoint linked to an echo server. The echo server receives a value and sends it back unchanged. Consider the program which creates a new channel, with endpoints $\tm{x}$ and $\tm{x'}$, spawns off an echo server listening on $\tm{x}$, and then returns $\tm{x'}$:
\[
  \begin{array}{ll}
    \tm{\main}
     & \tm{\letpair{x}{x'}{\new\;\unit}{}}
    \\
     & \tm{\andthen{\spawn\;(\lambda\unit.\echo_{x})}{x'}}
  \end{array}
  \hfill\qquad\hfill
  \begin{array}{lcl}
    \tm{\echo_x}
     & \defeq
     & \tm{\letpair{y}{x}{\recv\;{x}}{}}
    \\
     &
     & \tm{\letbind{x}{\send\;\pair{y}{x}}{\close\;x}}
  \end{array}
\]

If we reduce the above program, we get $\tm{\res{x}{x'}{(\ppar{\child\;\echo_{x}}{\main\;x'})}}$. Clearly, no more evaluation is possible, even though the configuration contains the thread $\tm{\child\;\echo_{x}}$, which is blocked on $\tm{x}$. In~\cref{cor:pgv-closed-progress} we will show that if a term does not return an endpoint, it must produce {only} a value.

\emph{Actions} are terms which perform communication actions and which synchronise between two threads.
\begin{defi}\label{def:pgv-actions}
  A~term \emph{acts on} an endpoint $\tm{x}$ if it is $\tm{\send\;\pair{V}{x}}$, $\tm{\recv\;{x}}$, $\tm{\close\;{x}}$, or $\tm{\wait\;{x}}$. A~term is an \emph{action} if it acts on some endpoint $\tm{x}$.
\end{defi}

\emph{Ready terms} are terms which perform communication actions, either by themselves, \eg creating a new channel or thread, or with another thread, \eg sending or receiving.
It is worth mentioning that the notion of readiness presented here is akin to {live} processes introduced by Caires and Pfenning \cite{cairespfenning10,DardhaP22}, and {poised} processes introduced by Pfenning and Griffith \cite{PfenningG15} and later used by Balzer \etal \cite{balzerpfenning17,balzertoninho19}. Ready processes like live/poised processes denote processes that are ready to communicate on their providing channel.
\begin{defi}\label{def:pgv-ready-actions}
  A~term $\tm{L}$ is \emph{ready} if it is of the form $\tm{\plug{E}{M}}$, where $\tm{M}$ is of the form $\tm{\new}$, $\tm{\spawn\;N}$, $\tm{\link\;\pair{x}{y}}$, or $\tm{M}$ acts on $\tm{x}$. In the latter case, we say that $\tm{L}$ is \emph{ready to act on} $\tm{x}$ or is \emph{blocked on}.
\end{defi}

\emph{Progress} for the term language is standard for GV, and deviates from progress for linear $\lambda$-calculus only in that terms may reduce to values or \emph{ready terms}, where the definition of ready terms encompasses all terms whose reduction is struck on some constant $\tm{K}$.

\begin{lem}\label{lem:pgv-open-progress-terms}
  If $\tseq[\cs{p}]{\ty{\Gamma}}{M}{T}$ and $\ty{\Gamma}$ contains only session types, then:
  \begin{enumerate*}[label= (\roman*) ]
    \item $\tm{M}$ is a value;
    \item $\tm{M}\tred\tm{N}$ for some $\tm{N}$; or
    \item $\tm{M}$ is ready.
  \end{enumerate*}
\end{lem}

With ``$\ty{\Gamma}$ contains only session types'' we mean that for every $\tmty{x}{T}\in\ty{\Gamma}$, $\ty{T}$ is a session type, \ie, is of the form $\ty{S}$.

\emph{Canonical forms} deviate from those for GV, in that we opt to move all $\nu$-binders to the top. The standard GV canonical form, alternating $\nu$-binders and their corresponding parallel compositions, does not work for PGV, since multiple channels may be split across a single parallel composition.

A~configuration either reduces, or it is equivalent to configuration in normal form. Crucial to the normal form is that each term $\tm{M_i}$ is blocked on the corresponding channel $\tm{x_i}$, and hence no two terms act on dual endpoints. Furthermore, no term $\tm{M_i}$ can perform a communication action by itself, since those are excluded by the definition of actions.
Finally, as a corollary, we get that well-typed terms which do not return endpoints return {just} a value:

\begin{defi}\label{def:pgv-canonical-forms}
  A~configuration $\tm{\conf{C}}$ is in canonical form if it is of the form
  \linebreak 
  $\tm{\res{x_1}{x'_1}{\dots\res{x_n}{x'_n}{(\child\;M_1\parallel\dots\parallel\child\;M_m\parallel\main\;N)}}}$
  where no term $\tm{M_i}$ is a value.
\end{defi}

\begin{lem}\label{lem:pgv-canonical-forms}
  If $\cseq[\main]{\ty{\Gamma}}{\conf{C}}$, there exists some $\tm{\conf{D}}$ such that $\tm{\conf{C}}\equiv\tm{\conf{D}}$ and $\tm{\conf{D}}$ is in canonical form.
\end{lem}
\proof
We move any $\nu$-binders to the top using \LabTirName{SC-ResExt}, discard any superfluous occurrences of $\tm{\child\;\unit}$ using \LabTirName{SC-ParNil}, and move the main thread to the rightmost position using \LabTirName{SC-ParComm} and \LabTirName{SC-ParAssoc}.
\qed

\begin{defi}
  A~configuration $\tm{\conf{C}}$ is in normal form if it is of the form
  \linebreak 
  $\tm{\res{x_1}{x'_1}{\dots\res{x_n}{x'_n}{(\child\;M_1\parallel\dots\parallel\child\;M_m\parallel\main\;V)}}}$
  where each $\tm{M_i}$ is ready to act on $\tm{x_i}$.
\end{defi}

\begin{lem}\label{lem:pgv-ready-priority}
  If $\tseq[\cs{p}]{\ty{\Gamma}}{L}{T}$ is ready to act on $\tmty{x}{S}\in\ty{\Gamma}$, then the priority bound $\cs{p}$ is some priority $\cs{o}$, \ie not $\cs{\pbot}$ or $\cs{\ptop}$.
\end{lem}
\proof
Let $\tm{L}=\tm{\plug{E}{M}}$. By induction on the structure of $\tm{E}$. $\tm{M}$ has priority $\pr({\ty{S}})$, and each constructor of the evaluation context $\tm{E}$ passes on the \emph{maximum} of the priorities of its premises. No rule introduces the priority bound $\cs{\ptop}$ on the sequent.
\qed

\begin{thm}\label{thm:pgv-closed-progress-confs}
  If $\cseq[\main]{\emptyenv}{\conf{C}}$ and $\tm{\conf{C}}$ is in canonical form, then either $\tm{\conf{C}}\cred\tm{\conf{D}}$ for some $\tm{\conf{D}}$; or $\tm{\conf{C}\equiv\conf{D}}$ for some $\tm{\conf{D}}$ in normal form.
\end{thm}
\proof
\label{prf:thm-pgv-closed-progress-confs}
Let $\tm{\conf{C}}=\tm{\res{x_1}{x'_1}{\dots\res{x_n}{x'_n}{(\child\;M_1\parallel\dots\parallel\child\;M_m\parallel\main\;N)}}}$.
We apply \cref{lem:pgv-open-progress-terms} to each $\tm{M_i}$ and $\tm{N}$. If for any $\tm{M_i}$ or $\tm{N}$ we obtain a reduction $\tm{M_i}\tred\tm{M'_i}$ or $\tm{N}\tred\tm{N'}$, we apply \LabTirName{E-LiftM} and \LabTirName{E-LiftC} to obtain a reduction on $\tm{\conf{C}}$.
Otherwise, each term $\tm{M_i}$ is ready, and $\tm{N}$ is either ready or a value.
Pick the \emph{ready} term $\tm{L}\in\{\tm{M_1},\dots,\tm{M_m},\tm{N}\}$ with the smallest priority bound.
\begin{enumerate}
  \item
        If $\tm{L}$ is a new $\tm{\plug{E}{\new\;\unit}}$, we apply \LabTirName{E-New}.
  \item
        If $\tm{L}$ is a spawn $\tm{\plug{E}{\spawn\;M}}$, we apply \LabTirName{E-Spawn}.
  \item
        If $\tm{L}$ is a link $\tm{\plug{E}{\link\;\pair{y}{z}}}$ or $\tm{\plug{E}{\link\;\pair{z}{y}}}$, we apply \LabTirName{E-Link}.
  \item
        Otherwise, $\tm{L}$ is ready to act on some endpoint $\tmty{y}{S}$. Let $\tmty{y'}{\co{S}}$ be the dual endpoint of $\tm{y}$. The typing rules enforce the linear use of endpoints, so there must be a term $\tm{L'}\in\{\tm{M_1},\dots,\tm{M_m},\tm{N}\}$ which uses $\tm{y'}$. $\tm{L'}$ must be either a ready term or a value:
        \begin{enumerate}
          \item
                $\tm{L'}$ is ready. By~\cref{lem:pgv-ready-priority}, the priority of $\tm{L}$ is $\pr(\ty{S})$. By duality, $\pr(\ty{\co{S}})=\pr(\ty{S})$.
                We cannot have $\tm{L}=\tm{L'}$, otherwise the action on $\tm{y'}$ would be guarded by the action on $\tm{y}$, requiring $\pr(\ty{\co{S}})<\pr(\ty{S})$.

                The term $\tm{L'}$ must be ready to act on $\tm{y'}$, otherwise the action $\tm{y'}$ would be guarded by another action with priority smaller than $\pr{(\ty{S})}$, which contradicts our choice of $\tm{L}$ as having the smallest priority.

                Therefore, we have two terms ready to act on dual endpoints. We apply the appropriate reduction rule, \ie \LabTirName{E-Send} or \LabTirName{E-Close}.
          \item
                $\tm{L'}=\tm{N}$ and is a value. We rewrite $\tm{\conf{C}}$ to put $\tm{L}$ in the position corresponding to the endpoint it is blocked on, using \LabTirName{SC-ParComm}, \LabTirName{SC-ParAssoc}, and optionally \LabTirName{SC-ResSwap}.
                We then repeat the steps above with the term with the next smallest priority, until either we find a reduction, or the configuration has reached the desired normal form.

                The argument based on the priority being the smallest continues to hold, since we know that neither $\tm{L}$ nor $\tm{L'}$ will be picked, and no other term uses $\tm{y}$ or $\tm{y'}$.
        \end{enumerate}
\end{enumerate}
\qed


\begin{cor}\label{cor:pgv-closed-progress}
  If $\cseq[\phi]{\emptyenv}{\conf{C}}$, $\tm{\conf{C}}\notcred$, and $\tm{\conf{C}}$ contains no endpoints, then $\tm{\conf{C}}\equiv\tm{\phi\;V}$ for some value $\tm{V}$.
\end{cor}

An immediate consequence of \cref{thm:pgv-closed-progress-confs} and \cref{cor:pgv-closed-progress} is that \emph{a term which does not return an endpoint will complete all its communication actions, thus satisfying deadlock freedom.}
}
{









\section{Relation to Priority CP}
\label{sec:pcp}

Thus far we have presented Priority GV (PGV) together with the relevant technical results. We remind the reader that this line of work of adding priorities, started with Priority CP (PCP) \cite{dardhagay18} where priorities are integrated in Wadler's Classical Processes (CP), which is a $\pi$-calculus leveraging the correspondence of session types as linear logic propositions \cite{wadler12}.
In his work, Wadler presents a connection (via encoding) of CP and GV. Following that work, we sat out to understand the connection between the priority versions of CP and GV, thus comparing PGV and PCP.
Before presenting our formal results, we will revisit PCP in the following section.

\begingroup
\usingnamespace{pcp}
\subsection{Revisiting Priority CP}
\label{app:revisiting-PCP}

\subsubsection*{Types}
Types ($\pcp{\ty{A}}, \pcp{\ty{B}}$) in PCP are based on classical linear logic propositions, and are defined by the following grammar:
\[
  \usingnamespace{pcp}
  \begin{array}{lcl}
    \ty{A}, \ty{B}
     & \Coloneqq & \ty{\tytens[\cs{o}]{A}{B}}
    \sep        \ty{\typarr[\cs{o}]{A}{B}}
    \sep        \ty{\tyone[\cs{o}]}
    \sep        \ty{\tybot[\cs{o}]}
    \sep        \ty{\typlus[\cs{o}]{A}{B}}
    \sep        \ty{\tywith[\cs{o}]{A}{B}}
    \sep        \ty{\tynil[\cs{o}]}
    \sep        \ty{\tytop[\cs{o}]}
  \end{array}
\]

Each connective is annotated with a priority $\cs{o}\in\mathbb{N}$.

Types $\pcp{\ty{\tytens[\cs{o}]{A}{B}}}$ and $\pcp{\ty{\typarr[\cs{o}]{A}{B}}}$ type the endpoints of a channel over which we send or receive a channel of type $\pcp{\ty{A}}$, and then proceed as type $\pcp{\ty{B}}$. Types $\pcp{\ty{\tyone[\cs{o}]}}$ and $\pcp{\ty{{\tybot}[\cs{o}]}}$ type the endpoints of a channel whose session has terminated, and over which we send or receive a \emph{ping} before closing the channel. These two types act as units for $\pcp{\ty{\tytens[\cs{o}]{A}{B}}}$ and $\pcp{\ty{\typarr[\cs{o}]{A}{B}}}$, respectively.

Types $\pcp{\ty{\typlus[\cs{o}]{A}{B}}}$ and $\pcp{\ty{\tywith[\cs{o}]{A}{B}}}$ type the endpoints of a channel over which we can receive or send a choice between two branches $\pcp{\ty{A}}$ or $\pcp{\ty{B}}$. We have opted for a simplified version of choice and followed the original Wadler's CP \cite{wadler14}, however types $\ty{\oplus}$ and $\ty{\with}$ can be trivially generalised to $\pcp{\ty{\oplus^{\cs{o}}\{l_i:A_i\}_{i\in I}}}$ and $\pcp{\ty{\with^{\cs{o}}\{l_i:A_i\}_{i\in I}}}$, respectively, as in the original PCP \cite{dardhagay18extended}.

Types $\pcp{\ty{\tynil[\cs{o}]}}$ and $\pcp{\ty{\tytop[\cs{o}]}}$ type the endpoints of a channel over which we can send or receive a choice between {no options}. These two types act as units for $\pcp{\ty{\typlus[\cs{o}]{A}{B}}}$ and $\pcp{\ty{\tywith[\cs{o}]{A}{B}}}$, respectively.

\subsubsection*{Typing Environments}
\label{sec:pcp-environments}
Typing environments $\pcp{\ty{\Gamma}}$, $\pcp{\ty{\Delta}}$ associate names to types. Environments are linear, so two environments can only be combined as $\pcp{\ty{\Gamma}}, \pcp{\ty{\Delta}}$ if their names are distinct, \ie $\pcp{\fv(\ty{\Gamma})\cap\fv(\ty{\Delta})=\varnothing}$.
\[
  \usingnamespace{pcp}
  \begin{array}{lcl}
    \ty{\Gamma}, \ty{\Delta}
     & \Coloneqq & \ty{\emptyenv}
    \sep        \ty{\Gamma}, \tmty{x}{A}
  \end{array}
\]

\subsubsection*{Type Duality}
\label{sec:pcp-duality}
Duality is an involutive function on types which preserves priorities:
\[
  \usingnamespace{pcp}
  \setlength{\arraycolsep}{1pt}
  \begin{array}{lcl}
    \ty{\co{(\tyone[\cs{o}])}} & = & \ty{\tybot[\cs{o}]} \\
    \ty{\co{(\tybot[\cs{o}])}} & = & \ty{\tyone[\cs{o}]}
  \end{array}
  \quad
  \begin{array}{lcl}
    \ty{\co{(\tytens[\cs{o}]{A}{B})}} & = & \ty{\typarr[\cs{o}]{\co{A}}{\co{B}}} \\
    \ty{\co{(\typarr[\cs{o}]{A}{B})}} & = & \ty{\tytens[\cs{o}]{\co{A}}{\co{B}}}
  \end{array}
  \quad
  \begin{array}{lcl}
    \ty{\co{(\tynil[\cs{o}])}} & = & \ty{\tytop[\cs{o}]} \\
    \ty{\co{(\tytop[\cs{o}])}} & = & \ty{\tynil[\cs{o}]}
  \end{array}
  \quad
  \begin{array}{lcl}
    \ty{\co{(\typlus[\cs{o}]{A}{B})}} & = & \ty{\tywith[\cs{o}]{\co{A}}{\co{B}}} \\
    \ty{\co{(\tywith[\cs{o}]{A}{B})}} & = & \ty{\typlus[\cs{o}]{\co{A}}{\co{B}}}
  \end{array}
\]

\subsubsection*{Priorities}
\label{sec:pcp-priorities}
The function $\pr(\cdot)$ returns smallest priority of a type. As with PGV, the type system guarantees that the top-most connective always holds the smallest priority.  The function $\minpr(\cdot)$ returns the \emph{minimum} priority of all types a typing context, or $\cs{\ptop}$ if the context is empty:
\[
  \usingnamespace{pcp}
  \setlength{\arraycolsep}{1pt}
  \begin{array}{lclclcl}
    \pr(\ty{\tyone[\cs{o}]}) & = & \cs{o} \\
    \pr(\ty{\tybot[\cs{o}]}) & = & \cs{o}
  \end{array}
  \qquad
  \begin{array}{lclclcl}
    \pr(\ty{\tytens[\cs{o}]{A}{B}}) & = & \cs{o} \\
    \pr(\ty{\typarr[\cs{o}]{A}{B}}) & = & \cs{o}
  \end{array}
  \qquad
  \begin{array}{lclclcl}
    \pr(\ty{\tynil[\cs{o}]}) & = & \cs{o} \\
    \pr(\ty{\tytop[\cs{o}]}) & = & \cs{o}
  \end{array}
  \qquad
  \begin{array}{lclclcl}
    \pr(\ty{\typlus[\cs{o}]{A}{B}}) & = & \cs{o} \\
    \pr(\ty{\tywith[\cs{o}]{A}{B}}) & = & \cs{o}
  \end{array}
\]
\[
  \minpr(\ty{\emptyenv})          = \cs{\ptop}
  \quad
  \minpr(\ty{\Gamma},\tmty{x}{T}) = \minpr(\ty{\Gamma})\sqcap\minpr(\ty{T})
\]

\subsubsection*{Terms}
Processes ($\pcp{\tm{P}}$, $\pcp{\tm{Q}}$) in PCP are defined by the following grammar.
\[
  \usingnamespace{pcp}
  \begin{array}[t]{lcl}
    \tm{P}, \tm{Q}
     & \Coloneqq & \tm{\link{x}{y}}
    \sep   \tm{\res{x}{y}{P}}
    \sep   \tm{(\ppar{P}{Q})}
    \sep   \tm{\halt}
    \\   & \sep & \tm{\send{x}{y}{P}}
    \sep   \tm{\close{x}{P}}
    \sep   \tm{\recv{x}{y}{P}}
    \sep   \tm{\wait{x}{P}}
    \\   & \sep & \tm{\inl{x}{P}}
    \sep   \tm{\inr{x}{P}}
    \sep   \tm{\offer{x}{P}{Q}}
    \sep   \tm{\absurd{x}}
  \end{array}
\]

Process $\pcp{\tm{\link{x}{y}}}$ links endpoints $\pcp{\tm{x}}$ and $\pcp{\tm{y}}$ and forwards communication from one to the other. $\pcp{\tm{\res{x}{y}{P}}}$, $\pcp{\tm{(\ppar{P}{Q})}}$ and $\pcp{\tm{\halt}}$ denote respectively the restriction processes where channel endpoints $\pcp{\tm{x}}$ and $\pcp{\tm{y}}$ are bound together and with scope $\pcp{\tm{P}}$, the parallel composition of processes $\pcp{\tm{P}}$ and $\pcp{\tm{Q}}$ and the terminated process.

Processes $\pcp{\tm{\send{x}{y}{P}}}$ and $\pcp{\tm{\recv{x}{y}{P}}}$ send or receive over channel $\pcp{\tm{x}}$ a value $\pcp{\tm{y}}$ and proceed as process $\pcp{\tm{P}}$. Processes $\pcp{\tm{\close{x}{P}}}$ and $\pcp{\tm{\wait{x}{P}}}$ send and receive an empty value---denoting the closure of channel $\pcp{\tm{x}}$, and continue as $\pcp{\tm{P}}$.

Processes $\pcp{\tm{\inl{x}{P}}}$ and $\pcp{\tm{\inr{x}{P}}}$ make a left and right choice, respectively and proceed as process $\pcp{\tm{P}}$. Dually, $\pcp{\tm{\offer{x}{P}{Q}}}$ offers both left and right branches, with continuations $\pcp{\tm{P}}$ and $\pcp{\tm{Q}}$, and $\pcp{\tm{\absurd{x}}}$ is the empty offer.

We write \emph{unbound} send as $\pcp{\tm{\usend{x}{y}{P}}}$, which is syntactic sugar for $\pcp{\tm{\send{x}{z}{(\ppar{\link{y}{z}}{P})}}}$. Alternatively, we could take $\pcp{\tm{\usend{x}{y}{P}}}$ as primitive, and let $\pcp{\tm{\send{x}{y}{P}}}$ be syntactic sugar for $\pcp{\tm{\res{y}{z}{(\usend{x}{z}{P})}}}$. CP takes \emph{bound} sending as primitive, as it is impossible to eliminate the top-level cut in terms such as $\pcp{\tm{\res{y}{z}{(\usend{x}{z}{P})}}}$, even with commuting conversions. In our setting without commuting conversions and with more permissive normal forms, this is no longer an issue, but, for simplicity, we keep bound sending as primitive.

\subsubsection*{On Commuting Conversions}
\label{app:commuting-conversions}

The main change we make to PCP is {removing commuting conversions}. Commuting conversions are necessary if we want our reduction strategy to correspond {exactly} to cut (or cycle in \cite{dardhagay18extended}) elimination. However, as Lindley and Morris~\cite{lindleymorris15} show, all communications that can be performed \emph{with} the use of commuting conversions, can also be performed \emph{without} them, but using structural congruence.

From the perspective of process calculi, commuting conversions behave strangely.
Consider the commuting conversion $(\kappa_{\parr})$ for $\pcp{\tm{\recv{x}{y}{P}}}$:
\begin{mathpar}
  (\kappa_{\parr})
  \quad
  \tm{\res{z}{z'}{(\ppar{\recv{x}{y}{P}}{Q})}}
  \red
  \tm{\recv{x}{y}{\res{z}{z'}{(\ppar{P}{Q})}}}
\end{mathpar}
As a result of $(\kappa_{\parr})$, $\pcp{\tm{Q}}$ becomes blocked on $\pcp{\tm{\labrecv{x}{y}}}$, and any actions $\pcp{\tm{Q}}$ was able to perform become unavailable. Consequently, CP is non-confluent:
\begin{mathpar}
  \setlength{\arraycolsep}{2em}
  \begin{array}{cc}
    \multicolumn{2}{c}{%
      \hspace*{10ex}
      {\tm{\res{x}{x'}{(\ppar{\recv{a}{y}{P}}{\res{z}{z'}{(\ppar{\close{z}{\halt}}{\wait{z'}{Q}})}})}}}}
    \\
    \qquad\rotatebox[origin=c]{270}{$\red\hphantom{{}^+}$}
     &
    \rotatebox[origin=c]{270}{$\red^+$}\qquad
    \\
    {\tm{\recv{a}{y}{\res{x}{x'}{(\ppar{P}{\res{z}{z'}{(\ppar{\close{z}{\halt}}{\wait{z'}{Q}})}})}}}}
     &
    {\tm{\recv{a}{y}{\res{x}{x'}{(\ppar{P}{Q})}}}}
  \end{array}
\end{mathpar}

In PCP, commuting conversions break our intuition that an action with lower priority occurs before an action with higher priority. To cite Dardha and Gay~\cite{dardhagay18extended} ``\emph{if a prefix on a channel endpoint $\pcp{\tm{x}}$ with priority $\cs{o}$ is pulled out at top level, then to preserve priority constraints in the typing rules [..], it is necessary to increase priorities of all actions after the prefix on $\pcp{\tm{x}}$}'' by $\cs{o+1}$.

\subsection{Operational Semantics}
The operational semantics for PCP, given in \cref{fig:pcp-operational-semantics}, is defined as a reduction relation $\pcp{\red}$ on processes (bottom) and uses structural congruence (top). Each of the axioms of structural congruence corresponds to the axiom of the same name for PGV. We write $\pcp{\red^+}$ for the transitive closures, and $\pcp{\red^\star}$ for the reflexive-transitive closures.

The reduction relation is given by a set of axioms and inference rules for context closure. Reduction occurs under restriction. $\LabTirName{E-Link}$ reduces a parallel composition with a link into a substitution. $\LabTirName{E-Send}$ is the main communication rule, where send and receive processes sychronise and reduce to the corresponding continuations. $\LabTirName{E-Close}$ follows the previous rule and it closes the channel identified by endpoints $\pcp{\tm{x}}$ and $\pcp{\tm{y}}$. $\LabTirName{E-Select-Inl}$ and $\LabTirName{E-Select-Inr}$ are generalised versions of $\LabTirName{E-Send}$. They state respectively that a left and right selection synchronises with a choice offering and reduces to the corresponding continuations. The last three rules state that reduction is closed under restriction, parallel composition and structural congruence, respectively.

\begin{figure}
  \usingnamespace{pcp}
  \textbf{Structural congruence.}
  \begin{mathpar}
    \begin{array}{llcl}
      \LabTirName{SC-LinkSwap}   & \tm{\link{x}{y}}
                                 & \equiv & \tm{\link{y}{x}}
      \\
      \LabTirName{SC-ResLink}    & \tm{\res{x}{y}{\link{x}{y}}}
                                 & \equiv & \tm{\halt}
      \\
      \LabTirName{SC-ResSwap}    & \tm{\res{x}{y}{P}}
                                 & \equiv & \tm{\res{y}{x}{P}}
      \\
      \LabTirName{SC-ResComm}    & \tm{\res{x}{y}{\res{z}{w}{P}}}
                                 & \equiv & \tm{\res{z}{w}{\res{x}{y}{P}}}
      \\
      \LabTirName{SC-ResExt}     & \tm{\res{x}{y}{(\ppar{P}{Q})}}
                                 & \equiv & \tm{\ppar{P}{\res{x}{y}{Q}}},
                                            \text{ if }{\tm{x},\tm{y}\notin\fv(\tm{P})}
      \\
      \LabTirName{SC-ParNil}     & \tm{\ppar{P}{\halt}}
                                 & \equiv & \tm{P}
      \\
      \LabTirName{SC-ParComm}    & \tm{\ppar{P}{Q}}
                                 & \equiv & \tm{\ppar{Q}{P}}
      \\
      \LabTirName{SC-ParAssoc}   & \tm{\ppar{P}{(\ppar{Q}{R})}}
                                 & \equiv & \tm{\ppar{(\ppar{P}{Q})}{R}}
    \end{array}
  \end{mathpar}
  \textbf{Reduction.}
  \begin{mathpar}
    \begin{array}{llcl}
      \LabTirName{E-Link}    & \tm{\res{x}{y}{(\ppar{\link{w}{x}}{P})}}
                             & \red & \tm{\subst{P}{w}{x}}
      \\
      \LabTirName{E-Send}    & \tm{\res{x}{y}{(\ppar{\send{x}{z}{P}}{\recv{x}{w}{Q}})}}
                             & \red & \tm{\res{x}{y}{\res{z}{w}{(\ppar{P}{Q})}}}
      \\
      \LabTirName{E-Close}   & \tm{\res{x}{y}{(\ppar{\close{x}{P}}{\wait{y}{Q}})}}
                             & \red & \tm{\ppar{P}{Q}}
      \\
      \LabTirName{E-Select-Inl}
                             & \tm{\res{x}{y}{(\ppar{\inl{x}{P}}{\offer{x}{Q}{R}})}}
                             & \red & \tm{\res{x}{y}{(\ppar{P}{Q})}}
      \\
      \LabTirName{E-Select-Inr}
                             & \tm{\res{x}{y}{(\ppar{\inr{x}{P}}{\offer{x}{Q}{R}})}}
                             & \red & \tm{\res{x}{y}{(\ppar{P}{R})}}
    \end{array}
    \\
    \inferrule*[lab=E-LiftRes]{
      \tm{P}\red\tm{P'}
    }{\tm{\res{x}{y}{P}}\red\tm{\res{x}{y}{P'}}}

    \inferrule*[lab=E-LiftPar]{
      \tm{P}\red\tm{P'}
    }{\tm{\ppar{P}{Q}}\red\tm{\ppar{P'}{Q}}}

    \inferrule*[lab=E-LiftSC]{
      \tm{P}\equiv\tm{P'}
      \\
      \tm{P'}\red\tm{Q'}
      \\
      \tm{Q'}\equiv\tm{Q}
    }{\tm{P}\red\tm{Q}}
  \end{mathpar}
  \caption{Operational Semantic for PCP.}%
  \label{fig:pcp-operational-semantics}
\end{figure}

\subsection{Typing Rules}
\Cref{fig:pcp-typing} gives the typing rules for our version of PCP. A typing judgement $\pcp{\seq{P}{\Gamma}}$ states that ``process $\pcp{\tm{P}}$ is well typed under the typing context $\pcp{\ty{\Gamma}}$''.

\begin{figure}
  \usingnamespace{pcp}
  \begin{mathpar}
    \inferrule*[lab=T-Link]{
    }{\seq{\link[\ty{A}]{x}{y}}{\tmty{x}{A},\tmty{y}{\co{A}}}}
    
    \inferrule*[lab=T-Res]{
      \seq{P}{\ty{\Gamma},\tmty{x}{A},\tmty{y}{\co{A}}}
    }{\seq{\res{x}{y}{P}}{\ty{\Gamma}}}

    \inferrule*[lab=T-Par]{
      \seq{P}{\ty{\Gamma}}
      \\
      \seq{Q}{\ty{\Delta}}
    }{\seq{\ppar{P}{Q}}{\ty{\Gamma},\ty{\Delta}}}
    
    \inferrule*[lab=T-Halt]{
    }{\seq{\halt}{\emptyenv}}
    \\
    \inferrule*[lab=T-Send]{
      \seq{P}{\ty{\Gamma},\tmty{y}{A},\tmty{x}{B}}
      \\
      \cs{o}<\minpr(\ty{\Gamma},\ty{A},\ty{B})
    }{\seq{\send{x}{y}{P}}{\ty{\Gamma},\tmty{x}{\tytens[\cs{o}]{A}{B}}}}
    
    \inferrule*[lab=T-Close]{
      \seq{P}{\ty{\Gamma}}
      \\
      \cs{o}<\minpr(\ty{\Gamma})
    }{\seq{\close{x}{P}}{\ty{\Gamma},\tmty{x}{\tyone[\cs{o}]}}}
    \\
    \inferrule*[lab=T-Recv]{
      \seq{P}{\ty{\Gamma},\tmty{y}{A},\tmty{x}{B}}
      \\
      \cs{o}<\minpr(\ty{\Gamma},\ty{A},\ty{B})
    }{\seq{\recv{x}{y}{P}}{\ty{\Gamma},\tmty{x}{\typarr[\cs{o}]{A}{B}}}}
    
    \inferrule*[lab=T-Wait]{
      \seq{P}{\ty{\Gamma}}
      \\
      \cs{o}<\minpr(\ty{\Gamma})
    }{\seq{\wait{x}{P}}{\ty{\Gamma},\tmty{x}{\tybot[\cs{o}]}}}
    \\
    \inferrule*[lab=T-Select-Inl]{
      \seq{P}{\ty{\Gamma},\tmty{x}{A}}
      \\
      \cs{o}<\minpr(\ty{\Gamma},\ty{A},\ty{B})
      \\
      \pr(\ty{A})=\pr(\ty{B})
    }{\seq{\inl{x}{P}}{\ty{\Gamma},\tmty{x}{\typlus[\cs{o}]{A}{B}}}}
    
    \inferrule*[lab=T-Select-Inr]{
      \seq{P}{\ty{\Gamma},\tmty{x}{B}}
      \\
      \cs{o}<\minpr(\ty{\Gamma},\ty{A},\ty{B})
      \\
      \pr(\ty{A})=\pr(\ty{B})
    }{\seq{\inr{x}{P}}{\ty{\Gamma},\tmty{x}{\typlus[\cs{o}]{A}{B}}}}
    
    \inferrule*[lab=T-Offer]{
      \seq{P}{\ty{\Gamma},\tmty{x}{A}}
      \\
      \seq{Q}{\ty{\Gamma},\tmty{x}{B}}
      \\
      \cs{o}<\minpr(\ty{\Gamma},\ty{A},\ty{B})
    }{\seq{\offer{x}{P}{Q}}{\ty{\Gamma},\tmty{x}{\tywith[\cs{o}]{A}{B}}}}
    
    \inferrule*[lab=T-Offer-Absurd]{
      \cs{o}<\pr(\ty{\Gamma})
    }{\seq{\absurd{x}}{\ty{\Gamma},\tmty{x}{\tytop[\cs{o}]}}}
  \end{mathpar}
  \caption{Typing Rules for PCP.}
  \label{fig:pcp-typing}
\end{figure}

\textsc{T-Link} states that the link process $\pcp{\tm{\link{x}{y}}}$ is well typed under channels $\pcp{\tm{x}}$ and $\pcp{\tm{y}}$ having dual types, respectively $\pcp{\ty{A}}$ and $\pcp{\ty{\co{A}}}$. \textsc{T-Res} states that the restriction process $\pcp{\tm{\res{x}{y}{P}}}$ is well typed under typing context $\pcp{\ty{\Gamma}}$ if process $\pcp{\tm{P}}$ is well typed in $\pcp{\ty{\Gamma}}$ augmented with channel endpoints $\pcp{\tm{x}}$ and $\pcp{\tm{y}}$ having dual types, respectively $\pcp{\ty{A}}$ and $\pcp{\ty{\co{A}}}$. \textsc{T-Par} states that the parallel composition of processes $\pcp{\tm{P}}$ and $\pcp{\tm{Q}}$ is well typed under the disjoint union of their respective typing contexts. \textsc{T-Halt} states that the terminated process $\pcp{\tm{\halt}}$ is well typed in the empty context.

\textsc{T-Send} and \textsc{T-Recv} state that the sending and receiving of a bound name $\pcp{\tm{y}}$ over a channel $\pcp{\tm{x}}$ is well typed under $\pcp{\ty{\Gamma}}$ and $\pcp{\tm{x}}$ of type $\pcp{\ty{\tytens[\cs{o}]{A}{B}}}$, respectively $\pcp{\typarr[\cs{o}]{A}{B}}$. Priority $\cs{o}$ is the smallest among all priorities of the types used by the output or input process, captured by the side condition $\pcp{\cs{o}<\minpr(\ty{\Gamma},\ty{A},\ty{B})}$.

Rules \textsc{T-Close} and \textsc{T-Wait} type the closure of channel $\pcp{\tm{x}}$ and are in the same lines as the previous two rules, requiring that the priority of channel $\pcp{\tm{x}}$ is the smallest among all priorities in $\pcp{\ty{\Gamma}}$.

\textsc{T-Select-Inl} and \textsc{T-Select-Inr} type respectively the left $\pcp{\tm{\inl{x}{P}}}$ and right $\pcp{\tm{\inr{x}{P}}}$ choice performed on channel $\pcp{\tm{x}}$. \textsc{T-Offer} and \textsc{T-Offer-Absurd} type the offering of a choice, or empty choice, on channel $\pcp{\tm{x}}$. In all the above rules the priority $\cs{o}$ of channel $\pcp{\tm{x}}$ is the smallest with respect to the typing context $\pcp{\cs{o}<\minpr(\ty{\Gamma})}$ and types involved in the choice $\pcp{\cs{o}<\minpr(\ty{\Gamma},\ty{A},\ty{B})}$.

\Cref{fig:pcp-typing-sugar} shows how syntactic sugar in PCP is well typed.
\begin{figure}
  \usingnamespace{pcp}
  \begin{mathpar}
    \inferrule*[lab=T-UnboundSend]{
      \inferrule*{
      }{\seq{P}{\ty{\Gamma},\tmty{x}{B}}}
      \\
      \cs{o}<\minpr{(\ty{\Gamma},\ty{A},\ty{B})}
    }{\seq{\usend{x}{y}{P}}{\ty{\Gamma},\tmty{x}{\tytens{A}{B}},\tmty{y}{\co{A}}}}
    \elabarrow
    \inferrule*{
      \inferrule*{
        \inferrule*{
        }{\seq{\link[\ty{A}]{z}{y}}{\tmty{y}{\co{A}},\tmty{z}{A}}}
        \\
        \inferrule*{
        }{\seq{P}{\ty{\Gamma},\tmty{x}{B}}}
      }{\seq{\ppar{\link[\ty{A}]{z}{y}}{P}}{\ty{\Gamma},\tmty{x}{B},\tmty{y}{\co{A}},\tmty{z}{A}}}
      \\
      \cs{o}<\minpr{(\ty{\Gamma},\ty{A},\ty{B})}
    }{\seq{\send{x}{z}{(\ppar{\link[\ty{A}]{z}{y}}{P})}}{\ty{\Gamma},\tmty{x}{\tytens{A}{B}},\tmty{y}{\co{A}}}}
  \end{mathpar}
  \caption{Typing Rules for Syntactic Sugar for PCP.}
  \label{fig:pcp-typing-sugar}
\end{figure}


Finally, since our reduction relation is a strict subset of the reduction relation in the original~\cite{dardhagay18extended}, we defer to their proof of subject reduction (Theorem 2 in~\cite{dardhagay18extended}). We prove progress for our version of PCP, see~\cref{prf:thm-pcp-closed-progress}.

\subsection{PCP and PLL}
In this subsection, we highlight the connection between PCP and linear logic.
Dardha and Gay \cite{dardhagay18} present PCP--consequently also PCP given in this paper--in a way which can be viewed both as Classical Processes with restriction (T-Res) and parallel composition (T-Par) typing rules, and as a new version of linear logic, which they call Priority Linear Logic (PLL).
PLL builds on CLL by replacing the cut rule with two logical rules: a mix and a cycle rule---here corresponding to T-Par and T-Res, respectively. Dardha and Gay \cite[\S 4]{dardhagay18} prove cycle-elimination, in the same lines as cut-elimination for CLL. As a corollary of cycle-elimination for PLL, we obtain deadlock freedom for PCP (Theorem 3 in \cite[\S 4]{dardhagay18}). In summary, PLL is an extension of CLL and the authors show the correspondence of PCP and PLL. Notice however, that PCP is not in correspondence with CLL itself, since processes in PCP are graphs, whether CLL induces trees.

\subsection{Technical Developments}
\begin{defi}
  A~process acts on an endpoint $\tm{x}$ if it is $\tm{\link{x}{y}}$, $\tm{\link{y}{x}}$, $\tm{\send{x}{y}{P}}$, $\tm{\recv{x}{y}{P}}$, $\tm{\close{x}{P}}$, $\tm{\wait{x}{P}}$, $\tm{\inl{x}{P}}$, $\tm{\inr{x}{P}}$, $\tm{\offer{x}{P}{Q}}$, or $\tm{\absurd{x}}$. A~process is an action if it acts on some endpoint $\tm{x}$.
\end{defi}
\begin{defi}\label{def:pcp-canonical-forms}
  A~process $\tm{P}$ is in canonical form if it is either $\tm{\halt}$ or of the form
  \linebreak 
  $\tm{\res{x_1}{x'_1}{\dots\res{x_n}{x'_n}{(P_1 \parallel\dots\parallel P_m)}}}$ where $m>0$ and each $\tm{P_j}$ is an action.
\end{defi}
\begin{lem}\label{lem:pcp-canonical-forms}
  If $\seq{P}{\ty{\Gamma}}$, there exists some $\tm{Q}$ such that $\tm{P}\equiv\tm{Q}$ and $\tm{Q}$ is in canonical form.
\end{lem}
\proof
If $\tm{P}=\tm{\halt}$, we are done. Otherwise, we move any $\nu$-binders to the top using \LabTirName{SC-ResExt}, and discard any superfluous occurrences of $\tm{\halt}$ using \LabTirName{SC-ParNil}.
\qed

The proof for progress (below) follows the same reasoning by Kobayashi~\cite{kobayashi06} used in the proof of deadlock freedom for closed processes (Theorem 2).

\begin{thm}\label{thm:pcp-closed-progress}
  If $\pcp{\seq{P}{\emptyenv}}$, then either $\pcp{\tm{P}=\tm{\halt}}$ or there exists a $\pcp{\tm{Q}}$ such that $\pcp{\tm{P}\red\tm{Q}}$.
\end{thm}
\begin{proof}[Proof (Sketch)]
\label{prf:thm-pcp-closed-progress}

This proof follows the exact same reasoning and proof sketch given by Kobayashi in \cite{kobayashi06} and later adopted by Dardha and Gay for PCP in their technical report \cite{dardhagay18extended}.

By~\cref{lem:pcp-canonical-forms}, we rewrite $\tm{P}$ to canonical form. If the resulting process is $\tm{\halt}$, we are done. Otherwise, it is of the form
\[
  \seq{\res{x_1}{x'_1}{\dots\res{x_n}{x'_n}{(P_1 \parallel\dots\parallel P_m)}}}{\emptyenv}
\]
where $m>0$ and each $\seq{P_i}{\ty{\Gamma_i}}$ is an action.

Consider processes $\tm {P_1 \parallel\dots\parallel P_m}$. Among them, we pick the process with the smallest priority $\minpr{(\ty{\Gamma_i})}$ for all $i$. Let this process be $\tm{P_i}$ and the priority of the top prefix be $\cs o$. $\tm{P_i}$ acts on some endpoint $\tmty{y}{A}\in\ty{\Gamma_i}$. We must have $\minpr{(\ty{\Gamma_i})}=\pr{(\ty{A})} = \cs o$, since the other actions in $\tm{P_i}$ are guarded by the action on $\tmty{y}{A}$, thus satisfying law (i) of priorities.

If $\tm{P_i}$ is a link $\tm{\link{y}{z}}$ or $\tm{\link{z}{y}}$, we apply \LabTirName{E-Link}.

Otherwise, $\tm{P_i}$ is an input/branching or output/selection action on endpoint $\tm{y}$ of type $\ty{A}$ with priority $\cs{o}$. Since process $\tm{P}$ is closed and consequently it respects law (ii) of priorities, there must be a co-action $\tm{y'}$ of type $\ty{\co{A}}$  where $\tm{y}$ and $\tm{y'}$ are dual endpoints of the same channel (by application of rule \textsc{T-Res}). By duality, $\pr{(\ty{A})}=\pr{(\ty{\co{A}})}= \cs{o}$. In the following we show that: $\tm{y'}$ is the subject of a top level action of a process $\tm{P_j}$ with $i\neq j$. This allows for the communication among $\tm{P_i}$ and $\tm{P_j}$ to happen immediately over channel endpoints $\tm{y}$ and $\tm{y'}$.

Suppose that $\tm{y'}$ is an action not in a different parallel process $\tm{P_j}$ but rather of $\tm{P_i}$ itself. That means that the action on $\tm{y'}$ must be prefixed by the action on $\tm{y}$, which is top level in $\tm{P_i}$. To respect law (i) of priorities we must have $\cs{o}<\cs{o}$, which is absurd. This means that $\tm{y'}$ is in another parallel process $\tm{P_j}$ for $i\neq j$.

Suppose that $\tm{y'}$ in $\tm{P_j}$ is not at top level. In order to respect law (i) of priorities, it means that $\tm{y'}$ is prefixed by actions that are smaller than its priority $\cs{o}$. This leads to a contradiction because stated that $\cs{o}$ is the smallest priority. Hence, $\tm{y'}$ must be the subject of a top level action.

We have two processes, acting on dual endpoints. We apply the appropriate reduction rule, \ie \LabTirName{E-Send}, \LabTirName{E-Close}, \LabTirName{E-Select-Inl}, or \LabTirName{E-Select-Inr}.
\end{proof}


\endgroup

\subsection{Correspondence between PGV and PCP}
\begingroup
We illustrate the relation between PCP and PGV by defining a translation $\cpgvC{\cdot}$ from PCP processes to PGV configurations which satisfies operational correspondence.

The translation $\tm{\cpgvC{\cdot}}$ translates PCP processes to PGV configurations by translating as much as possible of the $\pi$-calculus constructs from PCP to the identical configuration constructs in PGV, \ie, all top-level $\tm{\nu}$ and $\tm{\parallel}$ constructs. When it encounters the first action, it translates the remainder of the processs to a term using $\tm{\cpgvM{\cdot}}$.
The translation $\tm{\cpgvM{\cdot}}$ translates PCP processes to PGV terms by mapping the $\pi$-calculus constructs from PCP (\eg, $\tm{\nu}$, $\tm{\parallel}$, $\pcp{\tm{\send{\cdot}{\cdot}{\cdot}}}$, $\pcp{\tm{\send{\cdot}{\cdot}{\cdot}}}$, $\pcp{\tm{\recv{\cdot}{\cdot}{\cdot}}}$, \etc) to the corresponding constants in PGV (\eg, $\pgv{\tm{\new}}$, $\pgv{\tm{\spawn}}$, $\pgv{\tm{\send}}$, $\pgv{\tm{\recv}}$, \etc).
Translating a process with $\tm{\cpgvC{\cdot}}$ is the same as translating that process with $\tm{\cpgvM{\cdot}}$ followed by several steps of configuration reduction.
The translation $\tm{\cpgvT{\cdot}}$ tranlates session types from PCP to session types in PGV matching the translation on processes.

The translation on types is defined as follows:
\[
  \begin{array}{lcl}
    \ty{\cpgvT{\pcp{\tytens[\cs{o}]{A}{B}}}}
     & = & \pgv{\ty{\tysend[\cs{o}]{\co{\cpgvT{A}}}{\cpgvT{B}}}}
    \\
    \ty{\cpgvT{\pcp{\typlus[\cs{o}]{A}{B}}}}
     & = & \pgv{\ty{\tyselect[\cs{o}]{\cpgvT{A}}{\cpgvT{B}}}}
  \end{array}
  \hfill\quad\hfill%
  \begin{array}{lcl}
    \ty{\cpgvT{\pcp{\tyone[\cs{o}]}}}
     & = & \ty{\pgv{\tyends[\cs{o}]}}
    \\
    \ty{\cpgvT{\pcp{\tynil[\cs{o}]}}}
     & = & \pgv{\ty{\tyselectemp[\cs{o}]}}
  \end{array}
\]%
\[%
  \begin{array}{lcl}
    \ty{\cpgvT{\pcp{\typarr[\cs{o}]{A}{B}}}}
     & = & \pgv{\ty{\tyrecv[\cs{o}]{\cpgvT{A}}{\cpgvT{B}}}}
    \\
    \ty{\cpgvT{\pcp{\tywith[\cs{o}]{A}{B}}}}
     & = & \pgv{\ty{\tyoffer[\cs{o}]{\cpgvT{A}}{\cpgvT{B}}}}
  \end{array}
  \hfill\quad\hfill%
  \begin{array}{lcl}
    \ty{\cpgvT{\pcp{\tybot[\cs{o}]}}}
     & = & \ty{\pgv{\tyendr[\cs{o}]}}
    \\
    \ty{\cpgvT{\pcp{\tytop[\cs{o}]}}}
     & = & \pgv{\ty{\tyofferemp[\cs{o}]}}
  \end{array}
\]

The translation $\tm{\cpgvM{\cdot}}$ translates processes to \emph{terms} and maps the $\pi$-calculus constructs from PCP to the corresponding constants in PGV:
\begin{align*}
   & \pcp{\tm{\cpgvM{\link{x}{y}}}}
   &                                    & = \pgv{\tm{\link\;{\pair{x}{y}}}}                                                      \\
   & \pcp{\tm{\cpgvM{\res{x}{y}{P}}}}
   &                                    & = \pgv{\tm{\letpair{x}{y}{\new\;\unit}{\cpgvM{P}}}}                                    \\
   & \pcp{\tm{\cpgvM{\ppar{P}{Q}}}}
   &                                    & = \pgv{\tm{\andthen{\spawn\;{(\lambda\unit.\cpgvM{P})}}{\cpgvM{Q}}}}                   \\
   & \pcp{\tm{\cpgvM{\halt}}}
   &                                    & = \pgv{\tm{\unit}}                                                                     \\
   & \pcp{\tm{\cpgvM{\close{x}{P}}}}
   &                                    & = \pgv{\tm{\andthen{\close\;{x}}{\cpgvM{P}}}}                                          \\
   & \pcp{\tm{\cpgvM{\wait{x}{P}}}}
   &                                    & = \pgv{\tm{\andthen{\wait\;{x}}{\cpgvM{P}}}}                                           \\
   & \pcp{\tm{\cpgvM{\send{x}{y}{P}}}}
   &                                    & = \pgv{\tm{\letpair{y}{z}{\new\;\unit}{\letbind{x}{\send\;{\pair{z}{x}}}{\cpgvM{P}}}}} \\
   & \pcp{\tm{\cpgvM{\recv{x}{y}{P}}}}
   &                                    & = \pgv{\tm{\letpair{y}{x}{\recv\;{x}}{\cpgvM{P}}}}                                     \\
   & \pcp{\tm{\cpgvM{\inl{x}{P}}}}
   &                                    & = \pgv{\tm{\letbind{x}{\select{\labinl}\;{x}}{\cpgvM{P}}}}                             \\
   & \pcp{\tm{\cpgvM{\inr{x}{P}}}}
   &                                    & = \pgv{\tm{\letbind{x}{\select{\labinr}\;{x}}{\cpgvM{P}}}}                             \\
   & \pcp{\tm{\cpgvM{\offer{x}{P}{Q}}}}
   &                                    & = \pgv{\tm{\offer{x}{x}{\cpgvM{P}}{x}{\cpgvM{Q}}}}                                     \\
   & \pcp{\tm{\cpgvM{\absurd{x}}}}
   &                                    & = \pgv{\tm{\offeremp{x}}}
\end{align*}

Unfortunately, the operational correspondence along $\tm{\cpgvM{\cdot}}$ is unsound, as it translates $\nu$-binders and parallel compositions to $\pgv{\tm{\new}}$ and $\pgv{\tm{\spawn}}$, which can reduce to their equivalent configuration constructs using \LabTirName{E-New} and \LabTirName{E-Spawn}. The same goes for $\nu$-binders which are inserted when translating bound send to unbound send. For instance, the process $\pcp{\tm{\send{x}{y}{P}}}$ is blocked, but its translation uses $\pgv{\tm{\new}}$ and can reduce.
To address this issue, we introduce a second translation, $\tm{\cpgvC{\cdot}}$, which is equivalent to translating with $\tm{\cpgvM{\cdot}}$ then reducing with \LabTirName{E-New} and \LabTirName{E-Spawn}:
\begin{align*}
   & \pcp{\tm{\cpgvC{\res{x}{y}{P}}}}
   &                                   & = \pgv{\tm{\res{x}{y}{\cpgvC{P}}}}
  \\
   & \pcp{\tm{\cpgvC{\ppar{P}{Q}}}}
   &                                   & = \pgv{\tm{\ppar{\cpgvC{P}}{\cpgvC{Q}}}}
  \\
   & \pcp{\tm{\cpgvC{\send{x}{y}{P}}}}
   &                                   & = \pgv{\tm{\res{y}{z}{(\child\;\letbind{x}{\send\;\pair{z}{x}}{\cpgvM{P}})}}}
  \\
   & \pcp{\tm{\cpgvC{\inl{x}{P}}}}
   &                                   & = \pgv{\tm{\res{y}{z}{(\child\;\letbind{x}{\andthen{\close\;(\send\;\pair{\inl{y}}{x})}{z}}{\cpgvM{P}})}}}
  \\
   & \pcp{\tm{\cpgvC{\inr{x}{P}}}}
   &                                   & = \pgv{\tm{\res{y}{z}{(\child\;\letbind{x}{\andthen{\close\;(\send\;\pair{\inr{y}}{x})}{z}}{\cpgvM{P}})}}}
  \\
   & \pcp{\tm{\cpgvC{P}}}
   &                                   & = \pgv{\tm{\child{\cpgvM{P}}}},\quad\text{if none of the above apply}
\end{align*}
Typing environments are translated pointwise, and sequents $\pcp{\seq{P}{\ty{\Gamma}}}$ are translated as $\pgv{\cseq[\child]{\ty{\cpgvT{\ty{\Gamma}}}}{\cpgvC{P}}}$, where $\tm{\pgv{\child}}$ indicates a child thread, since translated processes do not have a main thread.
The translations $\tm{\cpgvM{\cdot}}$ and $\tm{\cpgvC{\cdot}}$ preserve typing, and the latter induces a sound and complete operational correspondence.

\begin{lem}\label{lem:pcp-to-pgv-terms-preservation}
  If $\pcp{\seq{P}{\ty{\Gamma}}}$, then $\pgv{\tseq[\cs{p}]{\ty{\cpgvT{\Gamma}}}{\cpgvM{P}}{\tyunit}}$.
\end{lem}
\proof
\label{prf:lem-pcp-to-pgv-terms-preservation}
By induction on the derivation of $\pcp{\seq{P}{\ty{\Gamma}}}$.
\begin{case*}[\LabTirName{T-Link}, \LabTirName{T-Res}, \LabTirName{T-Par}, and \LabTirName{T-Halt}]
  See \cref{fig:pcp-to-pgv-preservation}.
\end{case*}
\begin{case*}[\LabTirName{T-Close}, and \LabTirName{T-Wait}]
  See \cref{fig:pcp-to-pgv-preservation-close-and-wait}.
\end{case*}
\begin{case*}[\LabTirName{T-Send}]
  See \cref{fig:pcp-to-pgv-preservation-send}.
\end{case*}
\begin{case*}[\LabTirName{T-Recv}]
  See \cref{fig:pcp-to-pgv-preservation-recv}.
\end{case*}
\begin{case*}[\LabTirName{T-Select-Inl}, \LabTirName{T-Select-Inr}, and \LabTirName{T-Offer}]
  See \cref{fig:pcp-to-pgv-preservation-select-and-offer}.
\end{case*}
\qed
\begin{figure}
  \begin{mathpar}
    \pcp{\inferrule*[lab=T-Link]{
      }{\seq{\link[\ty{A}]{x}{y}}{\tmty{x}{A}, \tmty{y}{\co{A}}}}}
    \cpgvMarrow
    \pgv{\inferrule*{
    \inferrule*{
    }{\tmty{\link}{\tylolli[]{\typrod{\cpgvT{A}}{\co{\cpgvT{A}}}}{\tyunit}}}
    \\
    \inferrule*{
      \inferrule*{
      }{\tseq[\cs{\pbot}]
        {\tmty{x}{\cpgvT{A}}}
        {x}
        {\cpgvT{A}}}
      \\
      \inferrule*{
      }{\tseq[\cs{\pbot}]
        {\tmty{y}{\co{\cpgvT{A}}}}
        {y}
        {\co{\cpgvT{A}}}}
    }{\tseq[\cs{\pbot}]
      {\tmty{x}{\cpgvT{A}},\tmty{y}{\co{\cpgvT{A}}}}
      {\pair{x}{y}}
      {\typrod{\cpgvT{A}}{\co{\cpgvT{A}}}}}
    }{\tseq[\cs{\pbot}]
    {\tmty{x}{\cpgvT{A}},\tmty{y}{\co{\cpgvT{A}}}}
    {\link\;{\pair{x}{y}}}
    {\tyunit}}}

    \pcp{\inferrule*[lab=T-Res]{
        \seq{P}{\ty{\Gamma},\tmty{x}{A},\tmty{y}{\co{A}}}
      }{\seq{\res{x}{y}{P}}{\ty{\Gamma}}}}
    \cpgvMarrow
    \pgv{\inferrule*{
    \inferrule*{
      \inferrule*{
      }{\tmty
        {\new}
        {\tylolli{\tyunit}{{\typrod{\cpgvT{A}}{\co{\cpgvT{A}}}}}}}
      \\
      \inferrule*{
      }{\tseq[\cs{\pbot}]
        {\emptyenv}
        {\unit}
        {\tyunit}}
    }{\tseq[\cs{\pbot}]
      {\emptyenv}
      {\new\;\unit}
      {\typrod{\cpgvT{A}}{\co{\cpgvT{A}}}}}
    \\
    {\tseq[\cs{p}]
    {\ty{\cpgvT{\Gamma}},\tmty{x}{\cpgvT{A}},\tmty{y}{\co{\cpgvT{A}}}}
    {\cpgvM{P}}
    {\tyunit}}
    }{\tseq[\cs{p}]
    {\ty{\cpgvT{\Gamma}}}
    {\letpair{x}{y}{\new\;\unit}{\cpgvM{P}}}
    {\tyunit}}}

    \pcp{\inferrule*[lab=T-Par]{
        \seq{\tm{P}}{\ty{\Gamma}}
        \\
        \seq{\tm{Q}}{\ty{\Delta}}
      }{\seq{\tm{\ppar{P}{Q}}}{\ty{\Gamma},\ty{\Delta}}}}
    \cpgvMarrow
    \pgv{\inferrule*{
    \inferrule*{
    \inferrule*{
    }{\tmty{\spawn}{\tylolli{(\tylolli[\cs{\pr{(\ty{\Gamma})}},\cs{p}]{\tyunit}{\tyunit})}{\tyunit}}}
    \\
    \inferrule*{
      \tseq[\cs{p}]
      {\ty{\cpgvT{\Gamma}}}
      {\tm{\cpgvM{P}}}
      {\tyunit}
    }{\tseq[\cs{\pbot}]
      {\ty{\cpgvT{\Gamma}}}
      {\lambda\unit.\cpgvM{P}}
      {\tylolli[\cs{\pr{(\ty{\Gamma})}},\cs{p}]{\tyunit}{\tyunit}}}
    }{\tseq[\cs{\pbot}]
    {\ty{\cpgvT{\Gamma}}}
    {\tm{\spawn\;{(\lambda\unit.\cpgvM{P})}}}
    {\tyunit}}
    \\
    \tseq[\cs{q}]
    {\ty{\cpgvT{\Delta}}}
    {\tm{\cpgvM{Q}}}
    {\tyunit}
    }{\tseq[\cs{q}]
    {\ty{\cpgvT{\Gamma}},\ty{\cpgvT{\Delta}}}
    {\andthen{\spawn\;{(\lambda\unit.\cpgvM{P})}}{\cpgvM{Q}}}
    {\tyunit}}}
    \\
    \pcp{\inferrule*[lab=T-Halt]{
      }{\seq{\tm{\halt}}{\emptyenv}}}
    \cpgvMarrow
    \pgv{\inferrule*{
      }{\tseq[\cs{\pbot}]{\emptyenv}{\unit}{\tyunit}}}
  \end{mathpar}
  \caption{Translation $\cpgvM{\cdot}$ preserves typing (\LabTirName{T-Link}, \LabTirName{T-Res}, \LabTirName{T-Par}, and \LabTirName{T-Halt}).}
  \label{fig:pcp-to-pgv-preservation}
\end{figure}
\begin{figure}
  \begin{mathpar}
    \pcp{\inferrule*[lab=T-Close]{
        \seq{\tm{P}}{\ty{\Gamma}}
        \\
        \cs{o}<\pr(\ty{\Gamma})
      }{\seq{\tm{\close{x}{P}}}{\ty{\Gamma},\tmty{x}{\tyone[\cs{o}]}}}}
    \cpgvMarrow
    \pgv{\inferrule*{
    \inferrule*{
    \inferrule*{
    }{\tmty{\close}{\tylolli[\cs{\ptop},\cs{o}]{\tyends[\cs{o}]}{\tyunit}}}
    \\
    \inferrule*{
    }{\tseq[\cs{\pbot}]
      {\tmty{x}{\tyends[\cs{o}]}}
      {x}
      {\tyends[\cs{o}]}}
    }{\tseq[\cs{o}]
    {\tmty{x}{\tyends[\cs{o}]}}
    {\close\;{x}}
    {\tyunit}}
    \\
    \tseq[\cs{p}]{\ty{\cpgvT{\Gamma}}}{\cpgvM{P}}{\tyunit}
    \\
    \cs{o}<\pr(\ty{\cpgvT{\Gamma}})
    }{\tseq[\cs{o}\sqcup\cs{p}]
    {\ty{\cpgvT{\Gamma}},\tmty{x}{\tyends[\cs{o}]}}
    {\andthen{\close\;{x}}{\cpgvM{P}}}
    {\tyunit}}}
    \\
    \pcp{\inferrule*[lab=T-Wait]{
        \seq{\tm{P}}{\ty{\Gamma}}
        \\
        \cs{o}<\pr(\ty{\Gamma})
      }{\seq{\tm{\wait{x}{P}}}{\ty{\Gamma},\tmty{x}{\tyone[\cs{o}]}}}}
    \cpgvMarrow
    \pgv{\inferrule*{
    \inferrule*{
    \inferrule*{
    }{\tmty{\wait}{\tylolli[\cs{\ptop},\cs{o}]{\tyendr[\cs{o}]}{\tyunit}}}
    \\
    \inferrule*{
    }{\tseq[\cs{\pbot}]
      {\tmty{x}{\tyendr[\cs{o}]}}
      {x}
      {\tyendr[\cs{o}]}}
    }{\tseq[\cs{o}]
    {\tmty{x}{\tyendr[\cs{o}]}}
    {\wait\;{x}}
    {\tyunit}}
    \\
    \tseq[\cs{p}]{\ty{\cpgvT{\Gamma}}}{\cpgvM{P}}{\tyunit}
    \\
    \cs{o}<\pr(\ty{\cpgvT{\Gamma}})
    }{\tseq[\cs{o}\sqcup\cs{p}]
    {\ty{\cpgvT{\Gamma}},\tmty{x}{\tyendr[\cs{o}]}}
    {\andthen{\wait\;{x}}{\cpgvM{P}}}
    {\tyunit}}}
  \end{mathpar}
  \caption{Translation $\cpgvM{\cdot}$ preserves typing (\LabTirName{T-Close} and \LabTirName{T-Wait}).}
  \label{fig:pcp-to-pgv-preservation-close-and-wait}
\end{figure}
\begin{figure}
  \begin{mathpar}
    \pcp{\inferrule*[lab=T-Send]{
        \seq{\tm{P}}{\ty{\Gamma},\tmty{y}{A},\tmty{x}{B}}
        \\
        \cs{o}<\pr(\ty{\Gamma},\ty{A},\ty{B})
      }{\seq{\tm{\send{x}{y}{P}}}{\ty{\Gamma},\tmty{x}{\tytens[\cs{o}]{A}{B}}}}}
    \cpgvMarrow
    \\
    \pgv{\inferrule*[lab=(a)]{
        \inferrule*{
        }{\tmty
          {\new}
          {\tylolli{\tyunit}{{\typrod{\cpgvT{A}}{\co{\cpgvT{A}}}}}}}
        \\
        \inferrule*{
        }{\tseq[\cs{\pbot}]
          {\emptyenv}
          {\unit}
          {\tyunit}}
      }{\tseq[\cs{\pbot}]
        {\emptyenv}
        {\new\;\unit}
        {\typrod{\cpgvT{A}}{\co{\cpgvT{A}}}}}}
    \\
    \pgv{\inferrule*[lab=(b)]{
    \inferrule*{
    }{\tmty{\send}{\tylolli[\cs{\ptop},\cs{o}]
        {\typrod{\co{\cpgvT{A}}}{\tysend[\cs{o}]{\co{\cpgvT{A}}}{\cpgvT{B}}}}
        {\cpgvT{B}}}}
    \\
    \inferrule*{
      \inferrule*{
      }{\tseq[\cs{\pbot}]
        {\tmty{z}{\co{\cpgvT{A}}}}
        {x}
        {\co{\cpgvT{A}}}}
      \\
      \inferrule*{
      }{\tseq[\cs{\pbot}]
        {\tmty{x}{\tysend[\cs{o}]{\co{\cpgvT{A}}}{\cpgvT{B}}}}
        {x}
        {\tysend[\cs{o}]{\co{\cpgvT{A}}}{\cpgvT{B}}}}
    }{\tseq[\cs{\pbot}]
      {\tmty{x}{\tysend[\cs{o}]{\co{\cpgvT{A}}}{\cpgvT{B}}},\tmty{z}{\co{\cpgvT{A}}}}
      {\pair{z}{x}}
      {\typrod{\co{\cpgvT{A}}}{\tysend[\cs{o}]{\co{\cpgvT{A}}}{\cpgvT{B}}}}}
    }{\tseq[\cs{o}]
    {\tmty{x}{\tysend[\cs{o}]{\co{\cpgvT{A}}}{\cpgvT{B}}},\tmty{z}{\co{\cpgvT{A}}}}
    {\send\;{\pair{z}{x}}}
    {\cpgvT{B}}}}
    \\
    \pgv{\inferrule*{
    \LabTirName{(a)}
    \\
    \inferrule*{
    \LabTirName{(b)}
    \\
    {\tseq[\cs{p}]
    {\ty{\cpgvT{\Gamma}},\tmty{y}{\cpgvT{A}},\tmty{x}{\cpgvT{B}}}
    {\cpgvM{P}}
    {\tyunit}}
    \\
    \cs{o}<\pr(\ty{\cpgvT{\Gamma}},\ty{\cpgvT{A}},\ty{\cpgvT{B}})
    }{\tseq[\cs{o}\sqcup\cs{p}]
    {\ty{\cpgvT{\Gamma}},%
    \tmty{x}{\tysend[\cs{o}]{\co{\cpgvT{A}}}{\cpgvT{B}}},%
    \tmty{y}{\cpgvT{A}},\tmty{z}{\co{\cpgvT{A}}}}
    {\letbind{x}{\send\;{\pair{z}{x}}}{\cpgvM{P}}}
    {\tyunit}}
    }{\tseq[\cs{o}\sqcup\cs{p}]
    {\ty{\cpgvT{\Gamma}},\tmty{x}{\tysend[\cs{o}]{\co{\cpgvT{A}}}{\cpgvT{B}}}}
    {\letpair{y}{z}{\new\;\unit}{\letbind{x}{\send\;{\pair{z}{x}}}{\cpgvM{P}}}}
    {\tyunit}}}
  \end{mathpar}
  \caption{Translation $\cpgvM{\cdot}$ preserves typing (\LabTirName{T-Send}).}
  \label{fig:pcp-to-pgv-preservation-send}
\end{figure}
\begin{figure}
  \begin{mathpar}
    \pcp{\inferrule*[lab=T-Recv]{
        \seq{P}{\ty{\Gamma},\tmty{y}{A},\tmty{x}{B}}
        \\
        \cs{o}<\pr(\ty{\Gamma},\ty{A},\ty{B})
      }{\seq{\recv{x}{y}{P}}{\ty{\Gamma},\tmty{x}{\typarr[\cs{o}]{A}{B}}}}}
    \cpgvMarrow
    \\
    \pgv{\inferrule*[lab=(a)]{
    \inferrule*{
    }{\tmty{\recv}{\tylolli[\cs{\ptop},\cs{o}]
        {\tyrecv[\cs{o}]{{\cpgvT{A}}}{\cpgvT{B}}}
        {\typrod{{\cpgvT{A}}}{\cpgvT{B}}}}}
    \\
    \inferrule*{
    }{\tseq[\cs{\pbot}]
      {\tmty{x}{\tyrecv[\cs{o}]{{\cpgvT{A}}}{\cpgvT{B}}}}
      {x}
      {\tyrecv[\cs{o}]{{\cpgvT{A}}}{\cpgvT{B}}}}
    }{\tseq[\cs{o}]
    {\tmty{x}{\tyrecv[\cs{o}]{{\cpgvT{A}}}{\cpgvT{B}}}}
    {\recv\;{x}}{\typrod{\cpgvT{A}}{\cpgvT{B}}}}}
    \\
    \pgv{\inferrule*{
    \LabTirName{(a)}
    \\
    {\tseq[\cs{p}]
    {\ty{\cpgvT{\Gamma}},\tmty{y}{\cpgvT{A}},\tmty{x}{\cpgvT{B}}}
    {\cpgvM{P}}
    {\tyunit}}
    \\
    \cs{o}<\pr(\ty{\cpgvT{\Gamma}},\ty{\cpgvT{A}},\ty{\cpgvT{B}})
    }{\tseq[\cs{o}\sqcup\cs{p}]
    {\ty{\cpgvT{\Gamma}},%
      \tmty{x}{\tyrecv[\cs{o}]{{\cpgvT{A}}}{\cpgvT{B}}},\tmty{y}{\cpgvT{A}},\tmty{z}{{\cpgvT{A}}}}
    {\letbind{x}{\recv{x}}{\cpgvM{P}}}{\tyunit}}}
  \end{mathpar}
  \caption{Translation $\cpgvM{\cdot}$ preserves typing (\LabTirName{T-Recv}).}
  \label{fig:pcp-to-pgv-preservation-recv}
\end{figure}
\begin{figure}
  \begin{mathpar}
    \pcp{\inferrule*[lab=T-Select-Inl]{
        \seq{\tm{P}}{\ty{\Gamma},\tmty{x}{A}}
        \\
        \cs{o}<\pr(\ty{\Gamma})
      }{\seq{\inl{x}{P}}{\ty{\Gamma},\tmty{x}{\typlus[\cs{o}]{A}{B}}}}}
    \cpgvMarrow
    \pgv{\inferrule*{
    \inferrule*{
    \inferrule*{
    }{\tmty
      {\select{\labinl}}
      {\tylolli[\cs{\ptop},\cs{o}]{\tyselect[\cs{o}]{\cpgvT{A}}{\cpgvT{B}}}{\cpgvT{A}}}}
    \\
    \inferrule*{
    }{\tseq[\cs{\pbot}]
      {\tmty{x}{\tyselect[\cs{o}]{\cpgvT{A}}{\cpgvT{B}}}}
      {x}
      {\tyselect[\cs{o}]{\cpgvT{A}}{\cpgvT{B}}}}
    }{\tseq[\cs{o}]
    {\tmty{x}{\tyselect[\cs{o}]{\cpgvT{A}}{\cpgvT{B}}}}
    {\select{\labinl}\;{x}}
    {\cpgvT{A}}}
    \\
    {\tseq[\cs{p}]
    {\ty{\Gamma},\tmty{x}{\cpgvT{A}}}
    {\cpgvM{P}}
    {\tyunit}}
    \\
    \cs{o}<\pr(\ty{\Gamma})
    }{\tseq[\cs{o}\sqcup\cs{p}]
    {\ty{\Gamma},\tmty{x}{\tyselect[\cs{o}]{\cpgvT{A}}{\cpgvT{B}}}}
    {\letbind{x}{\select{\labinl}\;{x}}{\cpgvM{P}}}
    {\tyunit}}}
    \\
    \pcp{\inferrule*[lab=T-Select-Inr]{
        \seq{\tm{P}}{\ty{\Gamma},\tmty{x}{A}}
        \\
        \cs{o}<\pr(\ty{\Gamma})
      }{\seq{\inr{x}{P}}{\ty{\Gamma},\tmty{x}{\typlus[\cs{o}]{A}{B}}}}}
    \cpgvMarrow
    \pgv{\inferrule*{
    \inferrule*{
    \inferrule*{
    }{\tmty
      {\select{\labinr}}
      {\tylolli[\cs{\ptop},\cs{o}]{\tyselect[\cs{o}]{\cpgvT{A}}{\cpgvT{B}}}{\cpgvT{B}}}}
    \\
    \inferrule*{
    }{\tseq[\cs{\pbot}]
      {\tmty{x}{\tyselect[\cs{o}]{\cpgvT{A}}{\cpgvT{B}}}}
      {x}
      {\tyselect[\cs{o}]{\cpgvT{A}}{\cpgvT{B}}}}
    }{\tseq[\cs{o}]
    {\tmty{x}{\tyselect[\cs{o}]{\cpgvT{A}}{\cpgvT{B}}}}
    {\select{\labinr}\;{x}}
    {\cpgvT{B}}}
    \\
    {\tseq[\cs{p}]
    {\ty{\Gamma},\tmty{x}{\cpgvT{B}}}
    {\cpgvM{P}}
    {\tyunit}}
    \\
    \cs{o}<\pr(\ty{\Gamma})
    }{\tseq[\cs{o}\sqcup\cs{p}]
    {\ty{\Gamma},\tmty{x}{\tyselect[\cs{o}]{\cpgvT{A}}{\cpgvT{B}}}}
    {\letbind{x}{\select{\labinr}\;{x}}{\cpgvM{P}}}
    {\tyunit}}}
    \\
    \pcp{\inferrule*[lab=T-Offer]{
        \seq{P}{\ty{\Gamma},\tmty{x}{A}}
        \\
        \seq{Q}{\ty{\Gamma},\tmty{x}{B}}
        \\
        \cs{o}<\pr(\ty{\Gamma},\ty{A},\ty{B})
      }{\seq{\offer{x}{P}{Q}}{\ty{\Gamma},\tmty{x}{\tywith[\cs{o}]{A}{B}}}}}
    \cpgvMarrow
    \pgv{\inferrule*{
        \inferrule*{
        }{\tseq[\cs{\pbot}]
          {\tmty{x}{\tyoffer[\cs{o}]{\cpgvT{A}}{\cpgvT{B}}}}
          {x}
          {\tyoffer[\cs{o}]{\cpgvT{A}}{\cpgvT{B}}}}
        \\
        \tseq[\cs{p}]{\ty{\cpgvT{\Gamma}},\tmty{x}{\cpgvT{A}}}{\cpgvM{P}}{\tyunit}
        \\
        \tseq[\cs{p}]{\ty{\cpgvT{\Gamma}},\tmty{x}{\cpgvT{B}}}{\cpgvM{Q}}{\tyunit}
        \\
        \cs{o}<\pr(\ty{\cpgvT{\Gamma}},\ty{\cpgvT{A}},\ty{\cpgvT{B}})
      }{\tseq[\cs{o}\sqcup\cs{p}]
        {\ty{\cpgvT{\Gamma}},\tmty{x}{\tyoffer[\cs{o}]{\cpgvT{A}}{\cpgvT{B}}}}
        {\offer{x}{x}{\cpgvM{P}}{x}{\cpgvM{Q}}}
        {\tyunit}}}
  \end{mathpar}
  \caption{Translation $\cpgvM{\cdot}$ preserves typing (\LabTirName{T-Select-Inl}, \LabTirName{T-Select-Inr}, and \LabTirName{T-Offer}).}
  \label{fig:pcp-to-pgv-preservation-select-and-offer}
\end{figure}


\begin{thm}\label{thm:pcp-to-pgv-confs-preservation}
  If $\pcp{\seq{P}{\ty{\Gamma}}}$, then $\pgv{\cseq[\child]{\ty{\cpgvT{\Gamma}}}{\cpgvC{P}}}$.
\end{thm}
\begin{proof}
\label{prf:thm-pcp-to-pgv-confs-preservation}
By induction on the derivation of $\pcp{\seq{P}{\ty{\Gamma}}}$.

\begin{case*}[\LabTirName{T-Res}]
  Immediately, from the induction hypothesis.
  \begin{mathpar}
    \pcp{\inferrule*[lab=T-Res]{
        \seq{\ty{\Gamma},\tmty{x}{A},\tmty{y}{\co{A}}}{\tm{P}}
      }{\seq{\ty{\Gamma}}{\tm{\res{x}{y}{P}}}}}
    \cpgvCarrow
    \pgv{\inferrule*{
        \cseq[\child]{\ty{\cpgvT{\Gamma}},\tmty{x}{\cpgvT{A}},\tmty{y}{\cpgvT{B}}}{\tm{\cpgvC{P}}}
      }{\cseq[\child]{\ty{\cpgvT{\Gamma}}}{\tm{\res{x}{y}{\cpgvC{P}}}}}}
  \end{mathpar}
\end{case*}
\begin{case*}[\LabTirName{T-Par}]
  Immediately, from the induction hypotheses.
  \begin{mathpar}
    \pcp{\inferrule*[lab=T-Par]{
        \seq{\ty{\Gamma}}{\tm{P}}
        \\
        \seq{\ty{\Delta}}{\tm{Q}}
      }{\seq{\ty{\Gamma},\ty{\Delta}}{\ppar{P}{Q}}}}
    \cpgvCarrow
    \pgv{\inferrule*{
        \cseq[\child]{\ty{\cpgvT{\Gamma}}}{\tm{\cpgvC{P}}}
        \\
        \cseq[\child]{\ty{\cpgvT{\Delta}}}{\tm{\cpgvC{Q}}}
      }{\cseq[\child]{\ty{\cpgvT{\Gamma}},\ty{\cpgvT{\Delta}}}{\tm{\ppar{\cpgvC{P}}{\cpgvC{Q}}}}}}
  \end{mathpar}
\end{case*}
\begin{case*}[\LabTirName{*}]
  By \cref{lem:pcp-to-pgv-terms-preservation}
  \[
    \pcp{\seq{\ty{\Gamma}}{\tm{P}}}
    \cpgvCarrow
    \pgv{\inferrule*{
        \tseq[\cs{p}]{\ty{\cpgvT{\Gamma}}}{\tm{\cpgvM{P}}}{\tyunit}
      }{\cseq[\child]{\ty{\cpgvT{\Gamma}}}{\tm{\child\;\cpgvM{P}}}}}
  \tag*{\qedhere}
  \]
\end{case*}
\end{proof}


\begin{thm}\label{thm:pcp-to-pgv-operational-correspondence-soundness}
  If $\pcp{\seq{P}{\ty{\Gamma}}}$ and $\pgv{\tm{\cpgvC{P}}\cred\tm{\conf{C}}}$, there exists a $\tm{Q}$ such that $\pcp{\tm{P}\red^+\tm{Q}}$ and $\pgv{\tm{\conf{C}}\cred^\star\tm{\cpgvC{Q}}}$
\end{thm}
\begin{proof}
\label{prf:thm-pcp-to-pgv-operational-correspondence-soundness}
By induction on the derivation of $\pgv{\tm{\cpgvC{P}}\cred\tm{\conf{C}}}$.
We omit the cases which cannot occur as their left-hand side term forms are not in the image of the translation function, \ie \LabTirName{E-New}, \LabTirName{E-Spawn}, and \LabTirName{E-LiftM}.

\begin{case*}[\LabTirName{E-Link}]
  \[\pgv{%
      \tm{\res{x}{x'}{(\ppar{\plug{\conf{F}}{\link\;\pair{w}{x}}}{\conf{C}})}}
      \cred
      \tm{\ppar{\plug{\conf{F}}{\unit}}{\subst{\conf{C}}{w}{x'}}}
    }\]
  The source for $\pgv{\tm{\link\;\pair{w}{x}}}$ \emph{must} be $\pcp{\tm{\link{w}{x}}}$. None of the translation rules introduce an evaluation context around the recursive call, hence $\pgv{\tm{\conf{F}}}$ must be the empty context. Let $\pcp{\tm{P}}$ be the source term for $\pgv{\tm{\conf{C}}}$, \ie $\pgv{\tm{\cpgvC{P}}=\tm{\conf{C}}}$. Hence, we have:
  \begin{mathpar}
    \begin{tikzcd}[cramped, column sep=tiny]
      \pcp{\tm{\res{x}{x'}{(\ppar{\link{w}{x}}{P})}}}
      \arrow[r, "\pcp{\red}"]
      \arrow[d, "\cpgvC{\cdot}"]
      &
      \pcp{\tm{\subst{P}{w}{x'}}}
      \arrow[dd, "\cpgvC{\cdot}"]
      \\
      \pgv{\tm{\res{x}{x'}{(\ppar{\child\;\link\;\pair{w}{x}}{\cpgvC{P}})}}}
      \arrow[d, "\pgv{\cred^+}"]
      \\
      \pgv{\tm{\subst{\cpgvC{P}}{w}{x'}}}
      \arrow[r, "\pcp{=}"]
      &
      \pgv{\tm{\cpgvC{\subst{P}{w}{x'}}}}
    \end{tikzcd}
  \end{mathpar}
\end{case*}
\begin{case*}[\LabTirName{E-Send}]
  \[\pgv{%
    \tm{\res{x}{x'}{(\ppar{\plug{\conf{F}}{\send\;{\pair{V}{x}}}}{\plug{\conf{F'}}{\recv\;{x'}}})}}
    \cred
    \tm{\res{x}{x'}{(\ppar{\plug{\conf{F}}{x}}{\plug{\conf{F'}}{\pair{V}{x'}}})}}
    }\]
  There are three possible sources for $\pgv{\tm{\send}}$ and $\pgv{\tm{\recv}}$: $\pcp{\tm{\send{x}{y}{P}}}$ and $\pcp{\tm{\recv{x'}{y'}{Q}}}$; $\pcp{\tm{\inl{x}{P}}}$ and $\pcp{\tm{\offer{x'}{Q}{R}}}$; or $\pcp{\tm{\inr{x}{P}}}$ and $\pcp{\tm{\offer{x'}{Q}{R}}}$.
  \begin{subcase*}[$\pcp{\tm{\send{x}{y}{P}}}$ and $\pcp{\tm{\recv{x'}{y'}{Q}}}$]
    None of the translation rules introduce an evaluation context around the recursive call, hence $\pgv{\tm{\conf{F}}}$ must be $\pgv{\tm{\child\;\letbind{x}{\hole}{\cpgvM{P}}}}$. Similarly, $\pgv{\tm{\conf{F'}}}$ must be $\pgv{\tm{\child\;\letpair{y'}{x'}{\hole}{\cpgvM{Q}}}}$. The value $\pgv{\tm{V}}$ must be an endpoint $\tm{y}$, bound by the name restriction $\pgv{\tm{\res{y}{y'}{}}}$ introduced by the translation. Hence, we have:
    \begin{mathpar}
      \begin{tikzcd}[cramped, column sep=tiny]
        \pcp{\tm{\res{x}{x'}{(\ppar{\send{x}{y}{P}}{\recv{x'}{y'}{Q}})}}}
        \arrow[d, "\cpgvC{\cdot}"]
        \arrow[r, "\pcp{\red}"]
        &
        \pcp{\tm{\res{x}{x'}{\res{y}{y'}{(\ppar{P}{Q})}}}}
        \arrow[dd, "\cpgvC{\cdot}"]
        \\
        \pgv{\tm{\res{x}{x'}{\res{y}{y'}{}}\left(
            \begin{array}{l}
                \child\;\letbind{x}{\send\;{\pair{y}{x}}}{\cpgvM{P}}
                \parallel
                \\
                \child\;\letpair{y'}{x'}{\recv\;{x'}}{\cpgvM{Q}}
              \end{array}
            \right)}}
        \arrow[d, "\pgv{\equiv\cred^+}"]
        \\
        \pgv{\tm{\res{x}{x'}{\res{y}{y'}{(\ppar{\child\;\cpgvM{P}}{\child\;\cpgvM{Q}})}}}}
        \arrow[r, "\pgv{\cred^\star}", "\text{(by \cref{lem:pcp-to-pgv-cpgvM-to-cpgvC})}"']
        &
        \pgv{\tm{\res{x}{x'}{\res{y}{y'}{(\ppar{\cpgvC{P}}{\cpgvC{Q}})}}}}
      \end{tikzcd}
    \end{mathpar}
  \end{subcase*}
  \begin{subcase*}[$\pcp{\tm{\inl{x}{P}}}$ and $\pcp{\tm{\offer{x'}{Q}{R}}}$]
    None of the translation rules introduce an evaluation context around the recursive call, hence $\pgv{\tm{\conf{F}}}$ must be $$\pgv{\tm{\child\;\letbind{x}{\andthen{\close\;\hole}{y}}{\cpgvM{P}}}}.$$ Similarly, $\pgv{\tm{\conf{F'}}}$ must be $$\pgv{\tm{\child\;\letpair{y'}{x'}{\hole}{\andthen{\wait\;x'}{\casesum{y'}{y'}{\cpgvM{Q}}{y'}{\cpgvM{R}}}}}}.$$ Hence, we have:
    \begin{mathpar}
      \begin{tikzcd}[cramped, column sep=tiny]
        \pcp{\tm{\res{x}{x'}{(\ppar{\inl{x}{P}}{\offer{x}{Q}{R}})}}}
        \arrow[d, "\cpgvM{\cdot}"]
        \arrow[r, "\pcp{\red}"]
        &
        \pcp{\tm{\res{x}{x'}{(\ppar{P}{Q})}}}
        \arrow[dd, "\cpgvC{\cdot}"]
        \\
        \pgv{\tm{\res{x}{x'}{}\left(
            \begin{array}{l}
                \child\;\letbind{x}{\select{\labinl}\;{x}}{\cpgvM{P}}
                \parallel
                \\
                \child\;\offer{x'}{x'}{\cpgvM{Q}}{x'}{\cpgvM{R}}
              \end{array}
            \right)}}
        \arrow[d, "\pgv{\cred^+}"]
        \\
        \pgv{\tm{\res{x}{x'}{(\ppar{\child\;\cpgvM{P}}{\child\;\cpgvM{Q}})}}}
        \arrow[r, "\pgv{\cred^\star}", "\text{(by \cref{lem:pcp-to-pgv-cpgvM-to-cpgvC})}"']
        &
        \pgv{\tm{\res{x}{x'}{(\ppar{\cpgvC{P}}{\cpgvC{Q}})}}}
      \end{tikzcd}
    \end{mathpar}
  \end{subcase*}
  \begin{subcase*}[$\pcp{\tm{\inr{x}{P}}}$ and $\pcp{\tm{\offer{x'}{Q}{R}}}$]
    None of the translation rules introduce an evaluation context around the recursive call, hence $\pgv{\tm{\conf{F}}}$ must be $$\pgv{\tm{\child\;\letbind{x}{\andthen{\close\;\hole}{y}}{\cpgvM{P}}}}.$$ Similarly, $\pgv{\tm{\conf{F'}}}$ must be $$\pgv{\tm{\child\;\letpair{y'}{x'}{\hole}{\andthen{\wait\;x'}{\casesum{y'}{y'}{\cpgvM{Q}}{y'}{\cpgvM{R}}}}}}.$$ Hence, we have:
    \begin{mathpar}
      \begin{tikzcd}[cramped, column sep=tiny]
        \pcp{\tm{\res{x}{x'}{(\ppar{\inr{x}{P}}{\offer{x}{Q}{R}})}}}
        \arrow[d, "\cpgvM{\cdot}"]
        \arrow[r, "\pcp{\red}"]
        &
        \pcp{\tm{\res{x}{x'}{(\ppar{P}{Q})}}}
        \arrow[dd, "\cpgvC{\cdot}"]
        \\
        \pgv{\tm{\res{x}{x'}{}\left(
            \begin{array}{l}
                \child\;\letbind{x}{\select{\labinr}\;{x}}{\cpgvM{P}}
                \parallel
                \\
                \child\;\offer{x'}{x'}{\cpgvM{Q}}{x'}{\cpgvM{R}}
              \end{array}
            \right)}}
        \arrow[d, "\pgv{\cred^+}"]
        \\
        \pgv{\tm{\res{x}{x'}{(\ppar{\child\;\cpgvM{P}}{\child\;\cpgvM{R}})}}}
        \arrow[r, "\pgv{\cred^\star}", "\text{(by \cref{lem:pcp-to-pgv-cpgvM-to-cpgvC})}"']
        &
        \pgv{\tm{\res{x}{x'}{(\ppar{\cpgvC{P}}{\cpgvC{R}})}}}
      \end{tikzcd}
    \end{mathpar}
  \end{subcase*}
\end{case*}
\begin{case*}[\LabTirName{E-Close}]
  \[\pgv{%
    \tm{\res{x}{x'}{(\ppar{\plug{\conf{F}}{\wait\;{x}}}{\plug{\conf{F'}}{\close\;{x'}}})}}
    \cred
    \tm{\ppar{\plug{\conf{F}}{\unit}}{\plug{\conf{F'}}{\unit}}}
    }\]
  The source for $\pgv{\tm{\wait}}$ and $\pgv{\tm{\close}}$ \emph{must} be $\pcp{\tm{\wait{x}{P}}}$ and $\pcp{\tm{\close{x'}{Q}}}$.

  (The translation for $\pcp{\tm{\offer{x}{P}{Q}}}$ also introduces a $\pgv{\tm{\wait}}$, but it is blocked on another communication, and hence cannot be the first communication on a translated term. The translations for $\pcp{\tm{\inl{x}{P}}}$ and $\pcp{\tm{\inr{x}{P}}}$ also introduce a $\pgv{\tm{\close}}$, but these are similarly blocked.)

  None of the translation rules introduce an evaluation context around the recursive call, hence $\pgv{\tm{\conf{F}}}$ must be $\pgv{\tm{\andthen{\hole}{\cpgvM{P}}}}$. Similarly, $\pgv{\tm{\conf{F'}}}$ must be $\pgv{\tm{\andthen{\hole}{\cpgvM{Q}}}}$. Hence, we have:
  \begin{mathpar}
    \begin{tikzcd}
      \pcp{\tm{\res{x}{x'}{(\ppar{\close{x}{P}}{\wait{x'}{Q}})}}}
      \arrow[d, "\cpgvM{\cdot}"]
      \arrow[r, "\pcp{\red}"]
      &
      \pcp{\tm{\ppar{P}{Q}}}
      \arrow[dd, "\cpgvC{\cdot}"]
      \\
      \pgv{\tm{\res{x}{x'}{(\ppar
      {\child\;\andthen{\close\;{x}}{\cpgvM{P}}}
      {\child\;\andthen{\wait\;{x'}}{\cpgvM{Q}}}
      )}}}
      \arrow[d, "\pgv{\cred^+}"]
      \\
      \pgv{\tm{\ppar{\child\;\cpgvM{P}}{\child\;\cpgvM{Q}}}}
      \arrow[r, "\pgv{\cred^\star}", "\text{(by \cref{lem:pcp-to-pgv-cpgvM-to-cpgvC})}"']
      &
      \pgv{\tm{\ppar{\cpgvC{P}}{\cpgvC{Q}}}}
    \end{tikzcd}
  \end{mathpar}
\end{case*}
\begin{case*}[\LabTirName{E-LiftC}]
  By the induction hypothesis and \LabTirName{E-LiftC}.
\end{case*}
\begin{case*}[\LabTirName{E-LiftSC}]
  By the induction hypothesis, \LabTirName{E-LiftSC},
  and \cref{lem:pcp-to-pgv-confs-operational-correspondence-equiv}.
  \qedhere
\end{case*}
\end{proof}


\begin{lem}\label{lem:pcp-to-pgv-cpgvM-to-cpgvC}
  For any $\pcp{\tm{P}}$, either:
  \begin{itemize}
    \item $\pgv{\tm{\child\;\cpgvM{P}}=\tm{\cpgvC{P}}}$; or
    \item $\pgv{\tm{\child\;\cpgvM{P}}\cred^+\tm{\cpgvC{P}}}$, and for any $\pgv{\tm{\conf{C}}}$, if $\pgv{\tm{\child\;\cpgvM{P}}\cred\tm{\conf{C}}}$, then $\pgv{\tm{\conf{C}}\cred^\star\tm{\cpgvC{P}}}$.
  \end{itemize}
\end{lem}
\begin{proof}
\label{prf:lem-pcp-to-pgv-cpgvM-to-cpgvC}
By induction on the structure of $\pcp{\tm{P}}$.

\begin{case*}[$\pcp{\tm{\res{x}{y}{P}}}$]
  We have:
  \begin{mathpar}
    \begin{array}{lrll}
      \pcp{\tm{\cpgvM{\res{x}{y}{P}}}}
       & =
       & \pgv{\tm{\child\;\letpair{x}{y}{\new\;\unit}{\cpgvM{P}}}}
      \\
       & \pgv{\cred^+}
       & \pgv{\tm{\res{x}{y}{(\child\;\cpgvM{P})}}}
      \\
       & \pgv{\cred^\star}
       & \pgv{\tm{\res{x}{y}{\cpgvC{P}}}}
      \\
       & =
       & \pcp{\tm{\cpgvC{\res{x}{y}{P}}}}
    \end{array}
  \end{mathpar}
\end{case*}
\begin{case*}[$\pcp{\tm{\ppar{P}{Q}}}$]
  \begin{mathpar}
    \begin{array}{lrll}
      \pcp{\tm{\cpgvM{\ppar{P}{Q}}}}
       & =
       & \pgv{\tm{\child\;\andthen{\spawn\;(\lambda\unit.\cpgvM{P})}{\cpgvM{Q}}}}
      \\
       & \pgv{\cred^+}
       & \pgv{\tm{\ppar{\child\;\cpgvM{P}}{\child\;\cpgvM{Q}}}}
      \\
       & \pgv{\cred^\star}
       & \pgv{\tm{\ppar{\cpgvC{P}}{\cpgvC{Q}}}}
      \\
       & =
       & \pcp{\tm{\cpgvC{\ppar{P}{Q}}}}
    \end{array}
  \end{mathpar}
\end{case*}
\begin{case*}[$\pcp{\tm{\send{x}{y}{P}}}$]
  \begin{mathpar}
    \begin{array}{lrll}
      \pcp{\tm{\cpgvM{\send{x}{y}{P}}}}
       & =
       & \pgv{\tm{\letpair{y}{z}{\new\;\unit}{\letbind{x}{\send\;{\pair{z}{x}}}{\cpgvM{P}}}}}
      \\
       & \pgv{\cred^+}
       & \pgv{\tm{\res{y}{z}{(\child\;\letbind{x}{\send\;\pair{z}{x}}{\cpgvM{P}})}}}
      \\
       & =
       & \pcp{\tm{\cpgvC{\send{x}{y}{P}}}}
    \end{array}
  \end{mathpar}
\end{case*}
\begin{case*}[$\pcp{\tm{\inl{x}{P}}}$]
  \begin{mathpar}
    \begin{array}{lrll}
      \pcp{\tm{\cpgvM{\send{x}{y}{P}}}}
       & =
       & \pgv{\tm{\letbind{x}{\select{\labinl}\;{x}}{\cpgvM{P}}}}
      \\
       & \elabarrow
       & \pgv{\tm{\letbind{x}{\letpair{y}{z}{\new\;\unit}{\andthen{\close\;(\send\;{\pair{\inl{y}}{x}})}{z}}}{\cpgvM{P}}}}
      \\
       & \pgv{\cred^+}
       & \pgv{\tm{\res{y}{z}{(\child\;\letbind{x}{\andthen{\close\;(\send\;\pair{\inl{y}}{x})}{z}}{\cpgvM{P}})}}}
      \\
       & =
       & \pcp{\tm{\cpgvC{\send{x}{y}{P}}}}
    \end{array}
  \end{mathpar}
\end{case*}
\begin{case*}[$\pcp{\tm{\inr{x}{P}}}$]
  \begin{mathpar}
    \begin{array}{lrll}
      \pcp{\tm{\cpgvM{\send{x}{y}{P}}}}
       & =
       & \pgv{\tm{\letbind{x}{\select{\labinr}\;{x}}{\cpgvM{P}}}}
      \\
       & \elabarrow
       & \pgv{\tm{\letbind{x}{\letpair{y}{z}{\new\;\unit}{\andthen{\close\;(\send\;{\pair{\inr{y}}{x}})}{z}}}{\cpgvM{P}}}}
      \\
       & \pgv{\cred^+}
       & \pgv{\tm{\res{y}{z}{(\child\;\letbind{x}{\andthen{\close\;(\send\;\pair{\inr{y}}{x})}{z}}{\cpgvM{P}})}}}
      \\
       & =
       & \pcp{\tm{\cpgvC{\send{x}{y}{P}}}}
    \end{array}
  \end{mathpar}
\end{case*}
\begin{case*}[$*$]
  In all other cases, $\pgv{\tm{\child\;\cpgvM{P}}=\tm{\cpgvC{P}}}$.
  \qedhere
\end{case*}
\end{proof}


\begin{lem}\label{lem:pcp-to-pgv-confs-operational-correspondence-equiv}
  If $\pcp{\seq{P}{\ty{\Gamma}}}$ and $\pcp{\tm{P}\equiv\tm{Q}}$,
  then $\pgv{\tm{\cpgvC{P}}\equiv\tm{\cpgvC{Q}}}$.
\end{lem}
\proof
Every axiom of the structural congruence in PCP maps directly to the axiom of the same name in PGV.
\qed

\begin{thm}\label{thm:pcp-to-pgv-operational-correspondence-completeness}
  \hfill\\
  If $\pcp{\seq{P}{\ty{\Gamma}}}$ and $\pcp{\tm{P}\red\tm{Q}}$,
  then $\pgv{\tm{\cpgvC{P}}\cred^+\tm{\cpgvC{Q}}}$.
\end{thm}
\begin{proof}
\label{prf:thm-pcp-to-pgv-operational-correspondence-completeness}
By induction on the derivation of $\pcp{\tm{P}\red\tm{Q}}$.

\begin{case*}[\LabTirName{E-Link}]
  \begin{mathpar}
    \begin{tikzcd}[cramped]
      \pcp{\tm{\res{x}{x'}{(\ppar{\link{w}{x}}{P})}}}
      \arrow[r, "\pcp{\red}"]
      \arrow[d, "\cpgvC{\cdot}"]
      &
      \pcp{\tm{\subst{P}{w}{x'}}}
      \arrow[dd, "\cpgvC{\cdot}"]
      \\
      \pgv{\tm{\res{x}{x'}{(\ppar{\child\;\link\;\pair{w}{x}}{\cpgvC{P}})}}}
      \arrow[d, "\pgv{\cred^+}"]
      \\
      \pgv{\tm{\subst{\cpgvC{P}}{w}{x'}}}
      \arrow[r, "\pcp{=}"]
      &
      \pgv{\tm{\cpgvC{\subst{P}{w}{x'}}}}
    \end{tikzcd}
  \end{mathpar}
\end{case*}
\begin{case*}[\LabTirName{E-Send}]
  \begin{mathpar}
    \begin{tikzcd}
      \pcp{\tm{\res{x}{x'}{(\ppar{\send{x}{y}{P}}{\recv{x'}{y'}{Q}})}}}
      \arrow[d, "\cpgvM{\cdot}"]
      \arrow[r, "\pcp{\red}"]
      &
      \pcp{\tm{\res{x}{x'}{\res{y}{y'}{(\ppar{P}{Q})}}}}
      \arrow[dd, "\cpgvC{\cdot}"]
      \\
      \pgv{\tm{
          \setlength{\arraycolsep}{0pt}
          \res{x}{x'}{}\left(
          \begin{array}{l}
              \child\bigg(
              \begin{array}{l}
                  \letpair{y}{y'}{\new\;\unit}{}
                  \\
                  \letbind{x}{\send\;{\pair{y}{x}}}{\cpgvM{P}}
                \end{array}
              \bigg)
              \parallel
              \\
              \child\hphantom{\bigg(}
              \letpair{y'}{x'}{\recv\;{x'}}{\cpgvM{Q}}
            \end{array}
          \right)}}
      \arrow[d, "\pgv{\cred^+}"]
      \\
      \pgv{\tm{\res{x}{x'}{\res{y}{y'}{(\ppar{\child\;\cpgvM{P}}{\child\;\cpgvM{Q}})}}}}
      \arrow[r, "\pgv{\cred^\star}", "\text{(by \cref{lem:pcp-to-pgv-cpgvM-to-cpgvC})}"']
      &
      \pgv{\tm{\res{x}{x'}{\res{y}{y'}{(\ppar{\cpgvC{P}}{\cpgvC{Q}})}}}}
    \end{tikzcd}
  \end{mathpar}
\end{case*}
\begin{case*}[\LabTirName{E-Close}]
  \begin{mathpar}
    \begin{tikzcd}
      \pcp{\tm{\res{x}{x'}{(\ppar{\close{x}{P}}{\wait{x'}{Q}})}}}
      \arrow[d, "\cpgvM{\cdot}"]
      \arrow[r, "\pcp{\red}"]
      &
      \pcp{\tm{\ppar{P}{Q}}}
      \arrow[dd, "\cpgvC{\cdot}"]
      \\
      \pgv{\tm{\res{x}{x'}{(\ppar
      {\child\;\andthen{\close\;{x}}{\cpgvM{P}}}
      {\child\;\andthen{\wait\;{x'}}{\cpgvM{Q}}}
      )}}}
      \arrow[d, "\pgv{\cred^+}"]
      \\
      \pgv{\tm{\ppar{\child\;\cpgvM{P}}{\child\;\cpgvM{Q}}}}
      \arrow[r, "\pgv{\cred^\star}", "\text{(by \cref{lem:pcp-to-pgv-cpgvM-to-cpgvC})}"']
      &
      \pgv{\tm{\ppar{\cpgvC{P}}{\cpgvC{Q}}}}
    \end{tikzcd}
  \end{mathpar}
\end{case*}
\begin{case*}[\LabTirName{E-Select-Inl}]
  \begin{mathpar}
    \begin{tikzcd}
      \pcp{\tm{\res{x}{x'}{(\ppar{\inl{x}{P}}{\offer{x}{Q}{R}})}}}
      \arrow[d, "\cpgvM{\cdot}"]
      \arrow[r, "\pcp{\red}"]
      &
      \pcp{\tm{\res{x}{x'}{(\ppar{P}{Q})}}}
      \arrow[dd, "\cpgvC{\cdot}"]
      \\
      \pgv{\tm{\res{x}{x'}{}\left(
          \begin{array}{l}
              \child\;\letbind{x}{\select{\labinl}\;{x}}{\cpgvM{P}}
              \parallel
              \\
              \child\;\offer{x'}{x'}{\cpgvM{Q}}{x'}{\cpgvM{R}}
            \end{array}
          \right)}}
      \arrow[d, "\pgv{\cred^+}"]
      \\
      \pgv{\tm{\res{x}{x'}{(\ppar{\child\;\cpgvM{P}}{\child\;\cpgvM{Q}})}}}
      \arrow[r, "\pgv{\cred^\star}", "\text{(by \cref{lem:pcp-to-pgv-cpgvM-to-cpgvC})}"']
      &
      \pgv{\tm{\res{x}{x'}{(\ppar{\cpgvC{P}}{\cpgvC{Q}})}}}
    \end{tikzcd}
  \end{mathpar}
\end{case*}
\begin{case*}[\LabTirName{E-Select-Inr}]
  \begin{mathpar}
    \begin{tikzcd}
      \pcp{\tm{\res{x}{x'}{(\ppar{\inr{x}{P}}{\offer{x}{Q}{R}})}}}
      \arrow[d, "\cpgvM{\cdot}"]
      \arrow[r, "\pcp{\red}"]
      &
      \pcp{\tm{\res{x}{x'}{(\ppar{P}{R})}}}
      \arrow[dd, "\cpgvC{\cdot}"]
      \\
      \pgv{\tm{\res{x}{x'}{}\left(
          \begin{array}{l}
              \child\;\letbind{x}{\select{\labinr}\;{x}}{\cpgvM{P}}
              \parallel
              \\
              \child\;\offer{x'}{x'}{\cpgvM{Q}}{x'}{\cpgvM{R}}
            \end{array}
          \right)}}
      \arrow[d, "\pgv{\cred^+}"]
      \\
      \pgv{\tm{\res{x}{x'}{(\ppar{\child\;\cpgvM{P}}{\child\;\cpgvM{R}})}}}
      \arrow[r, "\pgv{\cred^\star}", "\text{(by \cref{lem:pcp-to-pgv-cpgvM-to-cpgvC})}"']
      &
      \pgv{\tm{\res{x}{x'}{(\ppar{\cpgvC{P}}{\cpgvC{R}})}}}
    \end{tikzcd}
  \end{mathpar}
\end{case*}
\begin{case*}[\LabTirName{E-LiftRes}]
  By the induction hypothesis and \LabTirName{E-LiftC}.
\end{case*}
\begin{case*}[\LabTirName{E-LiftPar}]
  By the induction hypotheses and \LabTirName{E-LiftC}.
\end{case*}
\begin{case*}[\LabTirName{E-LiftSC}]
  By the induction hypothesis, \LabTirName{E-LiftSC}, and \cref{lem:pcp-to-pgv-confs-operational-correspondence-equiv}.
  \qedhere
\end{case*}
\end{proof}


\endgroup

\section{Milner's Cyclic Scheduler}
\label{sec:milner}
As an example of a deadlock-free cyclic process, Dardha and Gay~\cite{dardhagay18extended} introduce an implementation of Milner's cyclic scheduler~\cite{milner89} in Priority CP. We reproduce that scheduler here, and show its translation to Priority GV.

\begingroup
\usingnamespace{pcp}
\begin{exa}\label{ex:pcp-scheduler}
  A set of processes $\tm{\proc{i}}$, $1\leq{i}\leq{n}$, is scheduled to perform some tasks in cyclic order, starting with $\tm{\proc1}$, ending with $\tm{\proc{n}}$, and notifying $\tm{\proc1}$ when all processes have finished.

  Our scheduler $\tm{\sched}$ consists of set of agents $\tm{\agent{i}}$, $1\leq{i}\leq{n}$, each representing their respective process. Each process $\tm{\proc{i}}$ waits for the signal to start their task on $\tm{a'_i}$, and signals completion on $\tm{b'_i}$. Each agent signals their process to start on $\tm{a_i}$, waits for their process to finish on $\tm{b_i}$, and then signals for the next agent to continue on $\tm{c_i}$. The agent $\tm{\agent1}$ initiates, then waits for every other process to finish, and signals $\tm{\proc1}$ on $\tm{d}$. Every other agent $\tm{\agent{i}}$, $2\leq{i}\leq{n}$ waits on $\tm{c'_{i-1}}$ for the signal to start. Each of the channels in the scheduler is of a terminated type, and is merely used to synchronise.

  Below is a diagram of our scheduler instantiated with three processes:
  
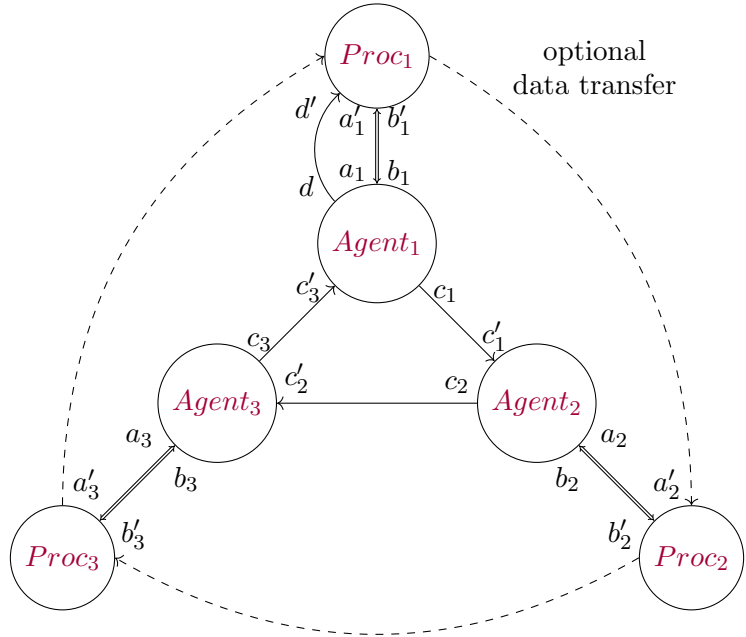
\begin{center}
    \begin{tikzpicture}
      \node[shape=circle,draw=black]                       (agent1) {\tm{\agent1}};
      \node[shape=circle,draw=black,below right=of agent1] (agent2) {\tm{\agent2}};
      \node[shape=circle,draw=black,below left =of agent1] (agent3) {\tm{\agent3}};

      \node[shape=circle,draw=black,above      =of agent1] (proc1) {\tm{\proc1}};
      \node[shape=circle,draw=black,below right=of agent2] (proc2) {\tm{\proc2}};
      \node[shape=circle,draw=black,below left =of agent3] (proc3) {\tm{\proc3}};

      \draw[->]
      (agent1) -- node[pos=0.35,above] {$c_1$}
      node[pos=1.0,above] {$c'_1$} ++ (agent2);
      \draw[->]
      (agent2) -- node[pos=0.1,above] {$c_2$}
      node[pos=0.9,above] {$c'_2$} ++ (agent3);
      \draw[->]
      (agent3) -- node[pos=0.0,above] {$c_3$}
      node[pos=0.65,above] {$c'_3$} ++ (agent1);

      \draw[implies-implies,double]
      (agent1) -- node[pos=0.15,left] {$a_1$} node[pos=0.85,left] {$a'_1$}
      node[pos=0.15,right] {$b_1$} node[pos=0.85,right] {$b'_1$} ++ (proc1);
      \draw[implies-implies,double]
      (agent2) -- node[pos=0.15,above right] {$a_2$} node[pos=0.85,above right] {$a'_2$}
      node[pos=0.15,below left] {$b_2$} node[pos=0.85,below left] {$b'_2$} ++ (proc2);
      \draw[implies-implies,double]
      (agent3) -- node[pos=0.15,above left] {$a_3$} node[pos=0.85,above left] {$a'_3$}
      node[pos=0.15,below right] {$b_3$} node[pos=0.85,below right] {$b'_3$} ++ (proc3);

      \draw[->]
      (agent1) to[bend left=45] node[pos=0.15,left] {$d$} node[pos=0.85,left] {$d'$} (proc1);

      \draw[->,dashed]
      (proc1.east) to[bend left] node[pos=0.15,align=center,above right] {optional\\data transfer} (proc2.north);
      \draw[->,dashed]
      (proc2.west) to[bend left] (proc3.east);
      \draw[->,dashed]
      (proc3.north) to[bend left] (proc1.west);
    \end{tikzpicture}
  \end{center}

  We implement the scheduler as follows, using $\tm{\prod_{I}P_i}$ to denote the parallel composition of the processes $\tm{P_i}$, $\tm{i}\in\tm{I}$, and $\tm{\plug{P}{Q}}$ to denote the plugging of $\tm{Q}$ in the one-hole process-context $\tm{P}$. The process-contexts $\tm{P_i}$ represent the computations performed by each process $\tm{\proc{i}}$. The process-contexts $\tm{Q_i}$ represent any post-processing, and any possible data transfer from $\tm{\proc{i}}$ to $\tm{\proc{i+1}}$. Finally, $\tm{Q_1}$ should contain $\tm{\labwait{d'}}$.
  \[
    \begin{array}{lrlr}
      \tm{\sched}
       & \defeq & \tm{\res{a_1}{a'_1}{\dots{\res{a_n}{a'_n}{}}}}
      \tm{\res{b_1}{b'_1}{\dots{\res{b_n}{b'_n}{}}}}
      \tm{\res{c_1}{c'_1}{\dots{\res{c_n}{c'_n}{}}}}
      \tm{\res{d}{d'}{}}
      \\ &     & \tm{(
        \ppar{\proc1}{\agent1}
        \parallel
        \prod_{2\leq{i}\leq{n}}(\ppar{\proc{i}}{\wait{c'_{i-1}}{\agent{i}}})
        )}
      \\
      \tm{\agent1}
       & \defeq & \tm{\close{a_i}{\wait{b_i}{\close{c_i}{\wait{c'_n}{\close{d}{\halt}}}}}}
      \\
      \tm{\agent{i}}
       & \defeq & \tm{\close{a_i}{\wait{b_i}{\close{c_i}{\halt}}}}
      \\
      \tm{\proc{i}}
       & \defeq & \tm{\wait{a'_i}{\plug{P_i}{\close{b'_i}{Q_i}}}}
    \end{array}
  \]
\end{exa}
\endgroup

\begingroup
\usingnamespace{pgv}
\begin{exa}\label{ex:pgv-scheduler}
  The PGV scheduler has exactly the same behaviour as the PCP version in~\cref{ex:pcp-scheduler}. It is implemented as follows, using $\tm{\prod_{I}\conf{C}_i}$ to denote the parallel composition of the processes $\tm{\conf{C}_i}$, $\tm{i}\in\tm{I}$, and $\tm{\plug{M}{N}}$ to denote the plugging of $\tm{N}$ in the one-hole term-context $\tm{M}$. For simplicity, we let $\tm{\sched}$ be a configuration. The terms $\tm{M_i}$ represent the computations performed by each process $\tm{\proc{i}}$. The terms $\tm{N_i}$ represent any post-processing, and any possible data transfer from $\tm{\proc{i}}$ to $\tm{\proc{i+1}}$. Finally, $\tm{N_1}$ should contain $\tm{\wait\;{d'}}$.
  \[
    \begin{array}{lrlr}
      \tm{\sched}
       & \defeq & \tm{\res{a_1}{a'_1}{\dots{\res{a_n}{a'_n}{}}}}
      \tm{\res{b_1}{b'_1}{\dots{\res{b_n}{b'_n}{}}}}
      \tm{\res{c_1}{c'_1}{\dots{\res{c_n}{c'_n}{}}}}
      \tm{\res{d}{d'}{}}
      \\ &     & \tm{\begin{array}{lll}
          (
           & \phi\;\proc1
          \parallel
          \child\;\andthen{\agent1}{\andthen{\wait\;{c'_n}}{\close\;{d}}}
          \\
          \parallel
           &
          \prod_{2\leq{i}\leq{n}}
          (\child\;\proc{i} \parallel \child\;\andthen{\wait\;{c'_{i-1}}}{\agent{i}})
           & )
        \end{array}}
      \\
      \tm{\agent{i}}
       & \defeq & \tm{\andthen{\close\;{a_i}}{\andthen{\wait\;{b_i}}{\close\;{c_i}}}}
      \\
      \tm{\proc{i}}
       & \defeq & \tm{\andthen{\wait\;{a'_i}}{\plug{M_i}{\andthen{\close\;{b'_i}}{N_i}}}}
    \end{array}
  \]
\end{exa}
\endgroup

If $\pcp{\tm{\cpgvM{P_i}}}=\pgv{\tm{M_i}}$ and $\pcp{\tm{\cpgvM{Q_i}}}=\pgv{\tm{N_i}}$, then the translation of $\pcp{\tm{\sched}}$ (\cref{ex:pcp-scheduler}), $\pcp{\tm{\cpgvC{\sched\,}}}$, is exactly $\pgv{\tm{\sched}}$ (\cref{ex:pgv-scheduler}).

}
{\section{Related Work and Conclusion}
\usingnamespace{pcp}

\subsubsection*{Deadlock freedom and progress}
Deadlock freedom and progress are well studied properties in the $\pi$-calculus.
For the ``standard'' typed $\pi$-calculus---types for channels used in input and output, an important line of work stems from Kobayashi's approach to deadlock freedom~\cite{kobayashi98}, where priorities are values from an abstract poset. Kobayashi~\cite{kobayashi06} simplifies the abstract poset to pairs of naturals, called \emph{obligations} and \emph{capabilities}. Padovani simplifies these further to a single natural, called a \emph{priority}~\cite{padovani14}, and adapts obligations/capabilities to session types~\cite{padovani13}. Later work by Kobayashi and co-authors \cite{GiachinoKL14,kobayashilaneve17} address deadlock detection for a value-passing CCS (a predecessor of the $\pi$-calculus) where the number of nodes in a network is arbitrary, namely modelling unbounded networks. The authors define a sound inference algorithm for their type system, which guarantees deadlock freedom for these more complex kinds of communication networks. This type system is more expressive than previous work by Kobayashi.

For the session-typed $\pi$-calculus, an important line of work stems from Dezani and co-authors. In their work, Dezani~\etal~\cite{dezani-ciancaglinimostrous06} guarantee progress by allowing only one active session at a time. As sessions do not interleave, consequently they do not deadlock. Later on, Dezani \etal~\cite{dezani-ciancagliniliguoro09progress} introduce a partial order on channels, similar to Kobayashi~\cite{kobayashi98} and produce a type system for progress.

Carbone and Debois~\cite{carbonedebois10} define progress for session-typed $\pi$-calculus in terms of a \emph{catalyser}, which provides the missing counterpart to a process. Intuitively,  either the process is deadlock-free in which case the catalyser is structurally congruent to the inaction process, or if the process is ``stuck'' for \eg, an input process $\tm{\recv{x}{y}{\halt}}$ (using CP syntax) then the catalyser simply provides the missing output required for communication. Carbone~\etal~\cite{carbonedardha14} use catalysers to show that progress is a compositional form of livelock freedom and can be lifted to session types via the encoding of session types to linear types~\cite{kobayashi07,dardhagiachino12,dardha14beat,dardhaetal17}. In this work, Carbone \etal \cite{carbonedardha14} sistematise and compare different notions of liveness properties: progress, deadlock freedom and livelock freedom. Their technique of using the encoding of session types combined with Kobayashi's type systems for deadlock and livelock freedom allows for a more flexible deadlock/livelock detection. It is worth noting that livelock freedom is a stronger property than deadlock freedom in the presence of recursion, as the former also discards useless divergent processes; for finite processes the two properties coincide.

Vieira and Vasconcelos~\cite{vieiravasconcelos13} use single priorities and an abstract partial order to guarantee deadlock freedom in a binary session-typed $\pi$-calculus and building on conversation types, which is akin to session types.

While our work focuses on {binary} session types, it is worth discussing related work on Multiparty Session Types (MPST), which describe communication among multiple agents in a distributed setting. The line of work on MPST starts with Honda~\etal~\cite{hondayoshida08}, which guarantees deadlock freedom {within a single session} by design, but the property does not hold for session interleaving.
Bettini~\etal~\cite{bettinicoppo08} follow a technique similar to Kobayashi's for MPST.
The main difference with our work is that we associate priorities with communication actions, where Bettini~\etal~\cite{bettinicoppo08} associate them with channels.
Coppo \etal \cite{coppoetal13,coppoetal16} present a deterministic, sound and complete, and compositional type inference algorithm for an \emph{interaction type system}, which is used to guarantee global progress for processes in a calculus based on asynchronous as well as dynamically interleaved and interfered multiparty sessions. The interaction type system allows to infer \emph{causalities}---much in the line of Kobayashi's priorities ---of channels, thus guaranteeing that session-typed processes do not get stuck at intermediate stages of their sessions.
Carbone and Montesi~\cite{carbonemontesi13} combine MPST with choreographic programming and obtain a formalism that satisfies deadlock freedom. In the same vein as MPST, choreographic programming specifies communication among all participants in a distributed system. While MPST target mainly protocol descriptions, choreographies have mainly targetted implementations and programming languages as they are suitable for describing concrete system implementations.
Deni\'{e}lou and Yoshida~\cite{DenielouY13} introduce \emph{multiparty compatibility}, which generalises the notion of duality in binary session types. They synthesise safe and deadlock-free \emph{global types}--specifying communication among {all} involved participants, from \emph{local session types}--specifying communication from the viewpoint of {one} participant. To do so, they leverage Labelled Transition Systems (LTSs) and communicating automata.
Castellani~\etal~\cite{CastellaniDGH20} guarantee livelock freedom, a stronger property than deadlock freedom, for MPST with {internal delegation}, where participants in the same session are allowed to delegate tasks to each other, and internal delegation is captured by the global type.
Scalas and Yoshida~\cite{scalasyoshida19} provide a revision of the foundations for MPST, and offer a less complicated and more general theory, by removing duality/consistency. The type systems is parametric and type checking is decidable, but allows for a novel integration of model checking techniques. More protocols and processes can be typed and are guaranteed to be free of deadlocks.

Neubauer and Thiemann~\cite{neubauert04} and Vasconcelos~\etal~\cite{vasconcelosravara04,vasconcelosgay06} introduce the first functional language with session types. Such works did not guarantee deadlock freedom until GV~\cite{lindleymorris15,wadler14}.
Toninho~\etal~\cite{toninhocaires12} present a translation of simply-typed $\lambda$-calculus into session-typed $\pi$-calculus, but their focus is not on deadlock freedom.
Fowler \etal \cite{fowleretal21} present Hypersequent GV (HGV), which is a variation of GV that uses hyper-environments, much in the same line as Hypersequent CP, and enjoys deadlock freedom, confluence, and strong normalisation.

\subsubsection*{Ties with logic}
The correspondence between logic and types lays the foundation for functional programming~\cite{wadler15}.
Since its inception by Girard~\cite{girard87}, linear logic has been a candidate for a foundational correspondence for concurrent programs.
A~correspondence with linear $\pi$-calculus was established early on by Abramsky~\cite{abramsky94} and Bellin and Scott~\cite{bellinscott94}. Many years later, several correspondences between linear logic and the $\pi$-calculus with binary session types were proposed. Caires and Pfenning~\cite{cairespfenning10} propose a correspondence with dual intuitionistic linear logic, while Wadler~\cite{wadler12} proposes a correspondence with classical linear logic. Both works guarantee deadlock freedom as a consequence of adopting cut elimination.
Building on a previous work \cite{cairespfenning10}, Toninho \etal \cite{toninhoetal13} present a Curry-Howard correspondence between session types and linear logic for functional language via linear contextual monads, which are first-class values, thus giving rise to a higher-order session-typed  language. In addition to the more standard results, the authors also prove a global progress theorem.
Qian \etal \cite{QianKB21} extend Classical Linear Logic with co-exponentials, which allows to model servers receiving requests from an arbitrary set of clients, yielding an extension to the Curry-Howard correspondence between logic and session typed processes.
Dardha and Gay~\cite{dardhagay18extended} define Priority CP by integrating Kobayashi and Padovani's work on priorities~\cite{kobayashi06,padovani14} with CP, which as described in the introduction, weakens its ties to linear logic in exchange for expressivity. However, they show how PCP can be also {viewed as} a an extension of linear logic, which they call Priority Linear Logic (PLL), and uses mix and cycle rules as opposed to the cut rule.
Dardha and P\'{e}rez~\cite{dardhaperez15extended,dardhaperez15,DardhaP22} compare priorities \`a la Kobayashi with tree restrictions \`a la CP, and show that the latter is a subsystem of the former. In addition, they give a detailed account of comparing several type systems for deadlock freedom spanning across session types, linear logic, and linear types.
Carbone~\etal~\cite{CarboneMSY15,carbonelindley16} give a logical view of MPST with a generalised duality.
Caires and P\'{e}rez~\cite{CairesP16} give a presentation of MPST in terms of binary session types and the use of a \emph{medium process} which guarantee protocol fidelity and deadlock freedom. Their binary session types are rooted in linear logic.
Ciobanu and Horne~\cite{CiobanuH15} give the first Curry-Howard correspondence between MPST and BV~\cite{Guglielmi07}, a conservative extension of linear logic with a non-commutative operator for sequencing.
Horne~\cite{Horne20} give a system for subtyping and multiparty compatibility where compatible processes are race free and deadlock free using a Curry-Howard correspondence, similar to the approach in~\cite{CiobanuH15}.
Balzer~\etal~\cite{balzerpfenning17} introduce sharing at the cost of deadlock freedom, which they restore using \emph{worlds}, an approach similar to priorities~\cite{balzertoninho19}.
Staying on sharing, Rocha and Caires \cite{rochacaires21} introduce an imperative feature, that of shared mutable states into a functional language based on Curry-Howard correspondence with linear logic. Their type system is thus able to capture programs which were not possible in previous works. The authors prove extensive technical results, including session fidelity, progress, confluence and
normalisation.
Lastly, Jacobs \etal \cite{JacobsBK22a} present a novel technique to guarantee deadlock freedom based on the notion of connectivity graph, which is an abstract representation of the topology of concurrent systems, and separation logic used to substructurally treat connectivity graph edges and labels.

\subsubsection*{Conclusion and Future Work}
We answered our research question by presenting Priority GV, a~session-typed functional language which allows cyclic communication structures and uses priorities to ensure deadlock freedom. We showed its relation to Priority CP~\cite{dardhagay18extended} via an operational correspondence.

Our formalism so far only captures the core of GV. In future work, we plan to explore recursion, following Lindley and Morris~\cite{lindleymorris16} and Padovani and Novara~\cite{padovaninovara15}, and sharing, following Balzer and Pfenning~\cite{balzerpfenning17} or Voinea~\etal~\cite{VoineaDG19}.

\subsubsection*{Acknowledgements}
The authors would like to thank the anonymous reviewers for their detailed feedback, which helped produce a more complete and polished work.
Also, the authors would like to thank Simon Fowler, April Gon\c{c}alves, and Philip Wadler for their comments on the manuscript.
}

\bibliographystyle{alphaurl}
\bibliography{main}

\end{document}